\documentclass[aps, pra, 10 pt, onecolumn, superscriptaddress, nofootinbib, longbibliography]{revtex4-2}
\usepackage{amsmath, amssymb, amsthm, mathtools, mathdots, dsfont, bm}
\usepackage{braket}
\usepackage{graphicx, epsfig, epstopdf,  placeins, caption, subcaption}
\usepackage[T1]{fontenc}
\usepackage[utf8]{inputenc}
\usepackage[english]{babel}
\usepackage{enumitem}
\usepackage{varwidth}
\usepackage[colorlinks=true, linkcolor=blue, citecolor=blue, urlcolor=blue]{hyperref}
\usepackage{orcidlink} 
 \usepackage{tikz}
\usetikzlibrary{arrows.meta, positioning, shapes.geometric, decorations.pathreplacing, shapes.misc, decorations.pathmorphing}
\usepackage{quantikz}
\usepackage{circuitikz}
 \usepackage{rotating}
 \usepackage{textcomp}
  \usepackage{comment}
 \usepackage[autostyle]{csquotes}
 \usepackage[titletoc,toc,title]{appendix}
 \usepackage{silence}
\makeatletter
\let\NR@oldlabel\label
\def\label#1{\NR@oldlabel{#1}}
\makeatother

\usepackage{placeins}
\FloatBarrier

\begin{document}
 \title{Decoherence Mitigation with Local NOT Gates in Multipartite Systems}

\author{Venkat Abhignan\,\orcidlink{0000-0002-6092-3670}}
\affiliation{Qdit Labs Pvt. Ltd., Bengaluru - 560092, India.}

\author{Raghav Sundararaman\,\orcidlink{0009-0002-2182-5776}}
\affiliation{Indian Institute of Technology Hyderabad, Sangareddy - 502284, Telangana, India.}
\affiliation{Qdit Labs Pvt. Ltd., Bengaluru - 560092, India.}

\author{Shriram Pragash M\,\orcidlink{0009-0003-2110-1118}}
\affiliation{Indian Institute of Technology Hyderabad, Sangareddy - 502284, Telangana, India.}
\affiliation{Qdit Labs Pvt. Ltd., Bengaluru - 560092, India.}

\author{R. Srikanth\,\orcidlink{0000-0001-7581-2546}}
\affiliation{Poornaprajna Institute of Scientific Research, Bengaluru - 562110, India.}

\author{Ashutosh Singh\,\orcidlink{0000-0001-9628-6127}}
\email[Corresponding author: ]{asinghrri@gmail.com}
\affiliation{Department of Physics and Astronomy and Institute of Quantum Science and Technology, University of Calgary, Calgary, AB, T2N 1N4, Canada.}

\begin{abstract} 
We study the entanglement dynamics of $n=2,3,4$‐qubit Bell- and GHZ‐type states under an amplitude‐damping channel (ADC).  
We quantify multipartite entanglement using the genuine multipartite concurrence (GMC) and evaluate its utility through the optimal teleportation fidelity. 
For $2$-qubit states, we analyze the standard (Bennett) teleportation protocol. For $3$- and $4$-qubit states, we study controlled quantum teleportation (CQT) with one and two \emph{controllers}, respectively. Entanglement sudden death (ESD) denotes the abrupt, finite-time disappearance of entanglement caused by decoherence in contrast to asymptotic decay. 
To counteract ESD, we apply local NOT ($\hat\sigma_x$) operations on $m$ of the $n$ qubits ($m \leq n$) and derive analytic formulae, revealing that a single-NOT operation often suffices to alter ESD into asymptotic decay when handling GMC. 
In contrast, teleportation fidelity can decay more rapidly for single-NOT flipped states, whereas flipping all qubits is more useful for preserving teleportation fidelity in certain regimes, highlighting that the amount of entanglement alone does not guarantee teleportation utility. 
Remarkably, in the case of GHZ-type states, ADC-evolved mixed biseparable states can be exploited successfully in the CQT protocol. 
Further, using the GHZ-symmetric parametrization, we map the 2- and 3-qubit ADC-evolved mixed states onto a $(x,y)$ plane, revealing their SLOCC (Stochastic Local Operations and Classical Communication) entanglement classes.
We also explicitly check the Bell–CHSH nonlocality hierarchy in the 2-qubit teleportation alongside localizable-entanglement diagnostics for 3-qubit CQT. 
Our results clarify the distinct roles of global versus localizable bipartite correlations and suggest simple, experimentally accessible unitary controls for preserving useful quantum resources in noisy channels.
\end{abstract}
\date{\today}
\maketitle

 \section{Introduction}
 \label{Sec_Introduction}

 One of the fundamental features of quantum theory and a key resource in quantum information tasks is entanglement \cite{QEnt_Horodecki2009}. It underpins quantum advantages in teleportation \cite{Teleport_Bennett1993}, computation \cite{Book_NielsenChuang2010, QComp_Sergey2018}, communication \cite{QCom_Cleve1997}, cryptography \cite{RNG-Bell_Pironio2010}, and metrology \cite{QMetrology_Demkowicz2014}. The durability of entanglement under decoherence and its robustness against environmental noise are crucial for the practical success of quantum information tasks. Environment-induced decoherence, causing finite-time disentanglement or ESD, is a well-known phenomenon \cite{ESD_Yu2004, ESD-PRL_Yu2006, ESD-OptCom_Yu2006, ESD_Yu2009} that has been experimentally demonstrated in atomic \cite{ESD_Laurat2007}, photonic \cite{ESD_Almeida2007}, and superconducting circuits \cite{PhysRevB.86.064502}, trapped-ion registers \cite{Barreiro2010}, solid-state spins in diamond (NV centers) \cite{PhysRevB.98.064306}, and NMR quantum processors \cite{Filgueiras2012}. Thus, techniques that preserve entanglement against the detrimental effects of decoherence are useful for quantum information tasks.
Decoherence-free subspaces \cite{DecohFreeSS-QComp_Lidar1998, DecohFreeSS_Kwiat2000, DecohFree-QM_Kielpinski2001, Noiseless-QIP_Viola2001}, quantum error correction \cite{QComp-Decoh_Shor1995, ErrorCorr_Steane1996}, dynamical decoupling \cite{DD_Viola1999, DD-QM_Biercuk2009, DD_Du2009}, the quantum Zeno effect \cite{DD-QZeno_Facchi2004, QZeno_Maniscalco2008, EntRevive_Oliveira2008}, quantum measurement reversal \cite{WM_Kim2009, QMR_Lee2011, WM_Sun2010, QMR_Kim2012, QMR_Lim2014, ESDM_Venkat2025}, and delayed-choice decoherence suppression \cite{DelayedChoice_Lee2014} are some of the methods that have been proposed for mitigating decoherence. Experimental evidence of the preservation of entanglement under ADC has been shown by quantum measurement reversal \cite{QMR_Kim2012, QMR_Lim2014}, and delayed-choice decoherence suppression \cite{DelayedChoice_Lee2014} protocols. However, both of these strategies are probabilistic, and a major drawback is that as the weak interaction strength increases, the probability of decoherence suppression (quantum measurement reversal) decreases. 

Beyond these cumbersome methods, entanglement manipulation using a local NOT operation has been proposed and discussed for two-qubit \cite{ESDM_Rau2008, ESDM_Ali2008, ESDM_Singh2017, ESDM-Thermal_Hunza2025} and higher-dimensional systems \cite{ESDM_Singh2022}. 
Entanglement preservation against a variant of ADC using local NOT operations has been experimentally demonstrated in a recent work \cite{ESDM_Behera2025}. The local NOT operation is a standard single-qubit primitive experimentally realizable as a resonant $\pi$-pulse ($X$ gate) in superconducting, trapped-ion, NV centers and atomic platforms, and as a half-wave plate or on-chip polarization rotator in photonic implementations. 
Scaling to multi-qubit NOT operations requires only site-selective $\pi$-pulses applied to the chosen qubits with pulse durations short compared to the amplitude-damping timescale. 
Quantitatively, typical control times validate the instantaneous-gate idealization where single-qubit $\pi$-pulse duration are typically $\tau_\pi\sim\!20$--$100\,$ns in superconducting transmons \cite{Barends2014} and are therefore $\gtrsim10^{3}$–$10^{4}$ times shorter than typical energy-relaxation times $T_1\sim 10$–$100\,\mu$s observed in the platform \cite{PhysRevB.86.100506, Kjaergaard2019}. 
In trapped-ion systems, single-qubit rotations are slower ($\sim 1$–$10\,\mu$s) but coherence and relaxation times are extremely long ($\sim$ seconds), so $\pi$-pulses remain effectively instantaneous relative to amplitude-damping timescales \cite{Harty2014, Wang2017}. 
For NV centers $\pi$-pulses of order $10$–$100\,$ns are commonly used while $T_1$ ranges from ms to s depending on environment, again giving $T_1/\tau_{\pi}\sim 10^{3}$–$10^{6}$ \cite{Jelezko2006, Toyli2012, Bar-Gill2013}. 
These orders-of-magnitude separations ($T_1/\tau_\pi\sim10^3-10^6$) justify treating local NOT operations ($\pi$-pulses) as instantaneous in our ADC model to leading order; finite-duration corrections can be included perturbatively if required for a specific experimental implementation.

Qubits of an entangled state subjected to an ADC lose population from the excited state to the ground state, which can lead to ESD in certain cases \cite{MultipartyESD_Xie2023}. For example, the two-qubit entangled state: $\alpha|00\rangle+\beta|11\rangle$, with $|\alpha|^2+|\beta|^2=1$, exhibits ESD when $|\beta|>|\alpha|$, whereas it undergoes asymptotic decay when $|\alpha|\geq|\beta|$ \cite{ESD_Yu2009}. We exploit this asymmetry to devise a strategy for protecting entanglement.
As shown previously in theoretical studies \cite{ESDM_Rau2008, ESDM_Singh2017, ESDM_Singh2022} and experimental demonstration \cite{ESDM_Behera2025}, inserting a local NOT gate within the ADC evolution (treated here as two sequential and independent ADC processes) effectively swaps the populations of the ground and excited states ($|0\rangle_g \leftrightarrow |1\rangle_e$), thereby modifying the disentanglement dynamics.
In practice, this population redistribution slows the depletion of coherence: the NOT operation transfers the remaining ground-state population into the excited subspace and vice versa, on which the second ADC subsequently acts. Consequently, the entanglement decay trajectory is altered. Depending on when the NOT operation is applied, parameterized by the damping strength accumulated during the first ADC stage, ESD can be hastened, delayed, or even completely avoided.

 Measures of entanglement for two-qubit states \cite{EntMeas_Hill1997, Wootters98, EntMeas_Vidal1999} have been examined exhaustively. 
 Beyond bipartite states, when three or more qubits are in the Greenberger-Horne-Zeilinger (GHZ) state or W-state \cite{ThreeQubitGME_Dur2000}, they are said to be genuinely multipartite entangled. Numerous applications such as distributed quantum computation \cite{CompPower_DHondt2006, QComp_Raussendorf2001}, multi-party quantum communication \cite{QSS_Hillery1999, MPQKD_Zhu2015, MPEnt_Yunfei2018}, and quantum sensor networks \cite{MPEnt-ClockSync_Ren2012, QSensorNet_Eldredge2018, QATelescope_Khabiboulline2019, QSensorNet_Qian2021} extend the bipartite paradigm and require multipartite entanglement. 
Despite the successful experimental generation of multipartite entanglement \cite{MPEnt_Yunfei2018, MPEnt_Friis2018, MPEntDetect_Saggio2019}, attempts to 
quantify entanglement in multipartite systems continue to face challenges \cite{TripartiteEntMeas_Xie2021}. For a measure to qualify as a genuine multipartite entanglement (GME) measure, it must satisfy the following two criteria: (a) for all product (fully separable) and bi-separable states, the measure must be zero; and (b) for all nonseparable states (GHZ-class and W-class in the three-qubit instance), the measure must be positive. We implement such a GME measure discussed by Hashemi et al., known as GMC \cite{GMC_Hashemi2012, GME_Ma2011}, which was recently found equivalent to another GME measure called ``Concurrence fill''  \cite{TripartiteEntMeas_Xie2021, MultipartyESD_Xie2023} for the mixed GHZ state.

Quantum teleportation \cite{Teleport_Bennett1993} is a communication protocol that enables the transfer of an unknown quantum state between two remote parties without physically transmitting the particle encoding the quantum information, utilizing preshared entanglement. In the context of GME states, a CQT protocol has been proposed ~\cite{CQT_Karlsson1998, CQT_Barasinski2018, CQT-PRA_Barasinski2019, CQT-PRL_Barasinski2019, CQT_Gangopadhyay2022, CQT_Rahmawati2019, CQT_Kumar2020, CQT-HD_Lv2024}, where third-party \emph{controllers} possessing one or more qubits of the entangled state (three-qubit GHZ or n-qubit GHZ state) influence the success of the teleportation process between two parties, Alice and Bob. Another related application employing a three-qubit GHZ state is quantum secret sharing~\cite{QSS_Hillery1999}. While several works have only employed CQT as the primitive, a recent work has implemented shared-secret teleportation via GHZ-encoded channels~\cite{Teleport-QSS_Lee2020}. These schemes essentially enable secure and distributed quantum information transfer without centralizing the complete information at any single location, thereby enhancing both scalability and security.

In this work, our scope is twofold: (i) to study the dynamics of GMC and teleportation fidelity of GHZ-type states under ADC, while also demonstrating the feasibility of CQT from mixed-biseparable multipartite entangled states, and (ii) to assess the efficacy of strategies such as local NOT operations for avoiding or delaying ESD in multipartite entanglement. For completeness, we also cover two-qubit Bell-type states. 
While teleportation fidelity is fundamentally linked to entanglement, previous works have shown that its behaviour under an ADC can differ significantly from entanglement measures \cite{NoisyTeleport_Fortes2015, NoisyTeleport_Bandyopadhyay2022}. 
For instance, when Bell states are subjected to ADC, the noisy states obtained from \(\ket{\Phi^+}=(\ket{00}+\ket{11})/\sqrt{2}\) can exhibit greater teleportation fidelity than those obtained from \(\ket{\Psi^+} =(\ket{01} + \ket{10})/\sqrt{2}\), even though entanglement measures may show the opposite ordering \cite{NoisyTeleport_Fortes2015}. Note that \(\ket{\Phi^+}\) and \(\ket{\Psi^+}\) are related by a local unitary \(\hat\sigma_x\) via \((\hat\sigma_x\otimes \mathbb{I}_2) \ket{\Phi^+} = \ket{\Psi^+}\). Thus, the application of NOT-gate on one qubit of $|\Phi^+\rangle$ state may preserve GMC but not teleportation fidelity. 
These observations indicate that teleportation fidelity may not be preserved in the same way as entanglement (discussed further in Sec. \ref{Sec_FidelityResults}). Nevertheless, we aim to identify potential similarities between GMC and teleportation fidelity in GHZ-type states, while also highlighting the key differences in their behaviour under the influence of ADC and NOT gates. Understanding these distinctions is essential for clarifying the role of entanglement in practical teleportation scenarios.

Beyond entanglement, stronger nonlocal correlations, such as Bell‐CHSH violation, the standard operational signature of Bell nonlocality, obey a strict hierarchy \cite{PhysRevLett.98.140402, PhysRevA.40.4277, Abo2023}. Under ADC, Bell nonlocality can suffer “sudden death” even when entanglement persists. This was experimentally investigated using the Horodecki criterion for CHSH violation \cite{HORODECKI1995}, and they found that for two‐qubit states under ADC, the Bell‐CHSH parameter drops below the classical bound at a finite damping strength, whereas entanglement persists \cite{PhysRevA.100.042311}. 
These findings fit into the well-known hierarchy of correlations and directly relate to teleportation performance, since teleportation fidelity provides an operational measure of a state's nonclassicality in this context \cite{PhysRevLett.72.797, HORODECKI1996, PhysRevA.64.042305, Hu2013}. In two‐qubit systems, if teleportation fidelity $\mathcal{F}>2/3$ then the state is nonclassical, and if additionally $\mathcal{F}>\mathcal{F}_{\rm LHV} \approx 0.87$ then it is guaranteed to exhibit Bell‐CHSH violation \cite{GISIN1996, Paulson2021}. 
We note that teleportation fidelity serves as an operational benchmark: purely classical ``measure-and-prepare'' schemes are bounded by $F = 2/3$~\cite{PhysRevLett.72.797}, while fidelity exceeding a stronger local-hidden-variable (LHV) threshold $F_{\rm LHV} \approx 0.87$ (under standard protocol assumptions) provides a sufficient (though not necessary) condition to certify correlations that cannot be simulated by LHV models (i.e., Bell nonlocality)~\cite{GISIN1996, HORODECKI1996}. Thus, teleportation fidelity establishes an experimentally relevant hierarchy between entanglement and Bell nonlocality.

Our primary interest lies in avoiding or at least delaying the occurrence of ESD using local NOT operations on GHZ-type states under ADC. Fundamentally, GHZ states play a central role in quantum foundations: they allow tests of quantum nonlocality without inequality, providing an “all-versus-nothing’’ contradiction with local realism~\cite{GHZ_1989}, and they exhibit the maximum possible violation of Mermin’s multipartite inequality~\cite{Mermin1990}. However, despite their importance, GHZ states are notoriously fragile in noisy environments. Notably, as the number of qubits increases in the GHZ-type state, they become more prone to sudden death due to the initial amplitude of ``ESD trigger-state'' ($|11\cdots1\rangle$) \cite{MultipartyESD_Xie2023}. Furthermore, we also investigate the possibility of CQT through a mixed biseparable multipartite entangled state obtained during ADC-evolution using LE \cite{CQT_Barasinski2018} and explicitly demonstrate that a tripartite entangled state is not a necessary resource for CQT. Notably, mixed biseparable states have already been experimentally verified as a resource for CQT \cite{CQT-PRL_Barasinski2019, CQT-PRA_Barasinski2019}. 

To further classify the behaviour of these states, we use the GHZ‐symmetric state $(x, y)$-parametrization framework \cite{GHZ-Sym_Eltschka2012, TripartiteEnt_Siewert2012}. GHZ‐symmetric states are invariant under qubit permutations, collective $\sigma_x$ flips, and correlated $Z$-rotations, and form a triangular convex set of states represented in an $(x,y)$ plane for two-qubit and three-qubit systems. Eltschka and Siewert \cite{GHZ-Sym_Eltschka2012} showed that this triangle is partitioned into GHZ, W, biseparable, and separable regions (entanglement classes) by specific boundaries for three-qubit systems.  Similarly, for two qubits, the GHZ‐symmetric family of states occupies a triangle with a separability line that divides entangled and separable states \cite{TripartiteEnt_Siewert2012}. The possibility of extending this type of parametrization for four-qubit systems was investigated \cite{GHZ-SymClass_Park2014}. In this work, by projecting noisy multipartite entangled state dynamics onto GHZ-symmetric families, we identify their entanglement classes. Furthermore, we associate this dynamics with the death of GMC, and identify the point at which the CQT fidelity drops below the classical limit, noting that the fidelity remains above classical threshold for a longer duration in the three- and four-qubit cases. We also verified that LE (an averaged conditional biseparable entanglement \cite{LocalizableEnt_Pop2005}) vanishes at the same point where CQT fidelity crosses the classical limit~\cite{CQT_Barasinski2018}.

The remainder of this paper is organized as follows. In Section~\ref{Sec_Background}, we introduce the theoretical background and key concepts required for the analysis. In particular, we discuss the dynamics of multiqubit states under the ADC and the effect of local NOT gates. We also review relevant concepts and definitions, including GMC, teleportation fidelity for two-qubit states, Bell nonlocality, and CQT schemes for GHZ states and Localizable entanglement through Sections~\ref{Sec_GMC}-\ref{Sec_LocEnt}.
 In section \ref{Sec_Results}, we apply these definitions and perform detailed calculations for GMC in Section~\ref{Sec_GMCResults} and teleportation fidelity in Section~\ref{Sec_FidelityResults}. Notably, prior works with NOT gates have discussed only entanglement dynamics for particular two-qubit systems \cite{ESDM_Rau2008, ESDM_Singh2017, ESDM_Behera2025} and bipartite higher-dimensional systems \cite{ESDM_Singh2022}. We conclude the paper with key takeaways in Section \ref{Sec_Conclusion}. 
 We provide the supporting calculations for $(x,y)$ GHZ‐symmetric classification in the Appendix \ref{Sec_Appendix}. The presented calculations for any GHZ-type states are a direct application of local NOT gates as a \emph{mitigation} strategy, demonstrating its utility in protecting GMC and teleportation fidelity in multi‐qubit systems. In effect, the results quantify the practical benefit of the NOT gates for GHZ-type states against ADC noise. Finally, we map these results onto the GHZ‐symmetric classification for a unified understanding of multipartite entanglement dynamics. 

\section{Theoretical background and key concepts}
\label{Sec_Background}

\begin{figure}[ht!]
    \centering
    \scalebox{0.8}{
        \begin{tikzpicture}[
            font=\small,
            ball/.style={circle, draw, shading=ball, ball color=gray!30, minimum size=1.5cm},
            qubit/.style={circle, draw=black, minimum size=1.2cm},
            box/.style={rectangle, draw=black, minimum height=1cm, minimum width=1.4cm, align=center},
            labelnode/.style={rectangle, draw=black!50, rounded corners, fill=black!5, minimum height=1.2cm, minimum width=1.6cm, align=center},
            snakearrow/.style={decorate, decoration={snake, amplitude=.5mm, segment length=2.5mm}, draw=red, thick}, thick
             ]
        
            
            \node[box] (rhon) at (-1.3, -0.5) {$\rho_n(\alpha)$};
            
            \node[qubit] (q1) at (0.5, 2) {Qubit 1};
            \node[box]   (not1) at (4.6, 2) {NOT};
            \node[box]   (not2) at (8.8, 2) {NOT};
            \node[qubit] (q2) at (10.5, 2) {Qubit 1};
            
            \node[qubit] (q3) at (0.5, 0) {Qubit 2};
            \node[box]   (not3) at (4.6, 0) {NOT};
            \node[box]   (not4) at (8.8, 0) {NOT};
            \node[qubit] (q4) at (10.5, 0) {Qubit 2};
            
            \node[box] (rhof) at (12.6, -0.5) {$\rho_{(m,n)}(\alpha,p,p')$};
            
            \node at (2.5, -4.2) {$p$};
            \node at (6.7, -4.2) {$p'$};
            \draw[->, thick] (0.5, -4.5) -- (4.2, -4.5);
            \draw[->, thick] (5, -4.5) -- (8.7, -4.5);
            \draw[snakearrow] (q1) -- (q3);
            \draw[snakearrow] (q2) -- (q4);
            \node at (0.5,-1.4){$\vdots$};
            \node at (2.5,-1.4){$\vdots$};
            \node at (4.6,-1.4){$\vdots$};
            \node at (6.7,-1.4){$\vdots$};
            \node at (8.8,-1.4){$\vdots$};
            \node at (10.5,-1.4){$\vdots$};
            \draw[snakearrow] (q3) -- (0.5,-1.25);
            \draw[snakearrow] (q4) -- (10.5,-1.25);
            
            \node[qubit] (q5) at (0.5, -3) {Qubit $n$};
            \draw[-,thick] (q1) -- (not1)node[pos=0.5, above] {ADC};
            \draw[-,thick] (not1) -- (not2)node[pos=0.5, above] {ADC};
            \draw[-,thick] (not2) -- (q2);
            \draw[-,thick] (q3) -- (not3)node[pos=0.5, above] {ADC};
            \draw[-,thick] (not3) -- (not4)node[pos=0.5, above] {ADC};
            \draw[-,thick] (not4) -- (q4);
            
            \node[box]   (not5) at (4.6, -3) {NOT};
            \node[box]   (not6) at (8.8, -3) {NOT};
            \node[qubit] (q6) at (10.5, -3) {Qubit $n$};
            
            \draw[-,thick] (q5) -- (not5)node[pos=0.5, above] {ADC};
            \draw[-,thick] (not5) -- (not6)node[pos=0.5, above] {ADC};
            \draw[-,thick] (not6) -- (q6);
            
            \draw[snakearrow] (q5) -- (0.5,-1.75);
            \draw[snakearrow] (q6) -- (10.5,-1.75);
            
            \draw[decorate, decoration={brace, amplitude=8pt}, thick](-0.2,-3.5) -- (-0.2,2.5) node[midway,xshift=-0.6cm]{};
            \draw[decorate, decoration={brace,mirror, amplitude=8pt}, thick](11.2,-3.5) -- (11.2,2.5) node[midway,xshift=-0.6cm]{}; 
        \end{tikzpicture}}
    \caption{Schematic illustrating entanglement preservation in multiqubit entangled states under ADC noise using local NOT operations applied to one or more qubits between successive ADC processes. The final NOT gate restores the state to the original population configuration.}
    \label{fig_NOTScheme}
\end{figure}
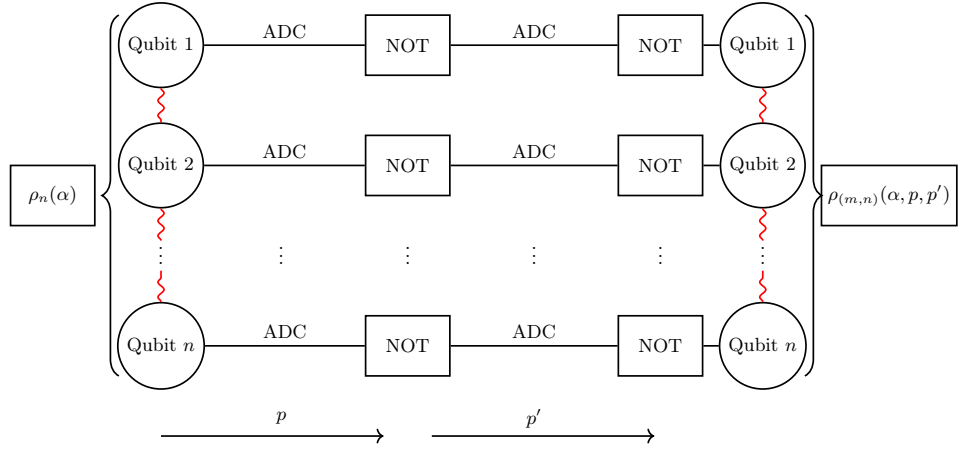

Fig.\,(\ref{fig_NOTScheme}) presents a schematic of the framework used to analyze ADC noise and the local NOT protocol in this work. We consider a system initially prepared in an $n$-qubit GHZ-type entangled state parameterized by $\alpha$, given by
\begin{equation}
|\mathrm{GHZ}_n(\alpha)\rangle = \alpha |0\rangle^{\otimes n} + \beta |1\rangle^{\otimes n},
\label{Eq_GHZState}
\end{equation}
where $|\beta|^2 = 1 - |\alpha|^2$. The corresponding density matrix of the state defined in Eq.~(\ref{Eq_GHZState}) is
\begin{equation}
\rho_n(\alpha) = |\mathrm{GHZ}_n(\alpha)\rangle \langle \mathrm{GHZ}_n(\alpha)|.
\label{Eq_GHZDM}
\end{equation}

Further, the single-qubit Kraus operators for ADC \cite{Book_NielsenChuang2010} are given by 
\begin{equation} 
M_1= \left( \begin{array}{cccc} 1 & 0 \\
0 & \sqrt{q} \end{array} \right),~~
M_2=\left( \begin{array}{cccc} 0 &\sqrt{p}\\
0 & 0  \end{array} \right),
\label{Eq_ADCKraus}
\end{equation} 
where $q=1-p$. These operators meet the completeness criterion; $M_1^\dag M_1+M_2^\dag M_2=\mathbb{\hat I}_2$, with $\mathbb{\hat I}_2$ being the identity matrix. By taking tensor products of single-qubit Kraus operators, we can obtain the Kraus operators for the two, three, and more qubit states as 
\begin{equation}
\begin{aligned}
M_{ij} &=M_i \otimes M_j~;~~i,j=1,2~,\\ 
M_{ijk} &=M_i \otimes M_j\otimes M_k~;~~i,j,k=1,2~,~\text{and}\\
M_{ijk\cdots} &=M_i \otimes M_j\otimes M_k\otimes\cdots ~;~~i,j,k,\cdots=1,2~.
\end{aligned}
\label{Eq_MultiPartyKraus}
\end{equation}

The Kraus operators of the second ADC are denoted as $M'_{ijk\cdots}$, where parameters $p$ and $q$ are replaced by $p'$ and $q'$, respectively, with their form identical to that in Eq.~(\ref{Eq_MultiPartyKraus}). In the presence of first ADC~(\ref{Eq_MultiPartyKraus}), the initial state $\rho_n(\alpha)$ evolves to the state
\begin{equation}
\rho_n(\alpha,p)=\sum _{i,j,k,\cdots} M_{ijk\cdots}~\rho_n(\alpha)~M_{ijk\cdots}^\dag~~;~~i,j,k,\cdots=1,2.
\end{equation} 

Next, we use the NOT operation $U_{(m,n)}$ that acts on only $m$ of the $n$ qubits ($m={0,1,2,\cdots,n}$) as 
\begin{equation}
\rho_{(m,n)}(\alpha,p)= U_{(m,n)}~ \rho_n(\alpha,p)~U_{(m,n)}^\dagger,
\label{Eq_06}
\end{equation}
where 
$U_{(0,2)} = (\hat{\mathbb{I}}_2 \otimes \hat{\mathbb{I}}_2)$, 
$U_{(1,2)} = (\hat{\sigma}_x \otimes \hat{\mathbb{I}}_2)$, 
$U_{(2,2)} = (\hat{\sigma}_x \otimes \hat{\sigma}_x)$, 
$U_{(0,3)} = (\hat{\mathbb{I}}_2 \otimes \hat{\mathbb{I}}_2 \otimes \hat{\mathbb{I}}_2)$, 
$U_{(1,3)} = (\hat{\sigma}_x \otimes \hat{\mathbb{I}}_2 \otimes \hat{\mathbb{I}}_2)$, 
$U_{(2,3)} = (\hat{\sigma}_x \otimes \hat{\sigma}_x \otimes \hat{\mathbb{I}}_2)$, and 
$U_{(3,3)} = (\hat{\sigma}_x \otimes \hat{\sigma}_x \otimes \hat{\sigma}_x)$. 
In general, the operation $U_{m,n}$ denotes the application of $\hat{\sigma}_x^{\otimes m}$ on $m$ qubits and $\hat{\mathbb{I}}_2^{\otimes (n-m)}$ on the remaining $(n-m)$ qubits.


We investigate the robustness of this state under the second ADC followed by the final NOT operation, resulting in
\begin{equation} 
\rho_{(m,n)}(\alpha,p,p') = U_{(m,n)}\left[\sum _{i,j,k,\cdots}  M'_{ijk\cdots}~\rho_{(m,n)}(\alpha,p)~ M_{ijk\cdots}^{\prime\dagger}\right] U_{(m,n)}^\dagger~~;~~i,j,k,\cdots=1,2.
\label{Eq_07}
\end{equation} 
Note that we apply the same set of NOT operators again to restore the state to its original population configuration.

\subsection{Genuine multipartite concurrence (GMC)}
\label{Sec_GMC}

To quantify the GME in multipartite systems, we utilize GMC \cite{GMC_Hashemi2012}, which has a closed-form expression for $n$-qubits with $X$-state density matrix
\begin{align} \label{Eq_08}
\rho_n=\left(  \begin{array}{cccccccc}
    a_{1} & \cdot & \cdot & \cdot & \cdot & \cdot & \cdot & z_{1} \\ 
    \cdot & a_{2} & \cdot  & \cdot & \cdot & \cdot & z_{2} & \cdot \\ 
    \cdot & \cdot & \ddots & \cdot & \cdot & \iddots & \cdot & \cdot \\ 
    \cdot & \cdot & \cdot & a_{N} & z_{N} & \cdot & \cdot & \cdot \\ 
    \cdot & \cdot & \cdot & z_{N}^{*} & b_{N} & \cdot & \cdot & \cdot \\ 
    \cdot & \cdot & \iddots & \cdot & \cdot & \ddots & \cdot & \cdot \\ 
    \cdot & z_{2}^{*} & \cdot & \cdot & \cdot & \cdot & b_{2} & \cdot \\ 
    z_{1}^{*} & \cdot & \cdot & \cdot & \cdot & \cdot & \cdot & b_{1} \\ 
  \end{array}
\right),
\end{align}
given by
\begin{align} \label{Eq_09}
\text{GMC}[\rho_n]= 2 \max\left[0,~|z_{i}|-w_{i}\right], ~ i=1,\dots,N,
\end{align} 
where $N=2^{n-1}$, $w_{i}=\sum_{j\neq i}^{N}\sqrt{a_{j} b_{j}}$. Note that this measure is useful for both pure and mixed states. We obtain the closed-form expressions for GMC for arbitrary state $\rho_{(m,n)}(\alpha,p,p')$ given in Eq.~(\ref{Eq_07}). Importantly, in the case of a two-qubit system, the GMC coincides with Wootters’ concurrence \cite{Wootters98}. Thus, throughout this work, any reference to the GMC for two qubits should be understood as referring to the bipartite entanglement measured by concurrence.

 \subsection{Teleportation fidelity using two-qubit Bell-type state}

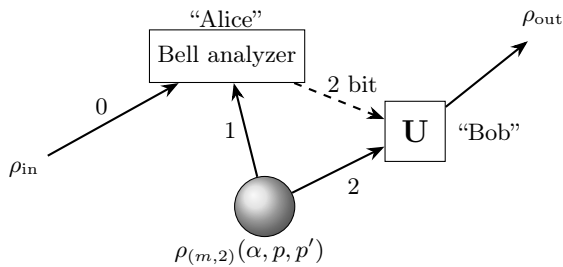
\begin{figure}[ht!]
    \centering
    \begin{tikzpicture}[>=Stealth, node distance=1.5cm and 2cm, every node/.style={font=\small}]
        \node (alice) at (0.3,-0.5) {$\rho_{\rm in}$};
        \node[draw, minimum width=1.4cm, minimum height=0.7cm] (bell) at (3,1) {Bell analyzer};
        \node at (3,1.5) {``Alice''};
        \node[circle, draw, shading=ball, ball color=gray!30, minimum size=0.8cm] (entangle) at (3.5,-1) {};
        \node[draw, minimum width=0.8cm, minimum height=0.8cm] (unitary) at (5.5,0) {\large $\mathbf{U}$};
        \node at (7.2,1.5) {$\rho_{\rm out}$};
        \node at (6.5,0) {``Bob''};
        \draw[->, thick] (alice) -- (bell) node[midway, above left=-2pt] {$0$};
        \draw[->, thick] (entangle) -- (bell) node[midway, left] {1};
        \draw[->, thick] (entangle) -- (unitary) node[midway, below right] {2};
        \draw[->, thick, dashed] (bell) -- (unitary) node[midway, above,xshift=5pt] {2 bit};
        \draw[->, thick] (unitary) -- ++(1.5,1.2);
        \node at (3.3,-1.6) {$\rho_{(m,2)}(\alpha, p, p')$};
    \end{tikzpicture}
    \caption{Schematic diagram of teleportation using a two-qubit entangled state.}
    \label{fig:teleportation-2}
\end{figure}

The schematic diagram in Fig.\,(\ref{fig:teleportation-2}) illustrates the teleportation protocol introduced by Bennett \emph{et al.} \cite{Teleport_Bennett1993} utilizing a two-qubit entangled state. For an input state $|\Psi_{\rm in}\rangle = a|0\rangle + b|1\rangle$, with $|a|^2+|b|^2=1$, and output state density matrix $\rho_{\rm out}$, the teleportation fidelity is defined as 
\begin{align}
    \mathcal{F} = \langle\Psi_{\rm in}|\rho_{\rm out}|\Psi_{\rm in}\rangle.
\end{align}
Thus, teleportation fidelity quantifies the closeness or overlap of the teleported qubit state $\rho_{\rm out}$ with the input state  $|\Psi_{in}\rangle$ (density matrix $\rho_{\rm in}=|\Psi_{\rm in}\rangle\langle\Psi_{\rm in}|$). In an ideal condition, if Alice and Bob share a Bell state $\rho_2(1/\sqrt{2})$ [see Eq.~(\ref{Eq_GHZDM})], one finds \(\mathcal{F}=1\) (perfect overlap).

More generally, for a Bell-type state transmitted through a noisy channel $\rho_{(m,2)}(\alpha, p, p')$ [see Eq.~(\ref{Eq_07})], the fully entangled fraction (FEF), denoted by $F$, provides a closed-form characterization of the teleportation fidelity. It is defined as
\begin{align}
F\!\left[\rho_{(m,2)}(\alpha,p,p')\right]
= \max_{\ket{\Phi}}
\langle \Phi | \rho_{(m,2)}(\alpha,p,p') | \Phi \rangle ,
\end{align}
where $\ket{\Phi}$ runs over all maximally entangled Bell states. The optimal teleportation fidelity is then given by
\begin{align}
\mathcal{F}\!\left[\rho_{(m,2)}(\alpha,p,p')\right]
= \frac{1 + 2\,F\!\left[\rho_{(m,2)}(\alpha,p,p')\right]}{3}.
\end{align}
As mentioned earlier, purely classical “measure-and-prepare” scheme cannot exceed teleportation fidelity \(\mathcal{F}=2/3\)  \cite{InfExtract_Massar1995}, where \(F(\rho)=1/2\) gives \(\mathcal{F}=2/3\), i.e.\ no quantum advantage \cite{CQT_Karlsson1998}. We obtained analytical closed-form expressions for teleportation fidelity in Section~(\ref{Sec_FidelityResults}) of an arbitrary state $\rho_{(m,2)}(\alpha,p,p')$ given in Eq.~(\ref{Eq_07}). 
 
\subsection{Bell Nonlocality for a two-qubit State}

Violation of the Bell–CHSH inequality \cite{clauser1969proposed} provides a decisive test of the incompatibility between quantum mechanics and local realism. Consider a two-qubit state shared between two parties, Alice and Bob, each performing measurements on one qubit. Alice (Bob) can choose between two dichotomic observables $A, A'$ ($B, B'$), each taking values in ${\pm1}$. The resulting measurement statistics admit a local hidden-variable (LHV) description whenever the CHSH parameter
$S=\langle A\otimes B+A\otimes B' + A'\otimes B-A'\otimes B' \rangle \leq 2$. Quantum mechanics, however, predicts violations of this bound for suitably chosen entangled states. In particular, every pure entangled two-qubit state violates the CHSH inequality for an appropriate choice of measurements \cite{gisin1991bell}. The observables can be parametrized as 
  $ A=\hat a\!\cdot\!\boldsymbol{\hat\sigma},~ 
   A'=\hat a'\!\cdot\!\boldsymbol{\hat\sigma},~
   B=\hat b\!\cdot\!\boldsymbol{\hat\sigma},~
   B'=\hat b'\!\cdot\!\boldsymbol{\hat\sigma}$,
where $\hat a,\hat a',\hat b,\hat b'$ are real unit vectors representing measurement directions on the Bloch sphere, and $\boldsymbol{\hat\sigma}=(\hat\sigma_x, \hat\sigma_y, \hat\sigma_z)$ is the vector of Pauli operators.  This is the standard operational benchmark for Bell nonlocality. For the sake of simplicity, here we consider the Horodecki criterion~\cite{HORODECKI1995} which links the CHSH bound directly to the correlation tensor $\boldsymbol{C}$ with components
\begin{equation}
   C_{ij}=\operatorname{Tr}\!\big[\rho_{(m,2)}(\alpha,p,p')~\,(\hat\sigma_i\otimes\hat\sigma_j)\big],~ i,j\in\{x,y,z\},
\end{equation} allowing us to compute the maximal CHSH value for a general two-qubit state $\rho_{(m,2)}(\alpha,p,p')$.
The maximal violation of Bell Nonlocality achievable over all measurement settings is determined by the Horodecki criterion~\cite{HORODECKI1995}:
\begin{equation} \label{eq:horodecki}
   \max_{\{A,A',B,B'\}} \text{Bell Nonlocality}
      = 2\sqrt{\kappa_1+\kappa_2},
\end{equation}
where $\kappa_1\geq \kappa_2\geq 0$ are the two largest eigenvalues of the real symmetric matrix $C^T C$, with $C^T$ denoting transpose of $C$. Hence, the state $\rho_{(m,2)}(\alpha,p,p')$ violates the CHSH inequality if and only if $\kappa_1+\kappa_2 > 1$. As discussed in Sec. \ref{Sec_Introduction}, including this measure alongside teleportation fidelity and GMC enables us to directly compare different quantum correlations and clarify the respective roles of no-, one-, and two-NOT gates ($m=0,1,2$) on qubits in noisy teleportation scenarios.

\subsection{Controlled quantum teleportation}

\subsubsection{CQT using three-qubit GHZ-type states}

\begin{figure}[ht!]
    \centering 
    \begin{minipage}[b]{0.52\textwidth}
        \centering
         \begin{tikzpicture}[>=Stealth, every node/.style={font=\small}]
        \node at (0,1) (alice) {$\rho_{\rm in}$};
        \node[draw, minimum width=1.4cm, minimum height=0.8cm] (bell) at (2,3) { Bell analyzer};
        \node at (2,3.6) {``Alice''};
        \node[circle, draw, shading=ball, ball color=gray!30, minimum size=0.8cm] (entangle) at (2,0) {};
        \node[draw, minimum width=0.8cm, minimum height=0.8cm] (b2) at (5,3) {\large $\mathbf{U}$};
        \node[draw, minimum width=0.8cm, minimum height=0.8cm] (b1) at (5,0) {\large $R_\theta$};
        \draw[->, thick] (alice) -- (bell) node[midway, above left=-2pt] {0};
        \draw[->, thick] (entangle) -- (bell) node[midway, left] {3};
        \draw[->, thick] (entangle) -- (b1) node[midway, below] {1};
        \draw[->, thick] (entangle) -- (b2) node[pos=0.5, right] {2};
        \draw[->, thick, dashed] (bell) -- (b2) node[pos=0.45, above, yshift=3pt] {2-bit};
        \draw[->, thick, dashed] (b1) -- (b2) node[midway, right] {1-bit};
        \draw[->, thick] (b2) -- ++(1,0.8) node[above right] {$\rho_{\rm out}$};
        \node at (1.5,-0.6) {$\rho_{(m,3)}(\alpha,p,p')$};
        \node at (6.4,3.2) {``Bob''};
        \node at (5.9,0.1) {``$B_1$''};
    \end{tikzpicture}
    \caption*{(a)}
    \end{minipage}
    \begin{minipage}[b]{0.7\textwidth}
        \centering
        \includegraphics[width=\textwidth]{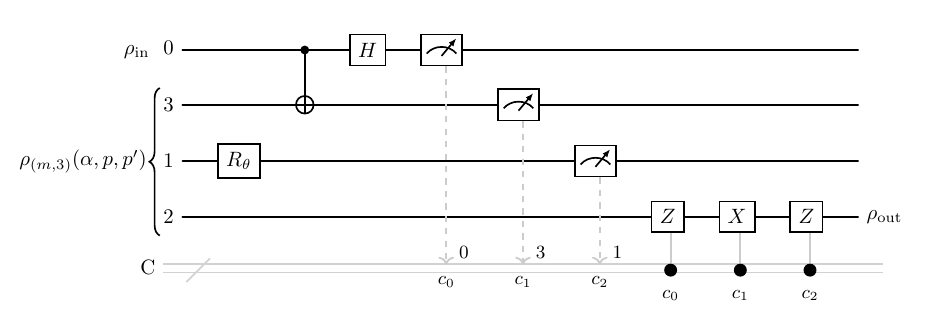} 
         \caption*{(b)}
    \end{minipage}
    \caption{CQT using three-qubit GHZ-type state with one \emph{controller}. (a) schematic diagram \cite{CQT_Karlsson1998}, and corresponding (b) circuit representation \cite{CQT_De2025}. The dotted lines represent the transfer of classical information, and the solid lines show the transfer of qubits from the entanglement source.}
    \label{fig_CQT-3}
\end{figure}

Fig.\,(\ref{fig_CQT-3}) shows (a) schematic diagram \cite{CQT_Karlsson1998}, and (b) corresponding circuit representation \cite{CQT_De2025} of CQT using a three-qubit GHZ-type state $\rho_{(m,3)}(\alpha, p, p')$ [see Eq.~(\ref{Eq_07})] shared over a noisy channel. The qubits 1, 2, and 3 are held by a \emph{controller} \(B_1\), Bob, and Alice, respectively.  
Alice’s Bell measurement on her qubit pairs `0' and `3' yields two bits of classical information $c_0$ and $c_1$, respectively, which are shared with Bob. The \emph{controller} $B_1$ applies the basis rotation $R_\theta$ on qubit `1' and measures it in the Z-basis \cite{CQT_Karlsson1998}, yielding one bit of information $c_2$ that is also shared with Bob. 
Bob then performs a local unitary operation $Z^{c_0}.X^{c_1}.Z^{c_2}$ on qubit `2' to obtain $\rho_\text{out}$. The output state has maximum fidelity with $\ket{\Psi_\text{in}}$ ($\rho_{\rm in}$) only if the \emph{controller} cooperates (and for a perfect GHZ-state). Optimal teleportation fidelity in this case occurs at $\theta=\pi/4$ in the \emph{controller's} $R_\theta$ \cite{CQT_Karlsson1998}.

For an arbitrary X–state $\rho_{(m,3)}(\alpha,p,p')$, the CQT fidelity for $R_\theta$ $(\theta=\pi/4)$ is given as \cite{CQT_Barasinski2018,CQT-PRA_Barasinski2019,CQT-PRL_Barasinski2019}
\begin{equation}
\mathcal{F}\left[\rho_{(m,3)}(\alpha,p,p')\right]
= \max_{i=1\ldots4}F_{\rm CQT}^{(i)}\left[\rho_{(m,3)}(\alpha,p,p')\right],\\
\label{Eq_15}
\end{equation}

\begin{equation}
\begin{aligned}
  F_{\rm CQT}^{(1)}\left[\rho_{(m,3)}(\alpha,p,p')\right] &= \frac{3 + |\Delta_1| + 4(|z_1| + |z_4|)}{6},
  && F_{\rm CQT}^{(2)}\left[\rho_{(m,3)}(\alpha,p,p')\right] = \frac{3 + |\Delta_1| + 4(|z_2| + |z_3|)}{6},\\
  F_{\rm CQT}^{(3)}\left[\rho_{(m,3)}(\alpha,p,p')\right] &= \frac{3 + \sqrt{|\Delta_2|^2 + 16(|z_1| + |z_4|)^2}}{6},
 && F_{\rm CQT}^{(4)}\left[\rho_{(m,3)}(\alpha,p,p')\right] = \frac{3 + \sqrt{|\Delta_2|^2 + 16(|z_2| + |z_3|)^2}}{6},
\end{aligned}
\label{Eq_16}
\end{equation}
where $\Delta_1 = a_1 - a_2 - a_3 + a_4 + b_1 - b_2 - b_3 + b_4$, and
$\Delta_2 = a_1 - a_2 + a_3 - a_4 - b_1 + b_2 - b_3 + b_4$.
Here, $a_i$, $b_i$, and $z_i$ are defined by Eq.~(\ref{Eq_08}), which is evolved through ADC by Eq.~(\ref{Eq_07}). 

Note that whenever we use the term `fidelity' $\mathcal{F}$ for the multipartite (bipartite) scenario, it refers to CQT (Bennet teleportation) fidelity. 
In the ideal case, for a pure-GHZ state with \(\theta=\pi/4\), all four expressions in Eq.~(\ref{Eq_16}) equal unity.  If the \emph{controller} withholds cooperation, the average fidelity drops to \(1/2\), consistent with the no quantum advantage benchmark. Also, fidelity obtained from the circuit matches the analytic formulae $\mathcal{F}\left[\rho_{(m,3)}(\alpha,p,p')\right]$ [see Eq.~(\ref{Eq_15})] for state $\rho_{(m,3)}(\alpha,p,p')$ in our study.  

\subsubsection{CQT using four-qubit GHZ-type states}

\begin{figure}[ht!]
    \centering  
         \begin{subfigure}[b]{0.52\textwidth}
       \begin{tikzpicture}[>=Stealth, every node/.style={font=\small}]
            \node at (0,1) (alice) {$\rho_{\rm in}$};
            \node[draw, minimum width=1.4cm, minimum height=0.8cm] (bell) at (1.8,3) {Bell analyzer};
            \node at (1.8,3.6) {``Alice''};
            \node[circle, draw, shading=ball, ball color=gray!30, minimum size=0.8cm] (entangle) at (2.3,0.5) {};
            \node[draw, minimum width=0.8cm, minimum height=0.8cm] (b3) at (4.5,3) {\large $\mathbf{U}$};
            \node[draw, minimum width=0.8cm, minimum height=0.8cm] (b2) at (6,1.25) {\large $\mathbf{R_{\theta_1}}$};
            \node[draw, minimum width=0.8cm, minimum height=0.8cm] (b1) at (4.5,-0.5) {\large $\mathbf{R_{\theta_2}}$};
            \draw[->, thick] (alice) -- (bell) node[midway, above left=-2pt] {0};
            \draw[->, thick] (entangle) -- (bell) node[midway, left] {1};
            \draw[->, thick] (entangle) -- (b1) node[midway, above] {2};
            \draw[->, thick] (entangle) -- (b2) node[midway, above] {3};
            \draw[->, thick] (entangle) -- (b3) node[midway, below] {4};
            \draw[->, thick, dashed] (bell) -- (b3) node[pos=0.5, above] {2-bit};
            \draw[->, thick, dashed] (b1) -- (b3) node[pos=0.6, left, xshift=1pt] {1-bit};
            \draw[->, thick, dashed] (b2) -- (b3) node[pos=0.5, right, xshift=1pt] {1-bit};
            \draw[->, thick] (b3) -- ++(1,0.8) node[above right] {$\rho_{\rm out}$};
            \node at (1.7,-0.1) {$\rho_{(m,4)}(\alpha,p,p')$};
            \node at (5.8,3.2) {``Bob''};
            \node at (6.9,1.3) {``$B_2$''};
            \node at (5.5,-0.5) {``$B_1$''};
        \end{tikzpicture}
           \caption*{(a)}
    \end{subfigure}
 \begin{subfigure}[b]{0.77\textwidth}
       \centering
    \includegraphics[width=\textwidth]{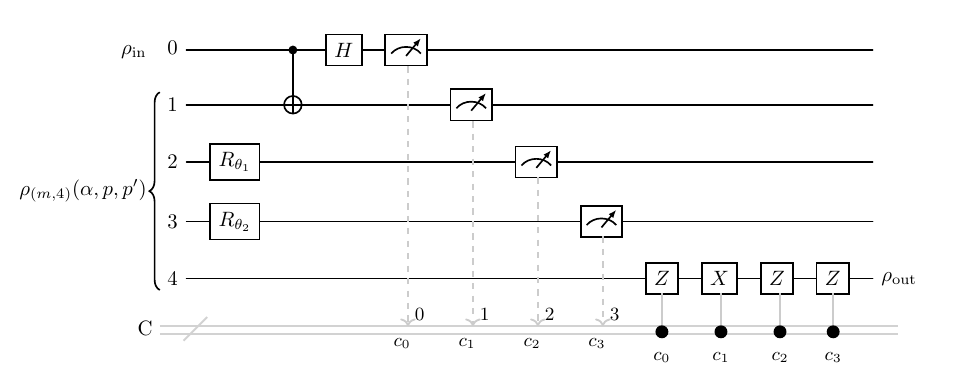} 
       \caption*{(b)}
    \end{subfigure}
    \caption{CQT using four-qubit GHZ-type state with two \emph{controllers}. (a) schematic diagram, and corresponding (b) circuit representation \cite{CQT_De2025}. The dotted lines represent the transfer of classical information, and the solid lines show the transfer of qubits from the entanglement source.}
    \label{fig_CQT-4}
\end{figure}

Similar to the three-qubit case, one can extend the CQT protocol to four parties using a four‑qubit entangled state as shown in Fig.\,(\ref{fig_CQT-4}). We consider Alice, \emph{two controllers} ($B_1$ and $B_2$), and Bob, each of whom holds a qubit from a four‑qubit GHZ-type state $\rho_{(m,4)}(\alpha,p,p')$ [see Eq.~(\ref{Eq_07})] shared over a noisy channel. Additionally, Alice has the unknown input qubit ($\rho_{\rm in}$) to be teleported.  
The CQT protocol proceeds as follows:  Alice performs a Bell measurement on her qubit pairs `0' and `1', yielding two bits of classical information $c_0$ and $c_1$, which are shared with Bob. Then \emph{controller} $B_1$ ($B_2$) applies the basis rotation $R_{\theta_1}$ ($R_{\theta_2}$) on qubit `2' (qubit `3') and measures it in the $Z$-basis, yielding one bit of information $c_2$ ($c_3$) that is also shared with Bob. Bob then performs a local unitary operation $Z^{c_0}.X^{c_1}.Z^{c_2}.Z^{c_3}$ on qubit `4' to obtain $\rho_\text{out}$. This state has maximum fidelity with $\ket{\Psi_\text{in}}$ ($\rho_{\rm in}$) only if the \emph{controllers} cooperate (and for a perfect GHZ-state). Once again, optimal fidelity occurs at $\theta_1 = \theta_2 = \pi/4$ in the \emph{controller's} basis.

The fidelity considerations for the four‑party CQT protocol are analogous to the three‑qubit case.  In particular, in the \emph{ideal} case, for a pure four‑qubit GHZ resource shared over a noise‑free channel and $\theta_1=\theta_2=\pi/4$, Bob recovers the input state with unit fidelity ($\mathcal{F}=1$). We also note from prior studies of noisy GHZ channels that multi-qubit GHZ states can show similar teleportation performance in some noise models. In particular, the Lindblad equation for $n$-qubit GHZ states ($n=3,4,5,6$) was solved with Pauli-type decoherence and found that the three-qubit GHZ is generally more robust than larger GHZ states under most noisy channels \cite{NoisyTeleport_Espoukeh2014}. However, it was also found that for certain correlated Pauli‑noise channels (Lindblad operators in the same Pauli basis), the average teleportation fidelity of three‑ and four‑qubit GHZ channels can be the same.  In general, however, three‑qubit GHZ channels tend to be slightly more robust than four‑qubit ones under uncorrelated noise. We obtained the analytical fidelity expressions in Section~(\ref{Sec_FidelityResults}) for the arbitrary state $\rho_{(m,4)}(\alpha,p,p')$ given in Eq.~(\ref{Eq_07}) and obtained similar results based on the circuit implementation.

\subsection{Averaged conditional biseparable entanglement measured through localizable entanglement}
\label{Sec_LocEnt}

We are further interested in identifying parameter regions of the mixed entangled states $\rho_{(m,n)}(\alpha,p,p')$ that are useful for a CQT protocol beyond losing GME as witnessed previously for a three-qubit GHZ channel~\cite{CQT_Barasinski2018}. 
Thus, we distinguish genuine $n$-partite entanglement ($\mathrm{GMC} > 0$) from entanglement confined to a proper bipartition of the system by considering an averaged conditional biseparable entanglement measure, namely localizable entanglement (LE).
In the latter case, regions with $\text{GMC} = 0$ can still exhibit nontrivial CQT fidelity $\mathcal{F} > 2/3$ \cite{CQT_Barasinski2018}.
LE was introduced originally in the context of interacting spin systems and spin chains ~\cite{EntCor_Verstraete2004, LocalizableEnt_Pop2005}. It is the amount of bipartite entanglement that can be \emph{localized}, on average, between two chosen qubits of a multipartite state by performing local measurements on the other qubits.  More precisely, fixing two target qubits \(A\) and \(B\), one maximizes the average bipartite entanglement between \(A\) and \(B\) over all possible local measurement strategies on the remaining parties \cite{EntCor_Verstraete2004, LocalizableEnt_Pop2005}.  For an $n$-qubit state, the LE between qubit pairs $(i,j)$ can be written operationally as the maximum average concurrence obtained by performing local measurements \(\{\Pi_k\}\) on the other qubits. 
Given an \(n\)-qubit state \(\rho_n\), the \emph{localizable concurrence} $\mathcal{C}_L[\rho_n^{(i,j)}]$ between qubits ($i,j$) is defined as \cite{EntCor_Verstraete2004, LocalizableEnt_Pop2005}
\begin{equation}\label{eq:CL_def}
\mathcal{C}_L[\rho_n^{(i,j)}]\;=\;\max_{\{\Pi_k\}}\sum_k p_k\,\text{GMC}\big[\rho_{ij}^{(k)}\big],
\end{equation}
where \(p_k=\text{Tr}\big[(\mathbb{I}_{ij} \otimes \Pi_k)\rho_n\big]\) and \(\rho_{ij}^{(k)} = \text{Tr}_{\overline{ij}} \big[(\mathbb{I}_{ij} \otimes\Pi_k) \rho_n(\mathbb{I}_{ij} \otimes\Pi_k)\big]/p_k\).
Here, $\overline{ij}$ denotes the subsystem composed of all qubits except $i$ and $j$, and $\mathrm{Tr}_{\overline{ij}}$ is the partial trace over the remaining $(n-2)$ qubits. 
Note that the maximization runs over admissible local measurement POVMs \(\{\Pi_k\}\) on the remaining ($n-2$) qubits.
 
A general three-qubit state $\rho_{ABC}$ is said to be \emph{biseparable} if it is separable across at least one bipartition, or a mixture thereof. That is, it can be expressed as a convex combination of states separable with respect to $AB|C$, $AC|B$, or $BC|A$.
One might think that `increasing separability' and loss of genuine multipartite entanglement always reduces LE, but surprisingly, mixed biseparable states can still yield nonzero LE. Barasiński \emph{et al.}~\cite{CQT_Barasinski2018} demonstrated that mixed three-qubit channels that are only biseparable (no genuine tripartite entanglement) enable high-fidelity CQT between two parties. These theoretical findings align with a more general analysis of LE in teleportation. Consiglio \emph{et al.}~\cite{LocalConc_Consiglio2021} showed that any teleportation protocol (with a Bell measurement) that beats the classical fidelity threshold requires a nonzero localizable concurrence between Alice and Bob. Importantly, CQT experiments \cite{CQT-PRL_Barasinski2019, CQT-PRA_Barasinski2019}  have explicitly shown that genuine tripartite entanglement is not a necessary resource for CQT. 

\section{Results and discussion}
\label{Sec_Results}

\subsection{GMC dynamics in the ADC with(out) NOT gates}
\label{Sec_GMCResults}

\subsubsection{GMC dynamics without NOT gates}

GMC for two-, three-, and four-qubit entangled states $\rho_{(0,n)}(\alpha,p,p')$ with $n=2,3,4$, respectively, in the presence of initial ADC (without NOT gates) is given below.
\begin{equation}
\begin{aligned}
\text{GMC}[\rho_{(0,2)}(\alpha,p,p')] &= 2 \max \left[0, ~ \alpha \beta  q q' -\beta^2 q q' (p+p' q)\right], \\
\text{GMC}[\rho_{(0,3)}(\alpha,p,p')] &= 2 \max \left[0, ~ \alpha  \beta  (q q')^{3/2}-3 \sqrt{\beta ^4 q^3 q'^3 (p+p' q)^3}\right], \\
\text{GMC}[\rho_{(0,4)}(\alpha,p,p')] &= 2 \max \left[0, ~ \alpha \beta  q^2 q'^2 -7 \beta^2 q^2 q'^2 (p+p' q)^2\right].
\end{aligned}
\end{equation}
Note that to keep the analytical expressions compact, we assume $\alpha, \beta \in \mathbb{R}$ throughout this work.

\begin{figure}[ht!]
    \centering
    \begin{minipage}[b]{0.32\textwidth}
        \centering
        \includegraphics[width=\textwidth]{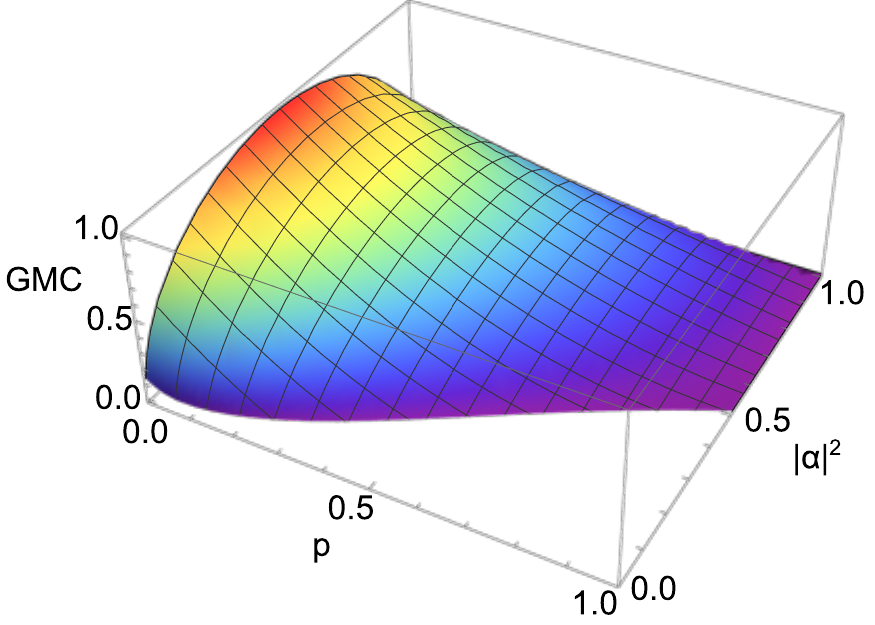} 
        \caption*{(a)}
    \end{minipage}
    \begin{minipage}[b]{0.32\textwidth}
        \centering
        \includegraphics[width=\textwidth]{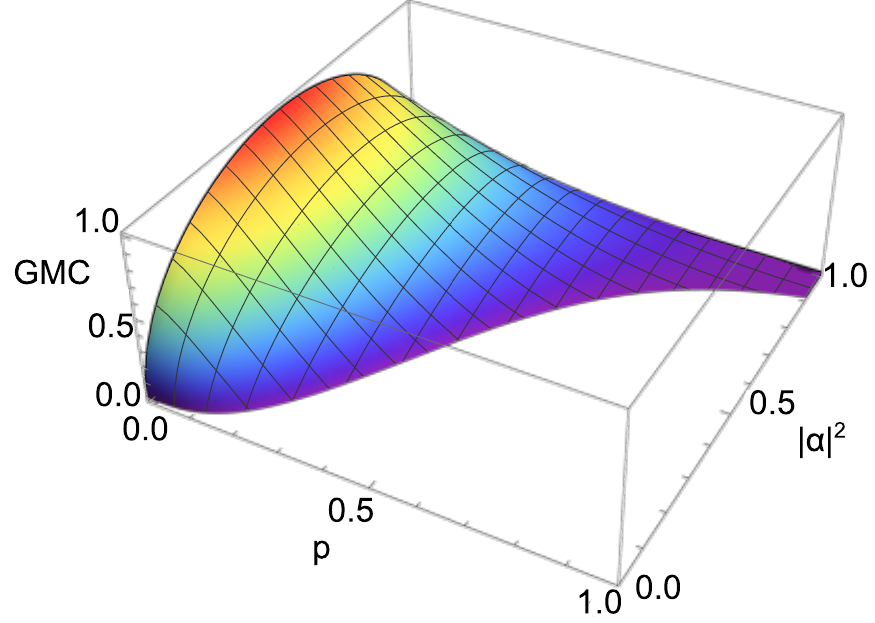} 
        \caption*{(b)}
    \end{minipage}
    \begin{minipage}[b]{0.32\textwidth}
        \centering
        \includegraphics[width=\textwidth]{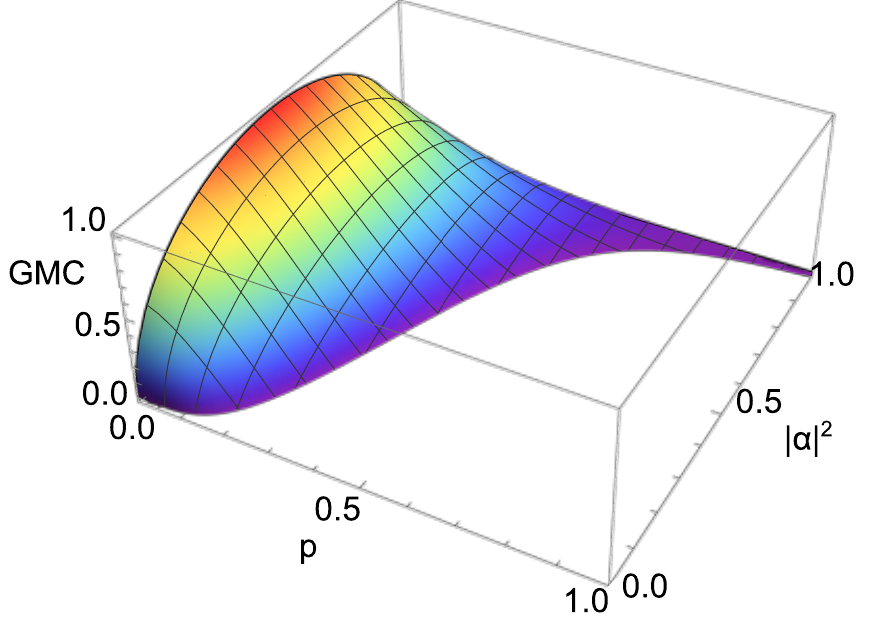} 
        \caption*{(c)}
    \end{minipage}
    \caption{Dynamics of two-, three-, and four-qubit entangled states under initial ADC. Evolution of $\rho_{(0,n)}(\alpha,p,0)$ for (a) $n=2$, (b) $n=3$, and (c) $n=4$. The states exhibit ESD (GMC death for $n\geq3$) for $|\alpha|^2 < 0.5, 0.9, 0.98$, respectively, and ADE otherwise.}
    \label{fig05}
\end{figure}

Fig.\,(\ref{fig05}) shows the GMC for $n$-qubit GHZ-type states under a single ADC with parameter $p$ (setting $p' = 0$) and state parameter $\alpha$. For $n = 2$, GMC exhibits ESD for $|\alpha|^2 < 0.5$ and ADE for $|\alpha|^2 \ge 0.5$. 
Increasing the number of qubits to $n = 3$ raises the ESD threshold to $|\alpha|^2 \leq 0.9$, and for $n = 4$ it further increases to $|\alpha|^2 \leq 0.98$. This demonstrates that ESD becomes more prevalent in GHZ-type states, and larger GHZ systems require a higher initial population imbalance to avoid ESD. 
Remarkably, maximally entangled GHZ-states [$\rho_n(\sqrt{0.5}), n\geq 3$] undergo ESD in the ADC in contrast to Bell-states [$\rho_2(\sqrt{0.5})$], which undergo ADE. These observations indicate the fragility of the GHZ and GHZ-type states under ADC.

\subsubsection{Two‐qubit GMC dynamics with NOT gates}

To protect entanglement against ADC noise, we study the impact of single- and two-qubit NOT operations ($m=1,2$) on a two-qubit entangled state $\rho_{(m,2)}(\alpha,p,p')$ as given below.

\begin{equation}
\begin{aligned}
\text{GMC}[\rho_{(1,2)}(\alpha,p,p')] &= 2 \max \left[0,\beta  q' \left(\alpha  q-\sqrt{p q \left(\beta ^2 (p p'+q) (p+p' q)+\alpha ^2 p'\right)}\right)\right], \\
\text{GMC}[\rho_{(2,2)}(\alpha,p,p')] &= 2 \max \left[0,\beta  q q' (\alpha -p)-p' q' \left(\alpha ^2+\beta  p^2\right)\right].
\end{aligned}
\end{equation}

\begin{figure}[ht!]
    \centering
    \begin{minipage}[b]{0.36\textwidth}
        \centering
        \includegraphics[width=\textwidth]{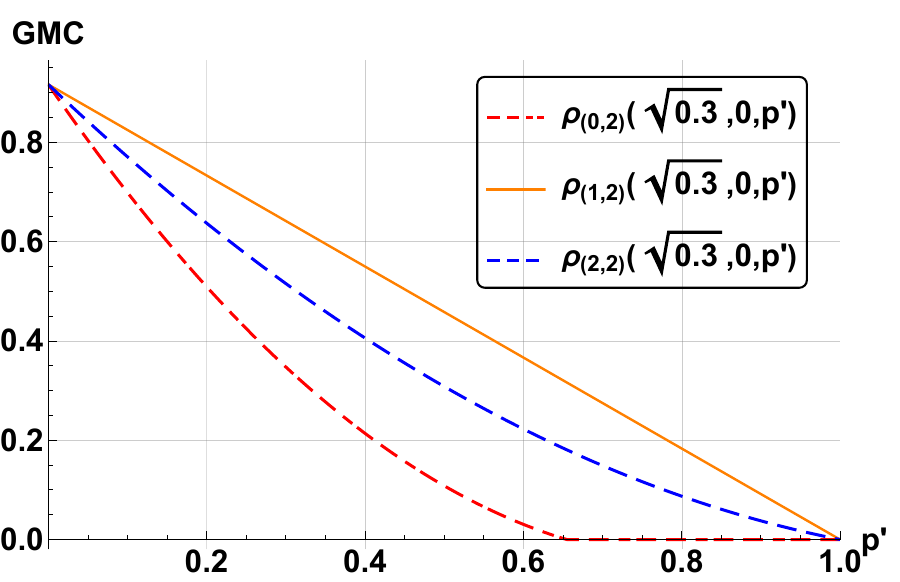} 
        \caption*{(a)}
    \end{minipage}
    \begin{minipage}[b]{0.36\textwidth}
        \centering
        \includegraphics[width=\textwidth]{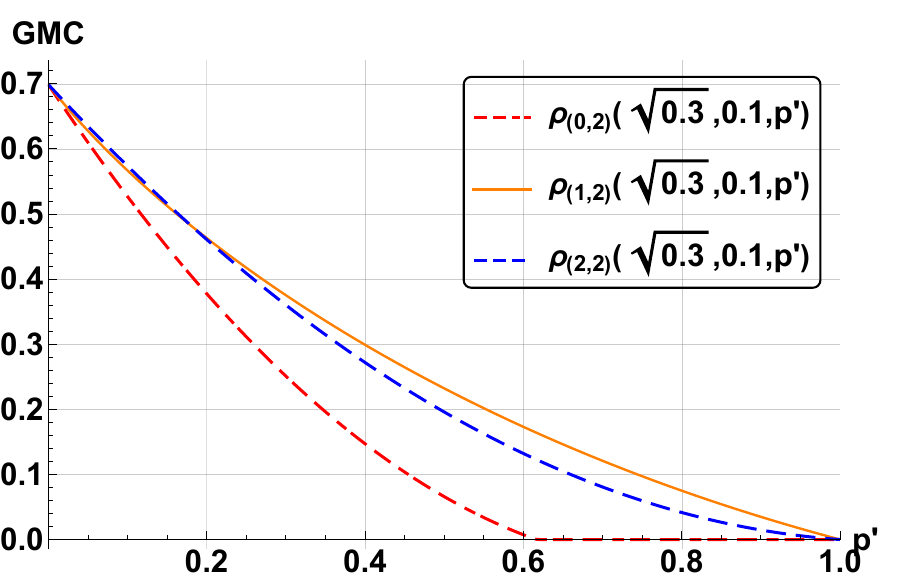} 
        \caption*{(b)}
    \end{minipage}
        \begin{minipage}[b]{0.36\textwidth}
        \centering
        \includegraphics[width=\textwidth]{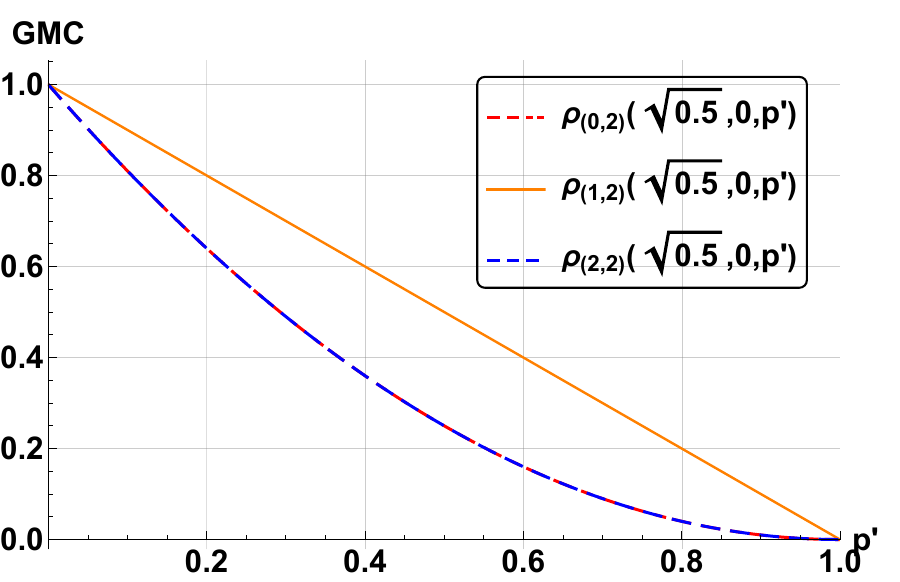} 
        \caption*{(c)}
    \end{minipage}
    \begin{minipage}[b]{0.36\textwidth}
        \centering
        \includegraphics[width=\textwidth]{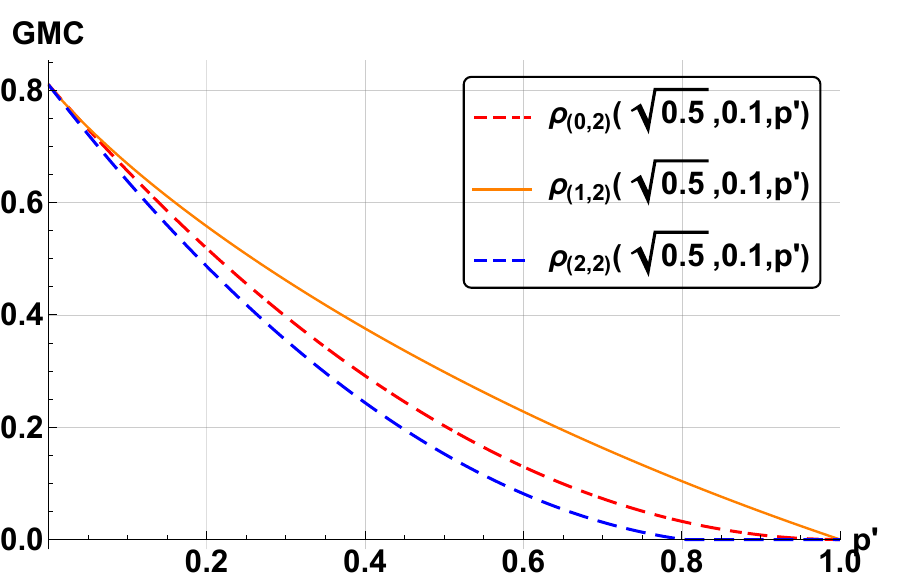}
        \caption*{(d)}
    \end{minipage}
   \caption{Dynamics of bipartite entangled state $\rho_2(\alpha)= (\alpha|00\rangle + \beta|11\rangle)(\alpha |00\rangle + \beta |11\rangle)^\dag$ under ADC. Initial states $\rho_2(\sqrt{0.3})$ undergoes ESD (dashed red curve) with $p'$ for (a) $p=0$, and (b) $p=0.1$. Initial maximally entangled state  $\rho_2(\sqrt{0.5})$ undergoes ADE with $p'$ for (c) $p=0$, and (d) $p=0.1$.} 
    \label{fig06}
\end{figure}

In Fig.\,(\ref{fig06}), we plot GMC vs.\ $p'$ for the two-qubit state $\rho_{(m,2)}(\alpha,p,p')$. We consider $|\alpha|^2=0.3$ [ESD regime; subfigures \ref{fig06}\,(a-b)], and $|\alpha|^2=0.5$ [maximally entangled state, ADE regime; subfigures \ref{fig06}\,(c-d)], for $p=0~\&~0.1$.  In each subfigure, the three curves correspond to zero ($m=0$), one ($m=1$), and two ($m=2$) NOT gates applied on the qubits between the two ADCs.  Across all subplots, the entanglement of a single-NOT state ($m=1$) decays slowest, maintains the highest amount of entanglement, and undergoes ADE, mitigating ESD most effectively.  Only in subfigure (b), the two-NOT state ($m=2$) briefly rises above the single-NOT state ($m=1$) for smaller values of $p'$, before dropping below it at $p'=0.19$. These results corroborate theoretical predictions that single‐qubit flips optimally protect Bell-type entanglement \cite{ESDM_Rau2008, ESDM_Singh2017}.

\subsubsection{Three‐qubit GMC dynamics with NOT gates}

The impact of one-, two-, and three-qubit NOT operations ($m=1,2,3$) on the GMC of a three-qubit entangled state $\rho_{(m,3)}(\alpha,p,p')$ is given below. These derivations are provided in Appendix \ref{Sec_Appendix-B} following the steps described in Section\,\ref{Sec_Background}.
\begin{equation}
\resizebox{\columnwidth}{!}{$
\begin{aligned}
\text{GMC}[\rho_{(1,3)}(\alpha,p,p')] &= 2 \max \left[0,~\alpha  \beta  (q q')^{3/2}-\beta  \left(q q'^{3/2} \sqrt{\alpha ^2 p p'+\beta ^2 p (p p'+q) (p+p' q)^2}\right.\right.  \left.\left.+2 \beta  q q'^{3/2} \sqrt{p (p p'+q)} (p+p' q)\right)\right], \\ 
\text{GMC}[\rho_{(2,3)}(\alpha,p,p')] &=2 \max \left[0,~\alpha  \beta  (q q')^{3/2}-\beta  \left(p \sqrt{q q'^3 \left(\beta ^2 (p p'+q)^2 (p+p' q)+\alpha ^2 p'^2\right)}\right.\right.  \left.\left.+2 \sqrt{p q q'^3 (p p'+q) \left(\beta ^2 p (p p'+q) (p+p' q)+\alpha ^2 p'\right)}\right)\right], \\ 
\text{GMC}[\rho_{(3,3)}(\alpha,p,p')] &= 2 \max \left[0,~\alpha  \beta  (q q')^{3/2}-3 \sqrt{q'^3 \left(\beta ^4 p^3 (p p'+q)^3+\alpha ^2 \beta ^2 p p' (p p'+q) (2 p p'+q)+\alpha ^4 p'^3\right)}\right].
\end{aligned}
$}
\end{equation}

\begin{figure}[ht!]
    \centering
    \begin{minipage}[b]{0.35\textwidth}
        \centering
        \includegraphics[width=\textwidth]{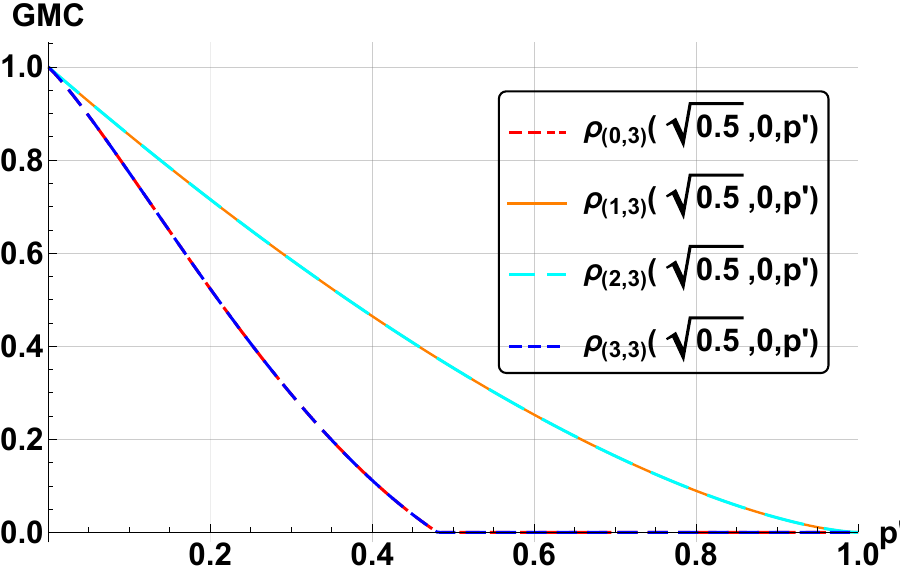} 
        \caption*{(a)}
    \end{minipage}
    \begin{minipage}[b]{0.35\textwidth}
        \centering
        \includegraphics[width=\textwidth]{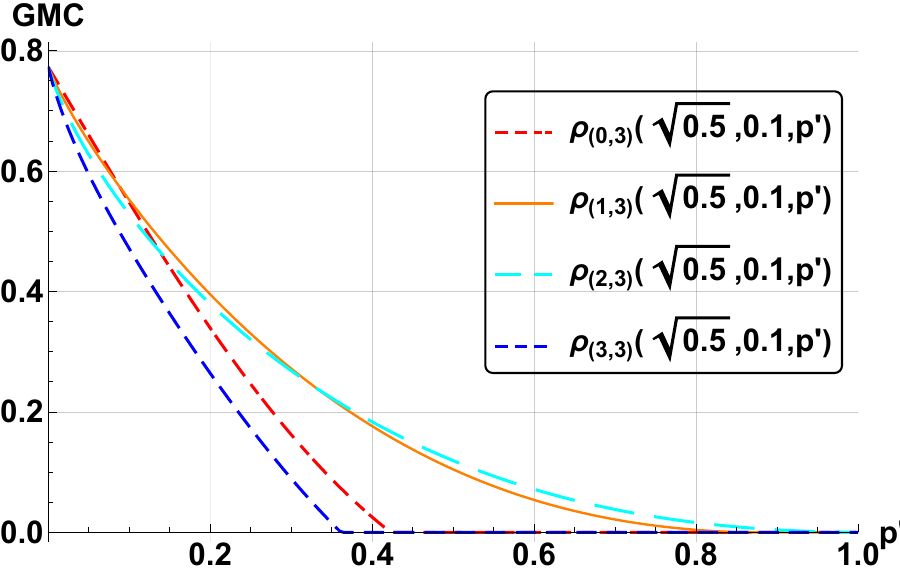} 
        \caption*{(b)}
    \end{minipage}
     \begin{minipage}[b]{0.35\textwidth}
        \centering
        \includegraphics[width=\textwidth]{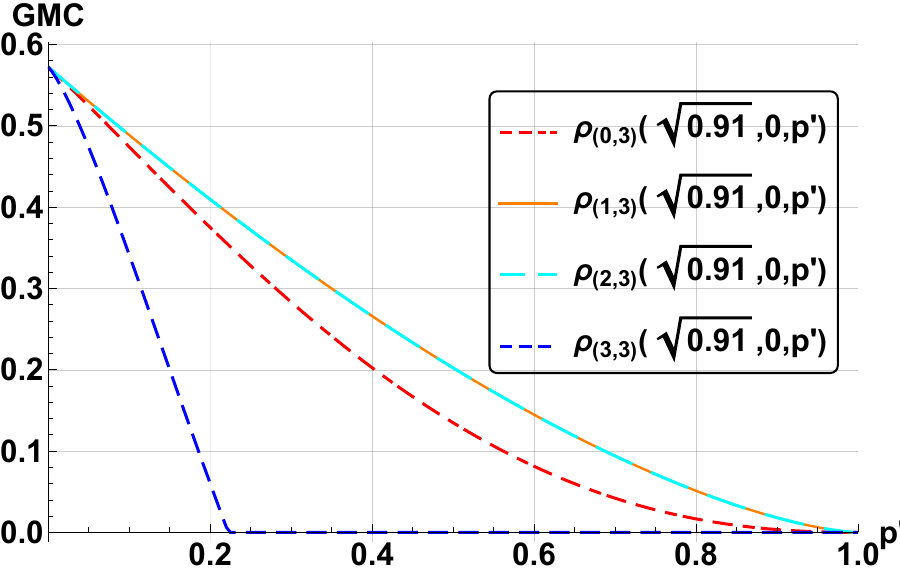} 
        \caption*{(c)}
    \end{minipage}
    \begin{minipage}[b]{0.35\textwidth}
        \centering
        \includegraphics[width=\textwidth]{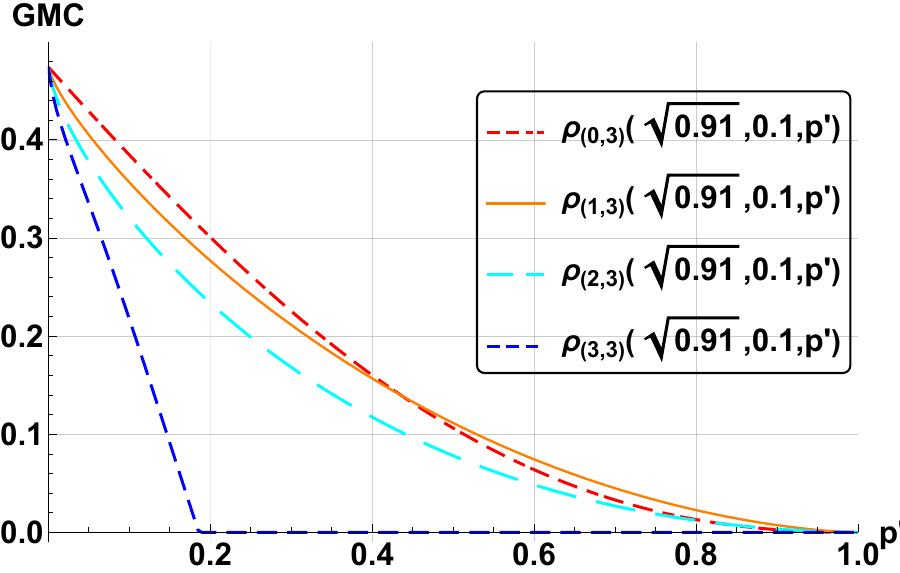} 
        \caption*{(d)}
    \end{minipage}
    \caption{Dynamics of tripartite entangled state  $\rho_3(\alpha)= (\alpha|000\rangle + \beta|111\rangle) (\alpha |000\rangle + \beta |111\rangle)^\dag$ under ADC. Initial maximally entangled state $\rho_3(\sqrt{0.5})$ undergoes ESD with $p'$ for (a) $p=0$, and (b) $p=0.1$. Initial state $\rho_3(\sqrt{0.91})$ undergoes ADE with $p'$ for (c) $p=0$, and (d) $p=0.1$.}
    \label{fig07}
\end{figure}

In Fig.\,(\ref{fig07}), we extend the analysis to three-qubit systems with $|\alpha|^2=0.5$ [maximally entangled state, ESD regime; subfigures \ref{fig07}\,(a-b)], and $|\alpha|^2=0.91$ [ADE regime; subfigures \ref{fig07}\,(c-d)], for $p = 0~\&~0.1$.
In the subfigure \ref{fig07}(a), one-NOT ($m=1$) and two-NOT ($m=2$) flipped states undergo ADE, whereas no-NOT ($m=0$) and all-NOT ($m=3$) flipped states suffer ESD. 
In the subfigure \ref{fig07}(b), two-NOT ($m=2$) flipped states undergo ADE, whereas no-NOT ($m=0$), one-NOT ($m=1$) and all-NOT ($m=3$) flipped states suffer ESD. 
In subfigure \ref{fig07}(b), $\rho_{(2,3)}(\sqrt{0.5},0.1,p')$ surpasses $\rho_{(1,3)}(\sqrt{0.5},0.1,p')$ around $p'=0.34$ and then retains higher concurrence asymptotically.
Also, in subfigure\,\ref{fig07}\,(c), all curves except $\rho_{(3,3)}(\sqrt{0.91},0,p')$ decay smoothly, but $\rho_{(2,3)}(\sqrt{0.91},0,p')$ and $\rho_{(1,3)}(\sqrt{0.91},0,p')$ tie for the slowest decay, indicating that partial flips on two qubits of a three-partite state are as effective as single flips in preserving GMC in the ADE regime. 
However, in subfigure\,\ref{fig07}\,(d), $\rho_{(1,3)}(\sqrt{0.91},0.1,p')$ surpasses $\rho_{(0,3)}(\sqrt{0.91},0.1,p')$ around $p'=0.43$, indicating that higher entanglement preservation for single‐qubit flips is only at higher $p'$. Notably, in subfigures\,\ref{fig07}\,(b-d), flipping all qubits hastens ESD.

\subsubsection{Four‐qubit GMC dynamics with NOT gates}

The impact of one-, two-, three-, and four-qubit NOT operations ($m=1,2,3,4$) on the GMC of a four-qubit entangled state $\rho_{(m,4)}(\alpha,p,p')$ is given below.
\begin{equation}
\resizebox{\columnwidth}{!}{$
\begin{aligned} 
\text{GMC}[\rho_{(1,4)}(\alpha,p,p')] &= 2 \max \left[0,~\alpha \beta q^2 q'^2 - \beta  q'^2 \sqrt{p q^3 \left(\beta ^2 (p p'+q) (p+p' q)^3+\alpha ^2 p'\right)}-6 \sqrt{\beta ^4 p q^3 q'^4 (p p'+q) (p+p' q)^3}\right],\\
\text{GMC}[\rho_{(2,4)}(\alpha,p,p')] &= 2 \max \left[0,~\alpha \beta q^2 q'^2 -\beta  q q'^2 \left(4 \beta  p^3 p'+4 \beta q p^2 \left(p'^2+1\right)+p \sqrt{\beta ^2 (p p'+q)^2 (p+p' q)^2+\alpha ^2 p'^2}\right.\right.  \\ &\left.\left.+4 \beta  p p' q^2+2 \sqrt{p (p p'+q) \left(\beta ^2 p (p p'+q) (p+p' q)^2+\alpha ^2 p'\right)}\right)\right], \\ 
\text{GMC}[\rho_{(3,4)}(\alpha,p,p')] & = 2 \max \left[0,~\alpha \beta q^2 q'^2 -\sqrt{\beta ^2 p^3 q q'^4 \left(\beta ^2 (p p'+q)^3 (p+p' q)+\alpha ^2 p'^3\right)}\right. \left. -3 \beta  q'^2 (p p'+q) \sqrt{\beta ^2 p^3 q (p p'+q) (p+p' q)+\alpha ^2 p p' q} \right.  \\ &\left.-3 \beta  p q'^2 \sqrt{\alpha ^2 p'^2 q (p p'+q)+\beta ^2 p q (p+p' q) (p p'+q)^3}\right], \\ 
\text{GMC}[\rho_{(4,4)}(\alpha,p,p')] &= 2 \max \left[0,~\alpha \beta q^2 q'^2 - q'^2 \left(3 \beta ^2 p^2 (p p'+q)^2\right.\right.  \left.\left.+4 \sqrt{\beta ^4 p^4 (p p'+q)^4+\alpha ^2 \beta ^2 p p' (p p'+q) \left(2 p^2 p'^2+2 p p' q+q^2\right)+\alpha ^4 p'^4}+3 \alpha ^2 p'^2\right)\right].
\end{aligned}
$}
\end{equation}

\begin{figure}[ht!]
    \centering
    \begin{minipage}[b]{0.36\textwidth}
        \centering
        \includegraphics[width=\textwidth]{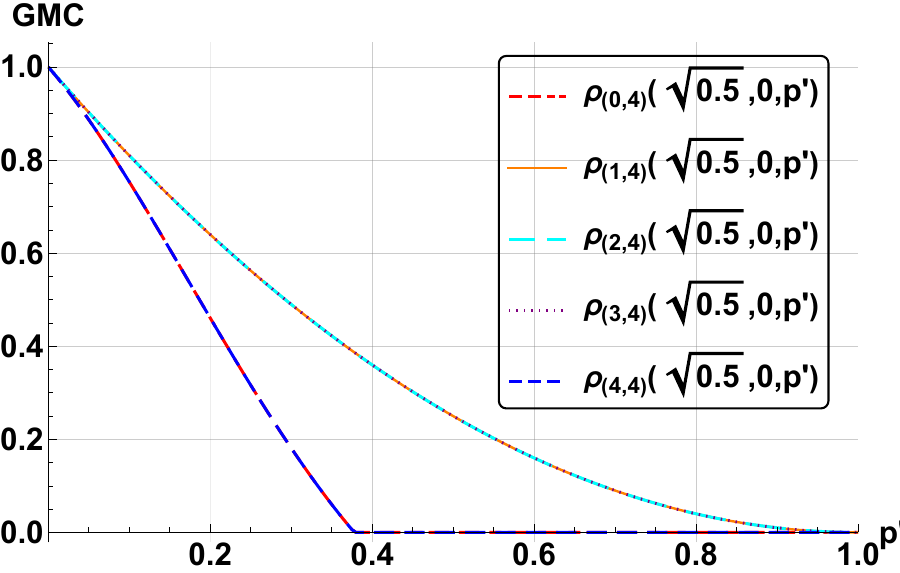} 
        \caption*{(a)}
    \end{minipage}
    \begin{minipage}[b]{0.36\textwidth}
        \centering
        \includegraphics[width=\textwidth]{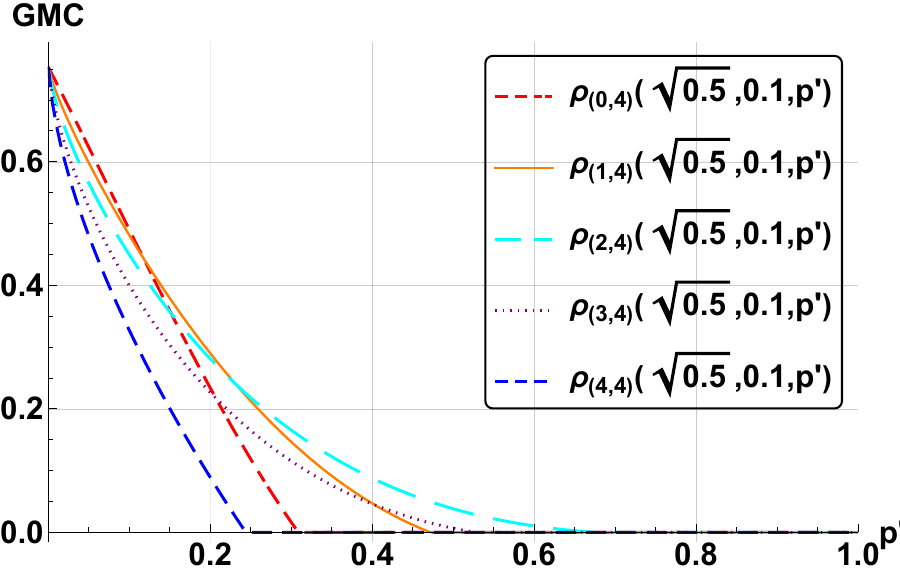} 
        \caption*{(b)}
    \end{minipage}
    \begin{minipage}[b]{0.36\textwidth}
        \centering
        \includegraphics[width=\textwidth]{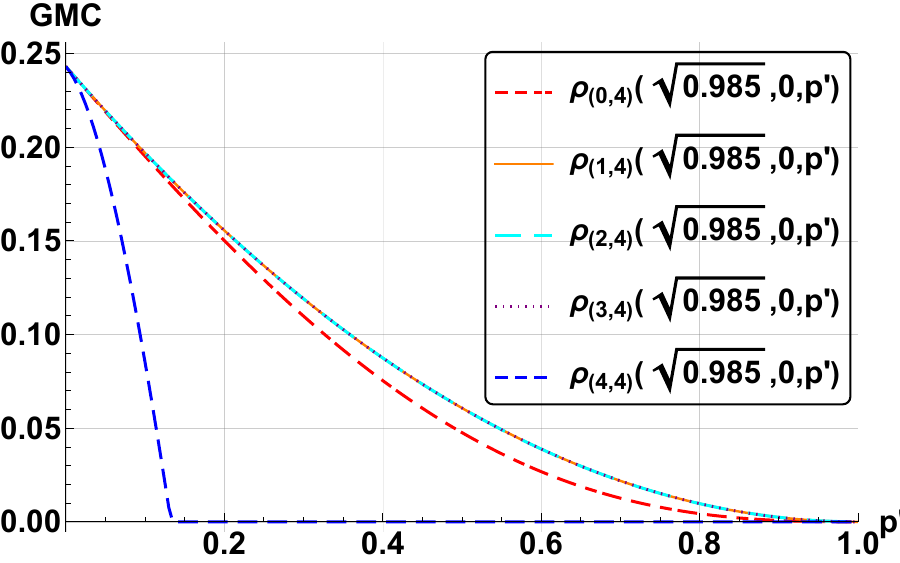} 
        \caption*{(c)}
    \end{minipage}
    \begin{minipage}[b]{0.36\textwidth}
        \centering
        \includegraphics[width=\textwidth]{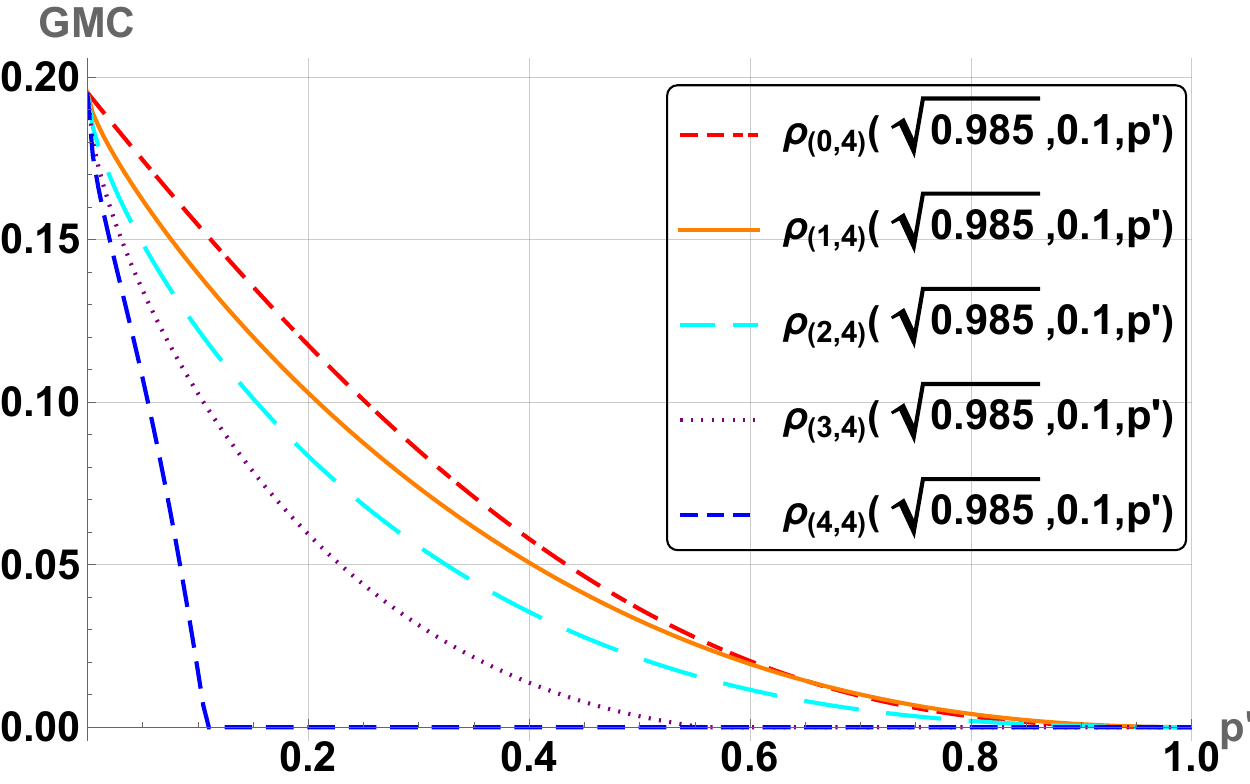} 
        \caption*{(d)}
    \end{minipage}
    \caption{Dynamics of four-party entangled state  $\rho_4(\alpha)= (\alpha|0000\rangle + \beta|1111\rangle)(\alpha |0000\rangle + \beta |1111\rangle)^\dag$. Initial maximally entangled state $\rho_4(\sqrt{0.5})$ undergoes ESD with $p'$ for (a) $p=0$, and (b) $p=0.1$. Initial state $\rho_4(\sqrt{0.985})$ undergoes ADE with $p'$ for (c) $p=0$, and (d) $p=0.1$.}
    \label{fig08}
\end{figure}

In Fig.\,(\ref{fig08}), we further extend the analysis to four-qubit systems for $|\alpha|^2=0.5$ [maximally entangled states, ESD regime; subfigures \ref{fig08}\,(a-b)], and $|\alpha|^2=0.91$ [ADE regime; subfigures \ref{fig08}\,(c-d)].   
In subfigure\,\ref{fig08}\,(a), GMC curves for $\rho_{(m,4)}(\sqrt{0.5},0,p')$ corresponding to partial flips $m=1,2,3$ coincide and decay asymptotically, while $m=0,4$ also coincide but undergo ESD. 
In the subfigure\,\ref{fig08}\,(b), all partial flips ($m=1,2,3$) delay ESD. Notably, $\rho_{(2,4)}(\sqrt{0.5},0.1,p')$ overtakes $\rho_{(1,4)}(\sqrt{0.5},p,p')$ at $p'=0.24$ and delays ESD maximally until $p'=0.67$.  
In the subfigure\,\ref{fig08}\,(c), all partial‐flip states ($m=1,2,3$) coincide and delay ESD, whereas the flip on all qubits ($m=4$) hastens ESD. 
In the subfigure\,\ref{fig08}\,(d), the advantage of NOT-gates at $p=0.1$ appears only after $p'=0.66$ when $\rho_{(1,4)}(\sqrt{0.985},0.1,p')$ surpasses $\rho_{(0,4)}(\sqrt{0.985},0.1,p')$ and delays ESD. In this case, only $m=1,2$ delay ESD, whereas $m=3,4$ hasten ESD.
Notably, NOT on all qubits gives rise to the hastening of ESD in subfigures \ref{fig08}\,(b-d).
The crossing behaviours of $\rho_{(2,4)}(\alpha,p,p')$ and $\rho_{(1,4)}(\alpha,p,p')$ in subfigure \ref{fig08}\,(b) illustrate that two bit-flips can sometimes outperform a single flip in four-qubit and three-qubit systems at higher $p'$ for $p>0$, which is investigated thoroughly for a range of $\alpha$ and $p$ parameters in Appendix\,(\ref{Sec_Appendix-A}). 

\subsubsection{\texorpdfstring{Single‐NOT for $n$-qubit states at $p=0$}{Single-NOT for n-qubit states at p = 0}}

\begin{figure}[ht!]
    \centering
    \begin{minipage}[b]{0.3\textwidth}
        \centering
        \includegraphics[width=\textwidth]{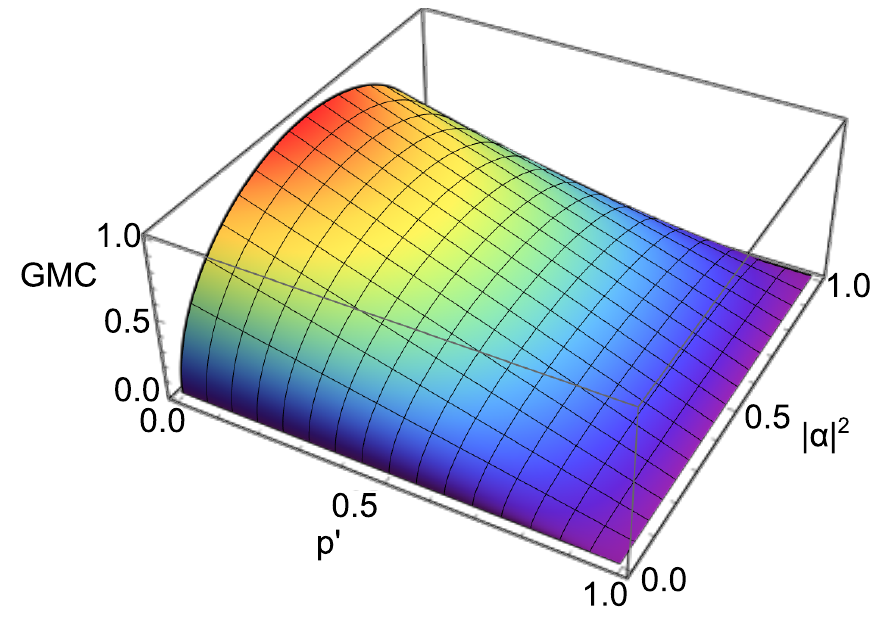} 
        \caption*{(a)}
    \end{minipage}
    \begin{minipage}[b]{0.32\textwidth}
        \centering
        \includegraphics[width=\textwidth]{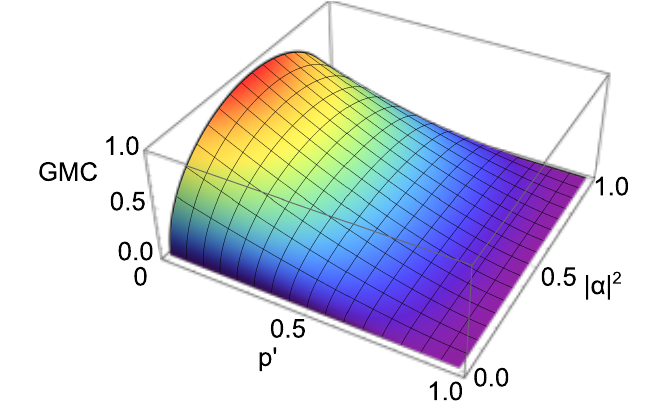} 
        \caption*{(b)}
    \end{minipage}
    \begin{minipage}[b]{0.32\textwidth}
        \centering
        \includegraphics[width=\textwidth]{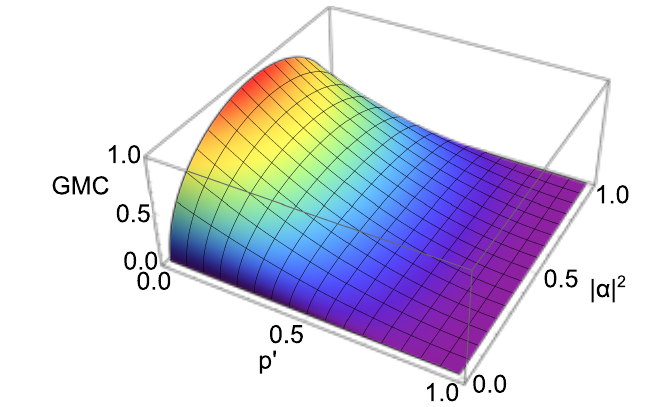} 
        \caption*{(c)}
    \end{minipage}
    \caption{Dynamics of 2,3,4-qubit entangled states with one-NOT gate (at $p=0$) followed by ADC $(p')$. Evolution of states (a) $\rho_{1,2}(\alpha,0,p')$, (b) $\rho_{1,3}(\alpha,0,p')$, and (c) $\rho_{1,4}(\alpha,0,p')$.}
    \label{fig09}
\end{figure}

The impact of a single NOT operation ($m=1$) on two-, three-, and four-qubit entangled states $\rho_{(1,n)}(\alpha,p,p')$ at $p=0$ is shown in Fig.,(\ref{fig09}), where the GMC is plotted as a function of $\alpha$ and $p'$ for $n=2,3,4$.
All three states decay asymptotically (ADE), confirming that even a single local flip suffices to avert ESD in GHZ states of these dimensions.  
Even though all single-NOT flipped states undergo ADE, the amount of entanglement (GME) gradually decreases as the number of qubits increases for a given state and damping parameter ($|\alpha|^2~\&~p'$),  reflecting the increasing susceptibility of larger multipartite systems.

\subsection{Teleportation fidelity dynamics in the ADC with NOT gates}
\label{Sec_FidelityResults}

In this section, we will study the performance of entangled states in a quantum communication task, namely, quantum teleportation or CQT. 
The teleportation fidelity for two-, three-, and four-qubit entangled states $\rho_{(0,n)}(\alpha,p,p')$ with $n=2,3,4$, respectively, in the presence of the two ADCs (without NOT gates) is given below.
\begin{equation}
\resizebox{\columnwidth}{!}{$
\begin{aligned}
    \mathcal{F}[\rho_{(0,2)}(\alpha,p,p')] &= \frac13 \max[1+2q(p+p'q)q'\beta^2,1+\alpha^2+2qq'\alpha\beta+(1+2qq'(-1+qq'))\beta^2], \\
    \mathcal{F}[\rho_{(0,3)}(\alpha,p,p')] &= \frac13 \max[2+2qq'\beta(\sqrt{qq'}\alpha +(-1+qq')\beta),\frac12(3+\sqrt{1+4qq'\beta^2(-1+qq'(4qq'\alpha^2+\beta^2))}),\frac12(3+|1-2qq'\beta^2|)], \\
    \mathcal{F}[\rho_{(0,4)}(\alpha,p,p')] &= \frac23 (\alpha^2 + 
   q^2 q'^2 \alpha \beta + ((p + p' q)^2 +  q (p + p' q) q' + q^2 q'^2) \beta^2).
\end{aligned}
$}
\end{equation}

\begin{figure}[ht!]
    \centering
    \begin{minipage}[b]{0.32\textwidth}
        \centering
        \includegraphics[width=\textwidth]{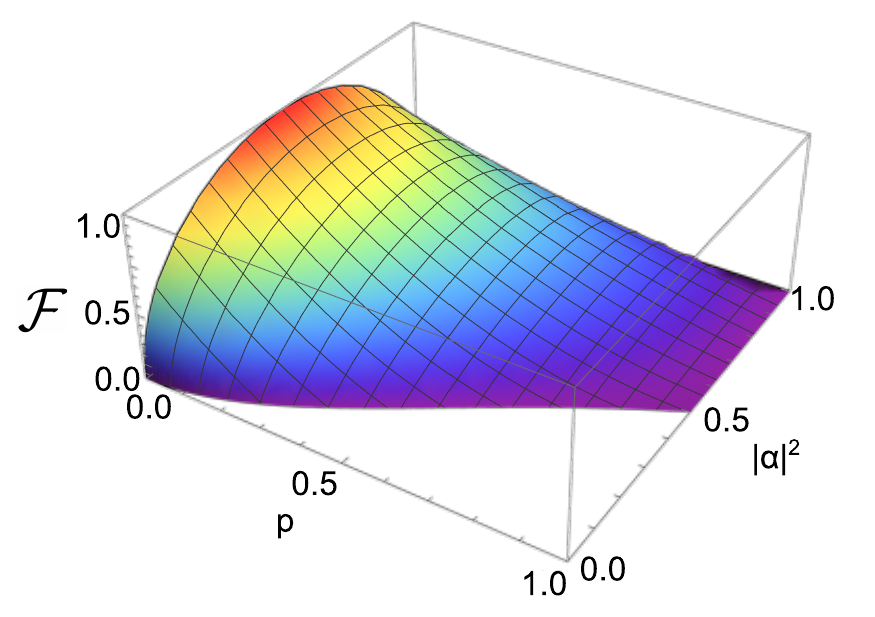} 
        \caption*{(a)}
    \end{minipage}
    \begin{minipage}[b]{0.32\textwidth}
        \centering
        \includegraphics[width=\textwidth]{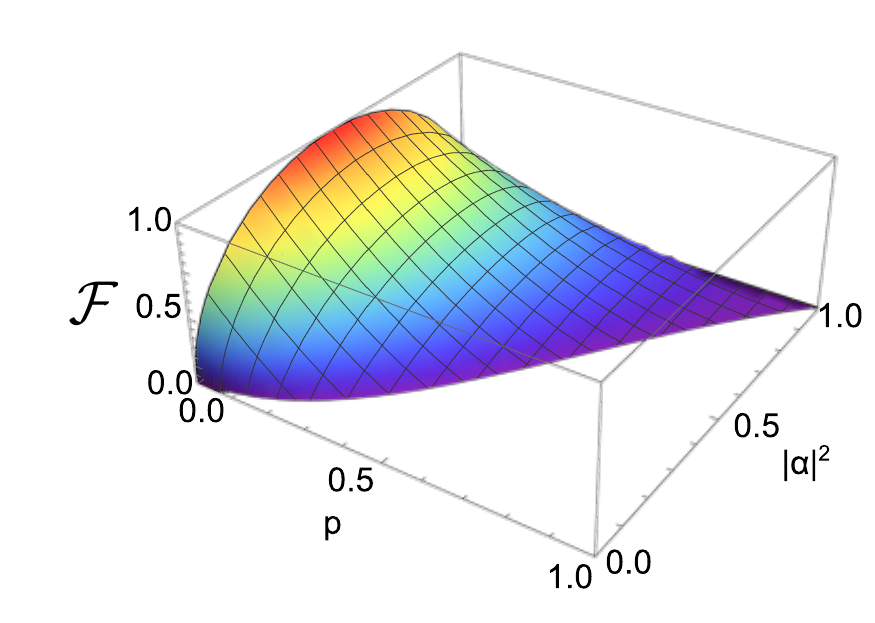} 
        \caption*{(b)}
    \end{minipage}
    \begin{minipage}[b]{0.32\textwidth}
        \centering
        \includegraphics[width=\textwidth]{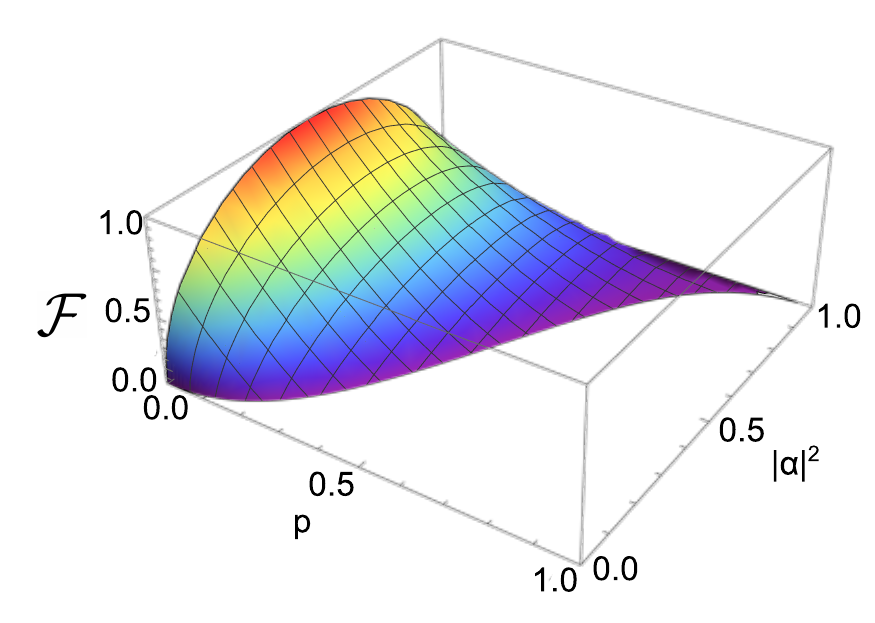} 
        \caption*{(c)}
    \end{minipage}
    \caption{Dynamics of teleportation fidelity ($\mathcal{F}$) using two-, three-, and four-qubit entangled states with initial ADC ($p$) and state parameter ($|\alpha|^2$). Only fidelity above the classical limit $\mathcal{F}=2/3$ is plotted. Teleportation fidelity vs. $p,~|\alpha|^2$ for (a) $\rho_{(0,2)}(\alpha,p,0)$, (b) $\rho_{(0,3)}(\alpha,p,0)$, and (c) $\rho_{(0,4)}(\alpha,p,0)$.}
    \label{fig9}
\end{figure}

The plots in Fig.\,(\ref{fig9}) show the teleportation fidelity of $n$-qubit GHZ-type states under a single ADC with parameter $p$ (setting $p'$ = 0). For $n = 2$, teleportation fidelity exhibits finite-time crossover to classical limit ($\mathcal{F}=2/3$) for $|\alpha|^2<0.5$, and approaches classical limit asymptotically for $|\alpha|^2\geq 0.5$, exhibiting the same behaviour as GMC. The teleportation plots show a different behaviour for $n = 3,4$ compared to GMC in Fig. (\ref{fig05}).
Increasing the qubits to $n=3$ raises the classical limit crossover threshold to $|\alpha|^2 \rightarrow 1$, and for $n = 4$, teleportation fidelity exhibits crossover to classical limit for all values of $|\alpha|^2\in(0,1)$. 
This qualitative behaviour of teleportation fidelity does not match that of GMC in Fig. (\ref{fig05}), but it in turn validates the similar points that larger GHZ systems require higher initial population imbalance to avoid teleportation fidelity crossover to classical limit, and that it is more common for GHZ-type states. However, when comparing the sudden death and asymptotic decay ranges, teleportation fidelity is more susceptible to decoherence than GMC.

\subsubsection{Teleportation using a two-qubit entangled resource shared through ADC}
\label{Sec_2QubitTeleportation}

The impact of one- and two-qubit NOT operations ($m=1,2$) on the teleportation fidelity of a two-qubit state $\rho_{(m,2)}(\alpha,p,p')$ used as an entangled resource is given below.
\begin{equation}
\begin{aligned}
    \mathcal{F}\left[\rho_{(1,2)}(\alpha,p,p')\right] &= \frac{1}{3}\max[1+p'\alpha^2-(-1+q'+2(-1+q)qq')\beta^2,1+q'(\alpha^2+2q\alpha\beta+(1+2(-1+q)qq')\beta^2)],\\
    \mathcal{F}\left[\rho_{(2,2)}(\alpha,p,p')\right] &= \frac{1}{3}\max[1+2p'q'\alpha^2+2p(pp'+q)q'\beta^2, 1+(p'^2+q'^2) \alpha^2+2qq'\alpha\beta + ((pp'+q)^2+p^2q'^2)\beta^2].
\end{aligned}
\end{equation}

\begin{figure}[ht!]
    \centering
    \begin{subfigure}[b]{0.36\textwidth}
        \centering
        \includegraphics[width=\textwidth]{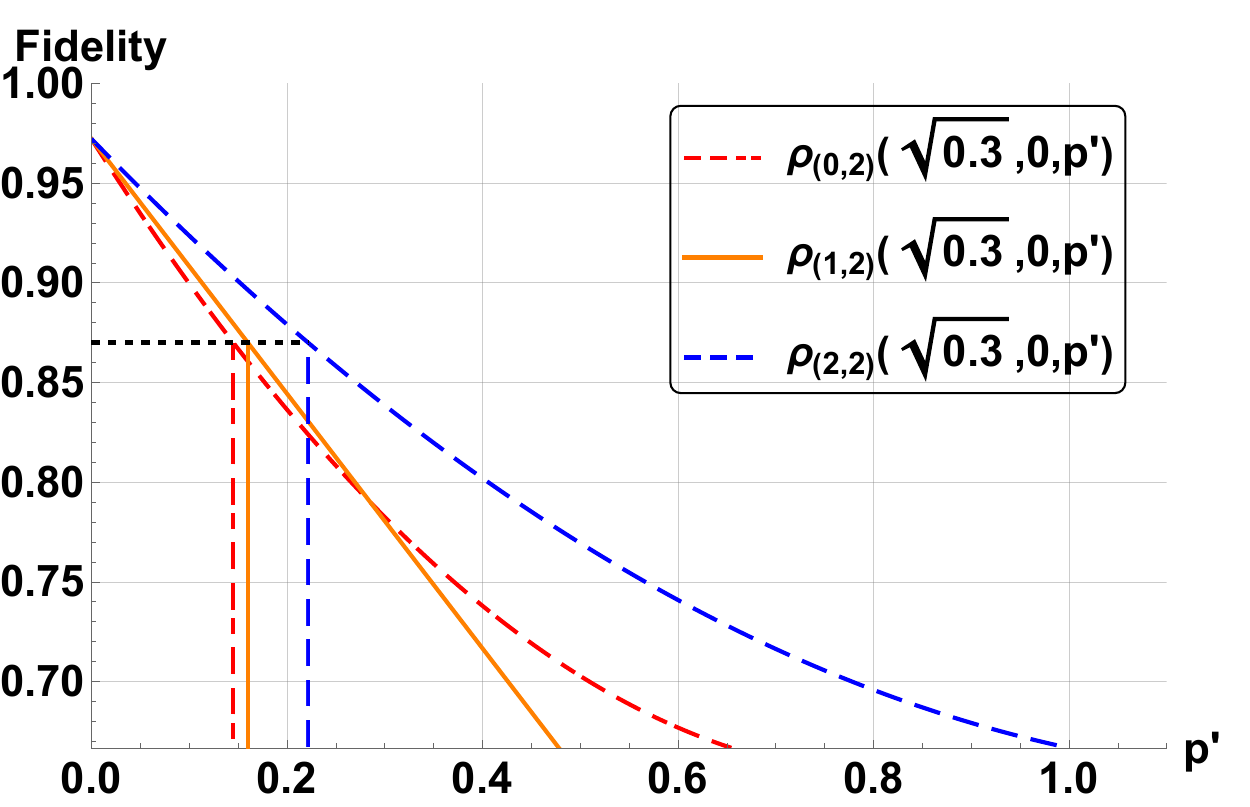} 
        \caption{}
        \label{fig_9a}
    \end{subfigure}
        \begin{subfigure}[b]{0.36\textwidth}
        \centering
        \includegraphics[width=\textwidth]{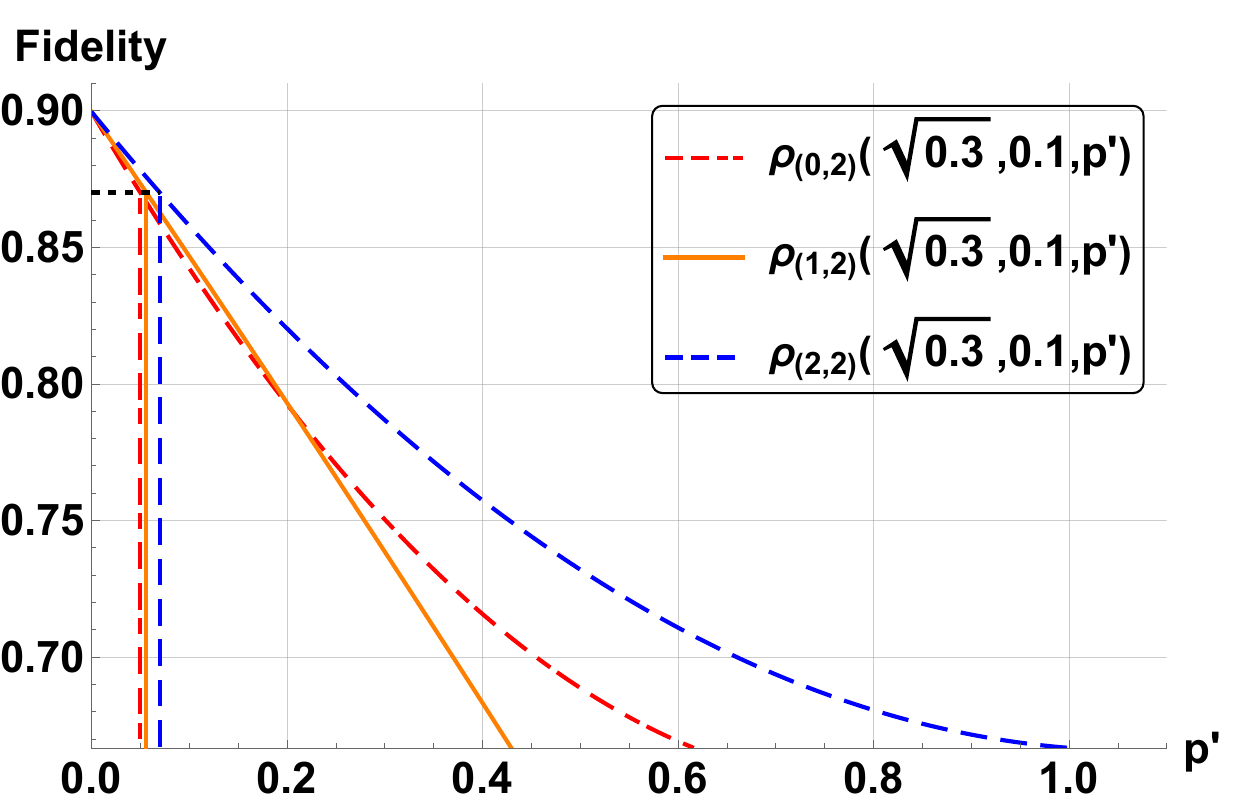} 
        \caption{}
        \label{fig_9b}
    \end{subfigure}
     \begin{subfigure}[b]{0.36\textwidth}
        \centering
        \includegraphics[width=\textwidth]{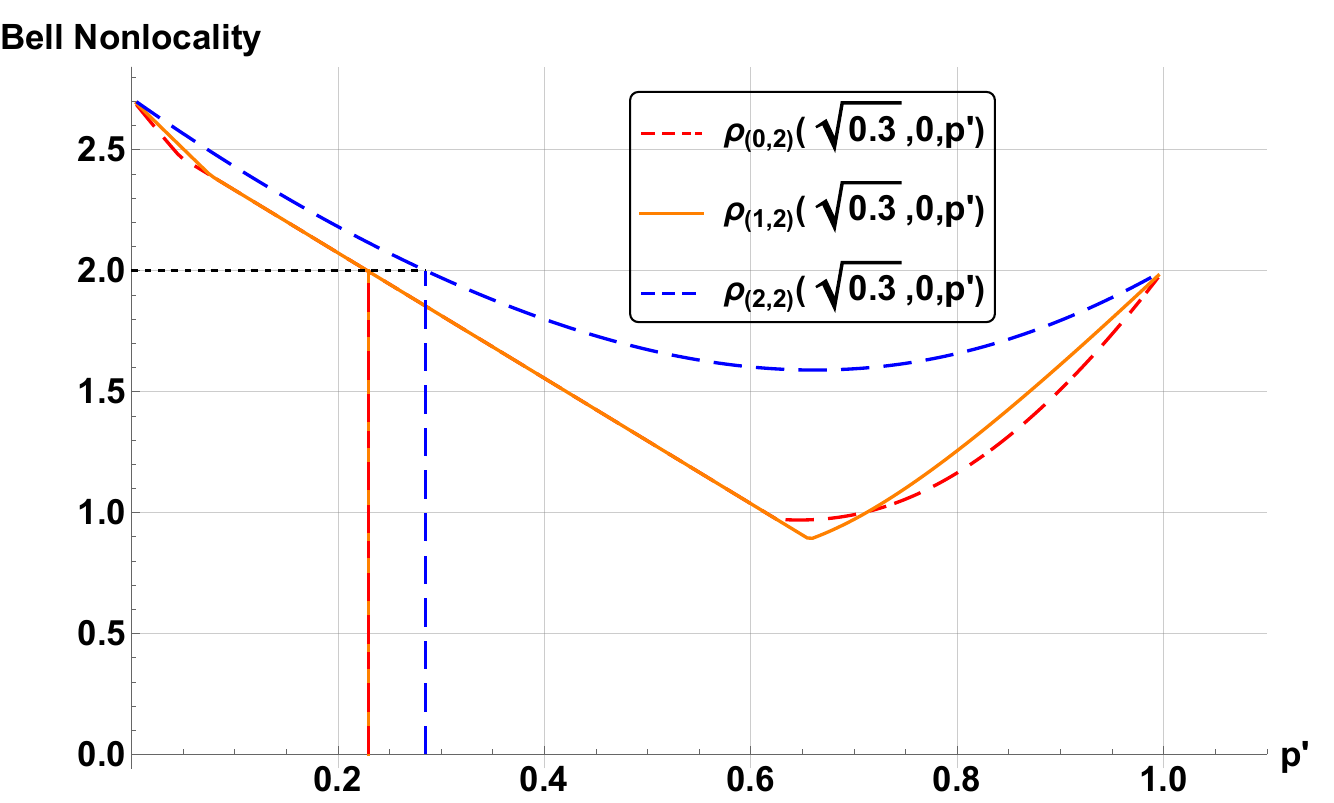} 
        \caption{}
        \label{fig_9c}
    \end{subfigure}
    \begin{subfigure}[b]{0.36\textwidth}
        \centering
        \includegraphics[width=\textwidth]{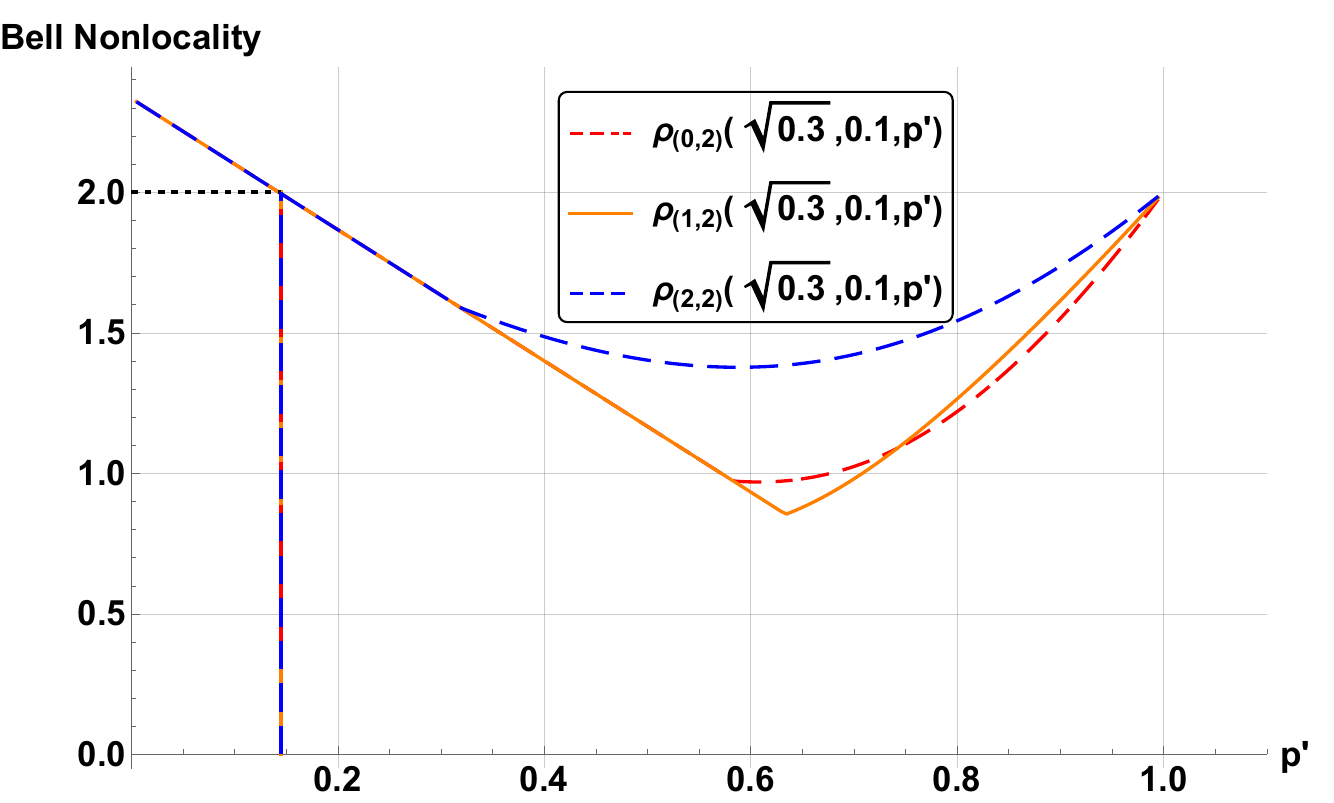} 
        \caption{}
        \label{fig_9d}
    \end{subfigure}
    \caption{Dynamics of teleportation fidelity and Bell nonlocality for a bipartite entangled state  $\rho_2(\alpha)= (\alpha|00\rangle + \beta|11\rangle) (\alpha |00\rangle + \beta |11\rangle)^\dag$ under ADC noise. Teleportation fidelity for $\rho_{(m,2)}(\sqrt{0.3},p,p')$ vs. $p'$ for (a) $p=0$, and (b) $p=0.1$. Bell nonlocality for  $\rho_{(m,2)}(\sqrt{0.3},p,p')$ vs. $p'$ for (c) $p=0$, and (d) $p=0.1$. We find that the teleportation fidelity decreases first below the nonlocal threshold $\mathcal{F}_{\rm LHV} \approx 0.87$, followed by the disappearance of Bell–CHSH violation at 2; subsequently, $\mathcal{F}$ drops below the classical limit of 2/3, and eventually the GMC goes to zero (compare subfigures a and c, or b and d).}
    \label{fig11}
\end{figure}

In Figs. (\ref{fig_9a}) and (\ref{fig_9b}), we show teleportation fidelity for state $\rho_{(m,2)}(\alpha,p,p')$ vs. $p'$ with $|\alpha|^2 = 0.3$, for $p=0$ and $p=0.1$, respectively. The three curves correspond to no- ($m=0$), one- ($m=1$), and two-qubit ($m=2$) NOT gates applied between ADCs.
Surprisingly, we observe that among all subplots, teleportation fidelity of one-NOT state decays the fastest and approaches the classical limit even before the no-NOT state, whereas the two-NOT state decays the slowest and asymptotically. This fidelity behaviour of $\rho_{(1,2)}(\alpha,p,p')$ contradicts the corresponding GMC as shown in Fig\,\ref{fig06}(a-b). 
To analyze this anomaly further, we show the Bell‐CHSH nonlocality for these states in subfigures (\ref{fig_9c}) and (\ref{fig_9d}). 
We observe that the teleportation fidelity first drops below the nonlocality threshold $\mathcal{F}_{\rm LHV} \approx 0.87$, then Bell‐CHSH violation ceases, then $\mathcal{F}$ falls below the classical limit ($2/3$), and finally GMC vanishes, consistent with the hierarchy of these correlations \cite{GISIN1996, Paulson2021}. In this section, we primarily focus on $\rho_{(m,n)}(\alpha,p,p')$ states with $|\alpha|^2 < 0.5$, for which applying NOT gates to all qubits under ADC yields improved teleportation fidelity.

The apparent contradiction that single-NOT flips producing better GMC preservation but worse teleportation fidelity than the no-NOT or two-NOT cases can be understood from two closely related physical facts: (i) the ADC acts asymmetrically on the computational basis, leaving $|0\rangle$ unchanged while decaying $|1\rangle\to|0\rangle$, and (ii) teleportation fidelity depends not only on the amount of entanglement but on the distribution of population/coherence across the maximally-entangled subspace used for teleportation. 
As mentioned, the Bell states $\ket{\Phi^+}=(\ket{00}+\ket{11})/\sqrt{2}$ and $\ket{\Psi^+}=(\ket{01}+\ket{10})/\sqrt{2}$ are related by a local $\sigma_x$ on one qubit. Under identical ADC acting on both qubits, the $\ket{\Phi^+}$ states retains the noise-free $\ket{00}$ component while only $\ket{11}$ is subject to decay; by contrast, each component of $\ket{\Psi^+}$ contains a single excitation and is thus more uniformly vulnerable to amplitude damping. 
As analyzed in the literature, these structural differences imply that the ADC-evolved $\ket{\Phi^+}$-type noisy states typically yield higher teleportation fidelity than the $\ket{\Psi^+}$-type states, even when straightforward entanglement measures (GMC) may rank them oppositely \cite{NoisyTeleport_Fortes2015}. In the NOT-gate protocol, applying a single NOT flip between ADC stages effectively transforms the $\ket{\Phi^+}$-type state into the $\ket{\Psi^+}$-type state, causing fidelity to degrade faster than in the no-NOT (or two-NOT) cases despite comparable or larger concurrence for the one-NOT state.

\subsubsection{CQT using a three-qubit entangled resource shared through ADC}

\begin{figure}[ht!]
    \centering 
    \begin{subfigure}[b]{0.36\textwidth}
        \centering
        \includegraphics[width=\textwidth]{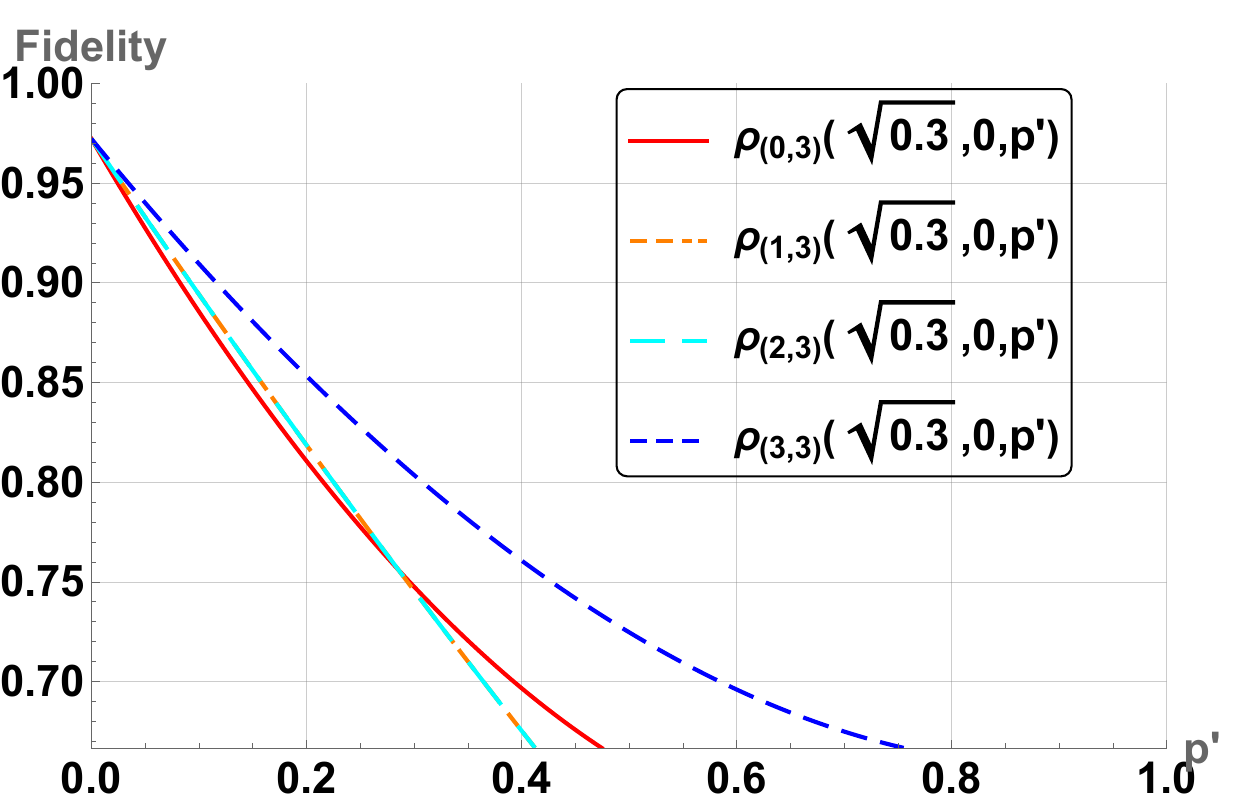} 
        \caption{}
        \label{fig12a}
    \end{subfigure} 
    \begin{subfigure}[b]{0.36\textwidth}
        \centering
        \includegraphics[width=\textwidth]{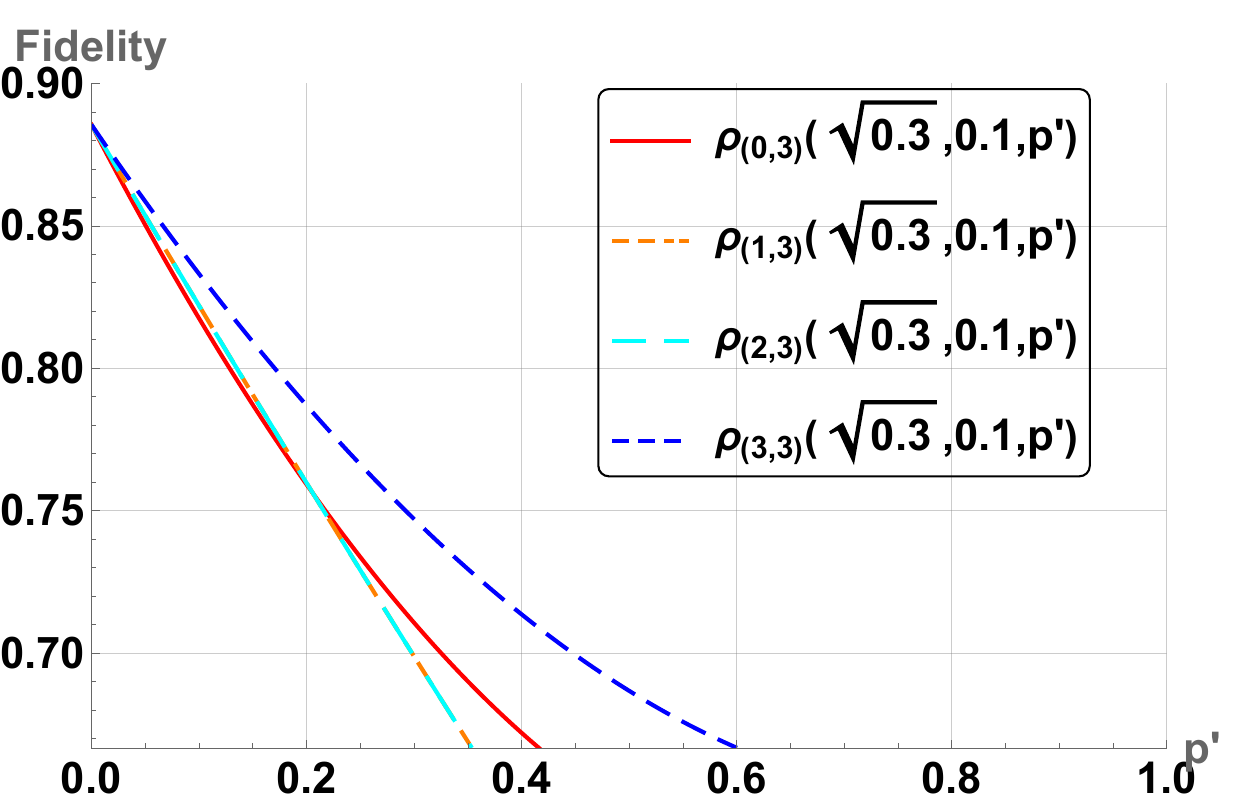} 
        \caption{}
        \label{fig12b}
    \end{subfigure}
    \vspace{0.5em}
    \begin{subfigure}[b]{0.36\textwidth}
        \centering
        \includegraphics[width=\textwidth]{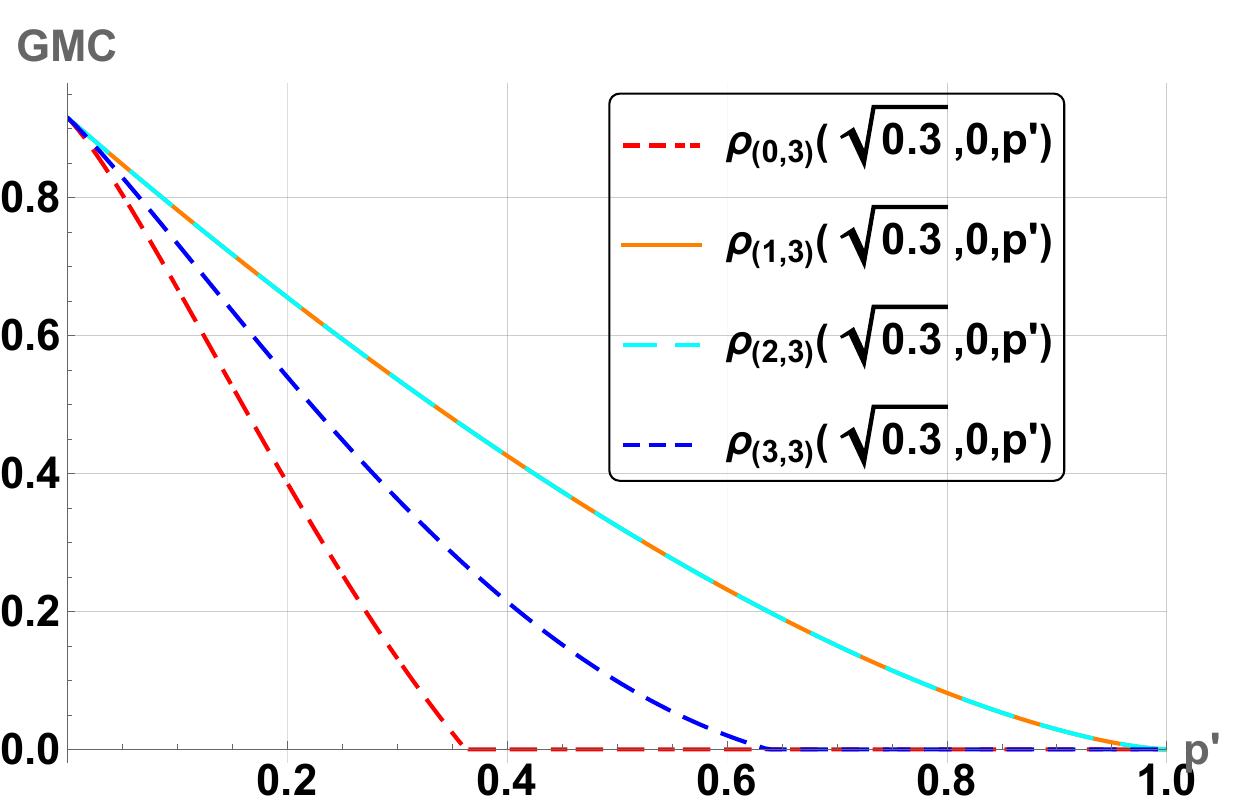} 
        \caption{}
         \label{fig12c}
    \end{subfigure} 
    \begin{subfigure}[b]{0.36\textwidth}
        \centering
        \includegraphics[width=\textwidth]{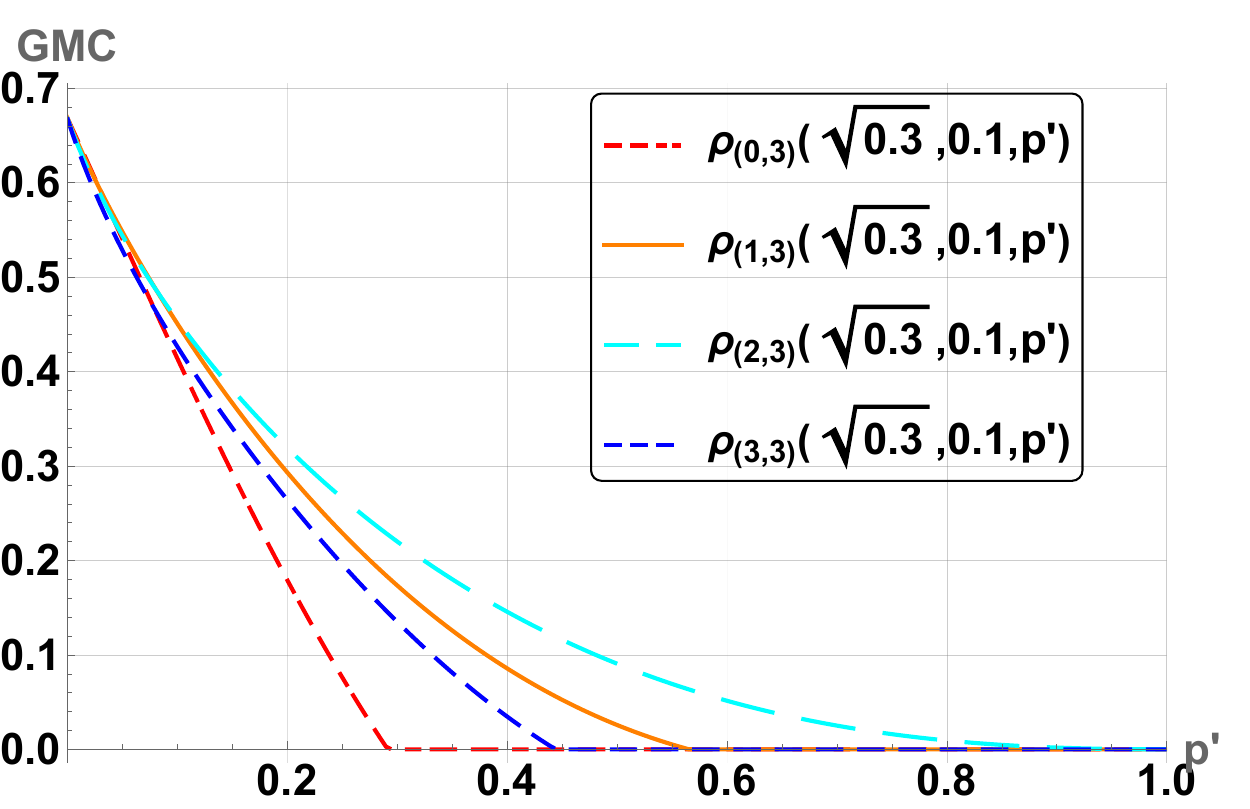} 
        \caption{}
         \label{fig12d}
    \end{subfigure}
    \caption{Dynamics of CQT fidelity and GMC for an initial entangled state  $\rho_3(\alpha)= (\alpha|000\rangle + \beta|111\rangle) \newline (\alpha|000\rangle + \beta |111\rangle)^\dag$ under ADC noise. CQT fidelities for state $\rho_{(m,3)}(\sqrt{0.3},p,p')$ vs. $p'$ for (a) $p=0$, and (b) $p=0.1$. GMC for state $\rho_{(m,3)}(\sqrt{0.3},p,p')$ vs. $p'$ for (c) $p=0$, and (d) $p=0.1$. The CQT fidelity for $p=0$ [$p=0.1$] with $m=0$ ($m=3$) approaches the classical limit at $p'=0.48$ ($p'=0.76$) [$p'=0.42$ ($p'=0.6$)], whereas the corresponding GMC vanishes at $p'=0.37$ ($p'=0.64$) [$p'=0.3$ ($p'=0.45$)]. The non-classical CQT fidelities beyond the zero GMC point indicate that the ADC-evolved mixed biseparable state can still be a useful resource for CQT.}
    \label{fig12}
\end{figure}

The impact of one-, two-, and three-qubit NOT operations ($m=1,2,3$) on the CQT fidelity of three-qubit states $\rho_{(m,3)}(\alpha,p,p')$ used as an entangled resource, is given below.
\begin{equation}
\resizebox{\columnwidth}{!}{$
\begin{aligned}
\mathcal{F}\left[\rho_{(1,3)}(\alpha,p,p')\right] &= \frac13 \max[(2 + 2qq' \beta (\sqrt{q q'} \alpha + (-1 + qq') \beta)), \frac12 (3 + \sqrt{1 + 4 q q' \beta^2 (-1 +  q q' (4 q q' \alpha^2 + \beta^2))}), \frac12 (3 + |1 - 2 q q' \beta^2]), \\
\mathcal{F}\left[\rho_{(2,3)}(\alpha,p,p')\right] &=\frac{1}{6} \max[3+\sqrt{1+4qq' \beta^2(-1+qq' (4qq'\alpha^2+\beta^2))}, 3+|1-2qq'\beta^2|, 3+4(qq')^{3/2}\alpha\beta + |-1+2q'+4(-1+q)qq'^2\beta^2], \\
\mathcal{F}\left[\rho_{(3,3)}(\alpha,p,p')\right] &= \frac13\max[(2 +  2 q' (-1 + q' + (-2 + q) q qq'\beta^2 +  q \beta (\sqrt{q q'} \alpha + \beta))), \frac12 (3 + \sqrt{1 + 4 q' (-1 + q \beta^2 +  q' (4 q^3 q' \alpha^2 \beta^2 + (-1 +  q \beta^2)^2))}), \\ & \frac12 (3 + |1 - 2 q' + 2 q q' \beta^2])].
\end{aligned}
$}
\end{equation}

In subfigures\,(\ref{fig12a}) and (\ref{fig12b}), we show the CQT fidelity of three-qubit entangled states $\rho_{(m,3)}(\alpha,p,p')$ with $|\alpha|^2 = 0.3$, for $p=0~\&~0.1$, respectively. 
In both subfigures, fidelity for all-NOT ($m=3$) exhibits a clear improvement over corresponding partial-NOT ($m = 0, 1,2$) cases, where $\mathcal{F}$ remains highest across the range of $p'$ values, and it approaches the classical limit only at $p'=0.76$ and $p'=0.6$, respectively. 
For comparison, corresponding GMC plots are shown in subfigures \,(\ref{fig12c}) and (\ref{fig12d}), where GMC is best preserved for $m=2~\&~1$.
These differences in GMC and fidelity behaviour highlight that the fidelity is not preserved in the same manner as entanglement, similar to the two-qubit case discussed in Section\,(\ref{Sec_2QubitTeleportation}). 
Moreover in subfigures\,(\ref{fig12a}) and (\ref{fig12c}), the CQT fidelity (GMC) for $m=0~\&~3$ cases approach the classical limit at $p'=0.48$ and $p'=0.76$ ($p'=0.36$ and $p'=0.63$), respectively. This demonstrates that an ADC-evolved mixed biseparable state, which persists beyond GME death, can still serve as a useful resource for CQT~\cite{CQT_Barasinski2018, CQT-PRL_Barasinski2019, CQT-PRA_Barasinski2019}. 
Notably, the computed localizable concurrence satisfies $\mathcal{C}_L[\rho_{(0,3)}^{(1,2)}(\sqrt{0.3},p,p')] = \mathcal{C}_L[\rho_{(0,3)}^{(2,3)}(\sqrt{0.3},p,p')] = \mathcal{C}_L[\rho_{(0,3)}^{(1,3)}(\sqrt{0.3},p,p')]$, and $\mathcal{C}_L[\rho_{(3,3)}^{(1,2)}(\sqrt{0.3},p,p')] = \mathcal{C}_L[\rho_{(3,3)}^{(2,3)}(\sqrt{0.3},p,p')] = \mathcal{C}_L[\rho_{(3,3)}^{(1,3)}(\sqrt{0.3},p,p')]$, and vanishes precisely when the fidelity of $\rho_{(0,3)}(\sqrt{0.3},p,p')$ and $\rho_{(3,3)}(\sqrt{0.3},p,p')$ state, respectively, falls below the classical limit for $p=0$ [see Figs. (\ref{fig12a}) and (\ref{fig12c})], $p=0.1$ [see Figs.\,(\ref{fig12b}) and (\ref{fig12d})] as discussed in Section\,(\ref{Sec_LocEnt}).

\subsubsection{CQT using a four-qubit entangled resource shared through ADC}

\begin{figure}[ht!]
    \centering  
    \begin{subfigure}[b]{0.36\textwidth}
        \centering
        \includegraphics[width=\textwidth]{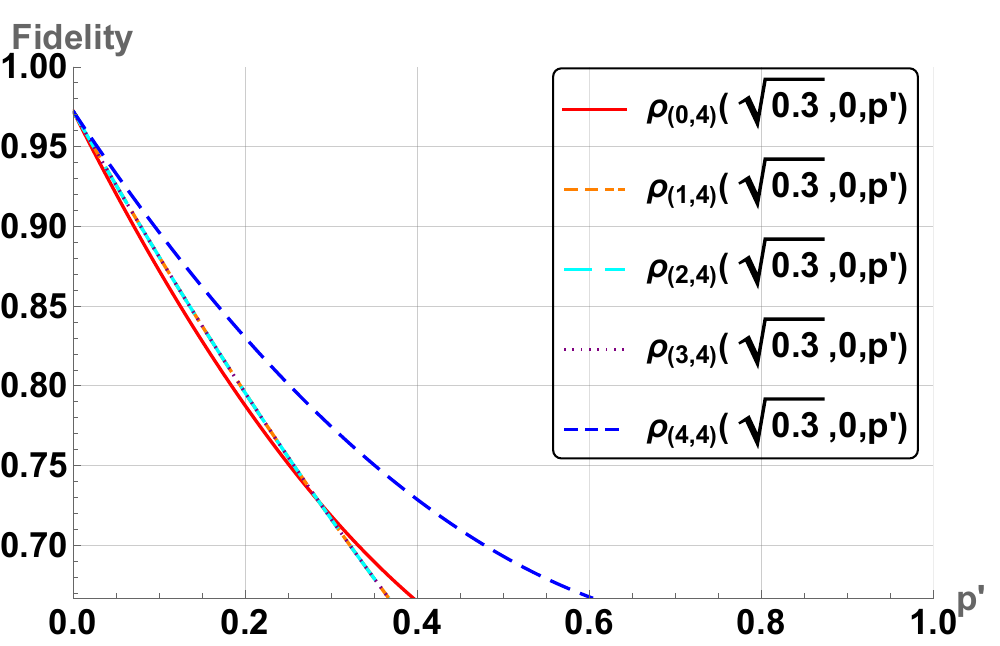} 
        \caption{}
         \label{fig13a}
    \end{subfigure} 
    \begin{subfigure}[b]{0.36\textwidth}
        \centering
        \includegraphics[width=\textwidth]{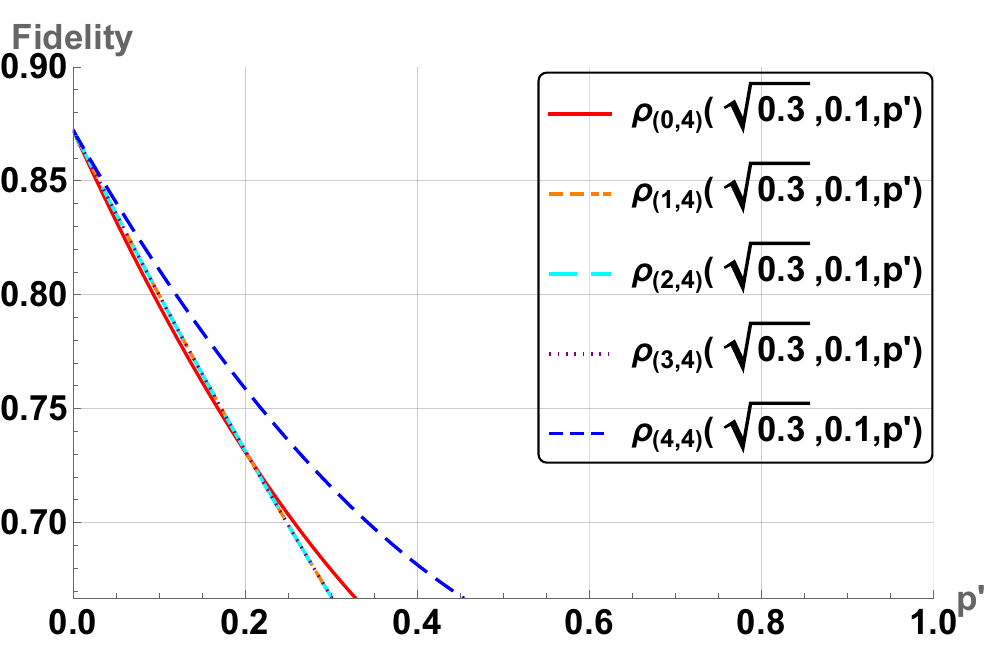} 
        \caption{}
         \label{fig13b}
    \end{subfigure}
   \vspace{0.5em} 
    \begin{subfigure}[b]{0.36\textwidth}
        \centering
        \includegraphics[width=\textwidth]{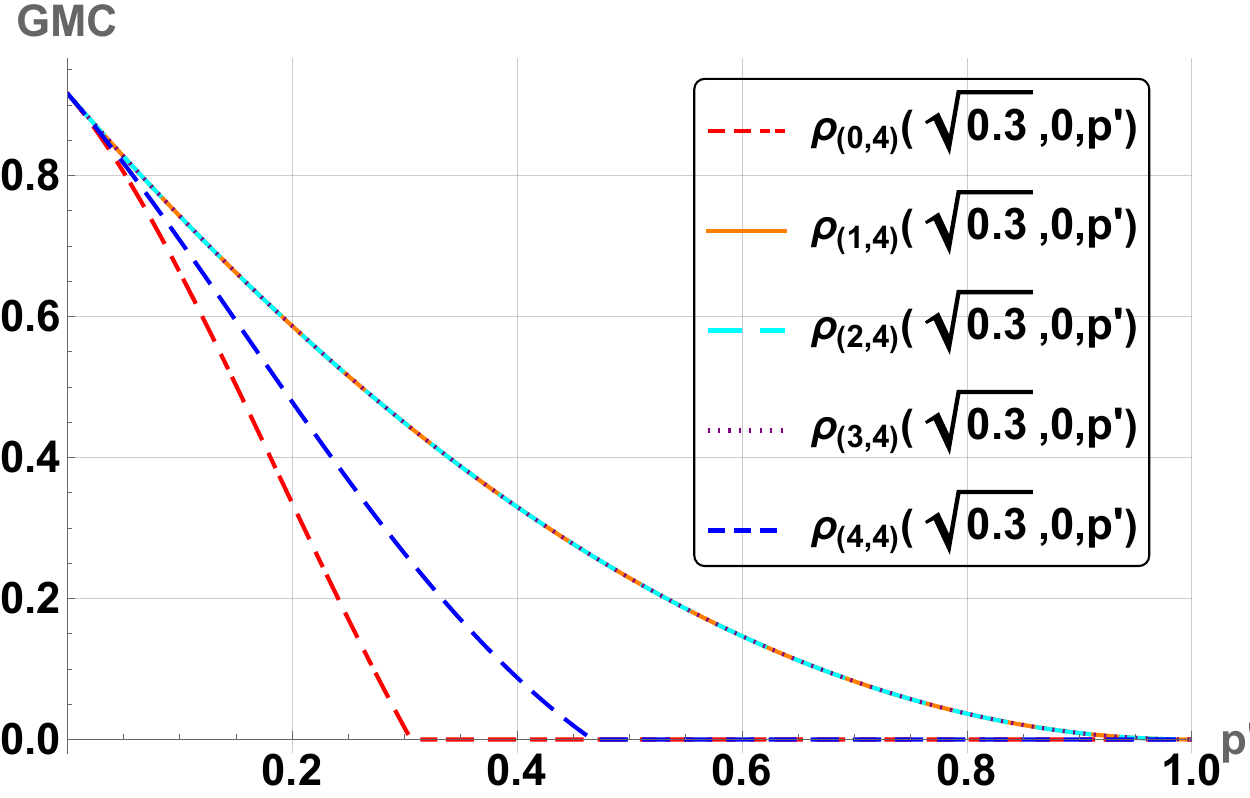} 
        \caption{}
         \label{fig13c}
    \end{subfigure} 
    \begin{subfigure}[b]{0.36\textwidth}
        \centering
        \includegraphics[width=\textwidth]{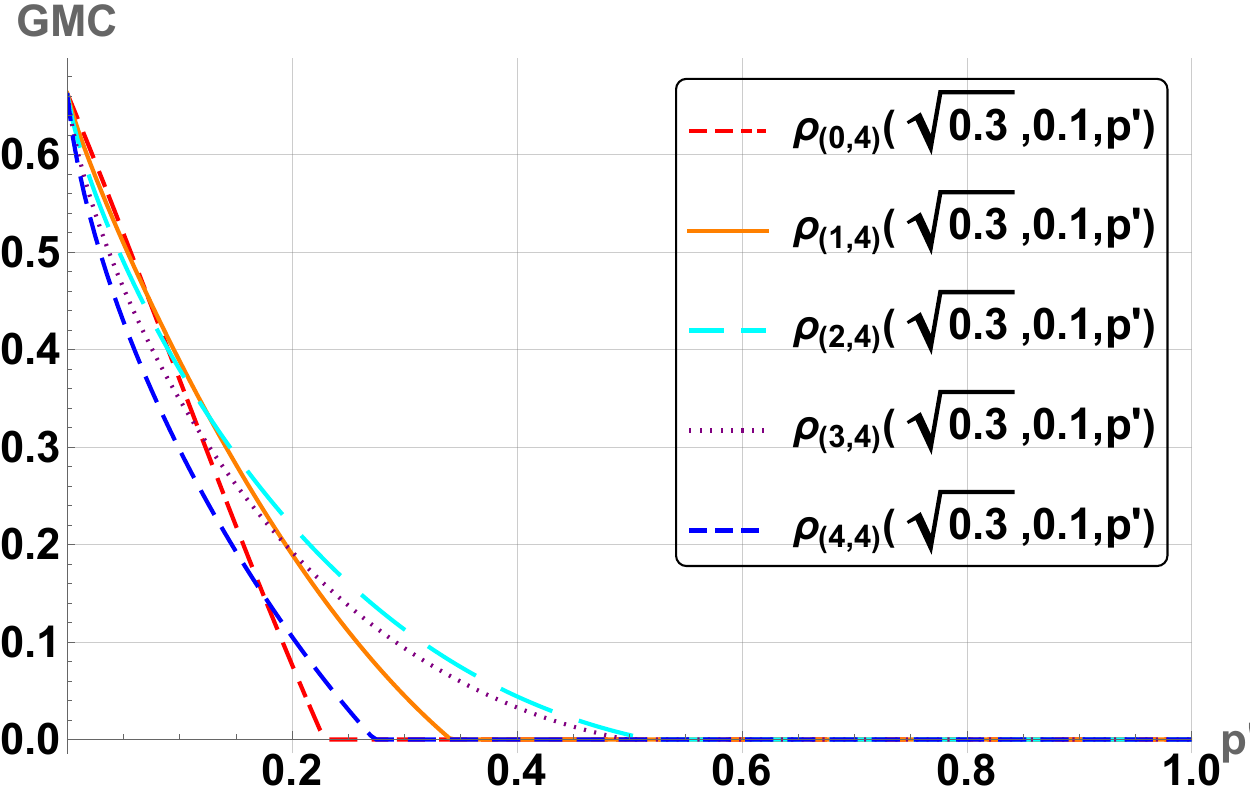} 
        \caption{}
         \label{fig13d}
    \end{subfigure}
    \caption{Dynamics of CQT fidelity and GMC using a four-party entangled state  $\rho_4(\alpha) = (\alpha|0000\rangle + \beta|1111\rangle) \newline (\alpha |0000\rangle + \beta | 1111\rangle)^\dag$ as resource shared over an ADC. CQT fidelity for state $\rho_{(m,4)}(\sqrt{0.3},p,p')$ vs. $p'$ for (a) $p=0$, and (b) $p=0.1$. GMC for state $\rho_{(m,4)}(\sqrt{0.3},p,p')$ vs. $p'$ for (c) $p=0$, and (d) $p=0.1$.} 
    \label{fig13}
\end{figure}

The impact of one-, two-, and three-, and four-qubit NOT operations ($m=1,2,3,4$) on the CQT fidelity of a four-qubit state $\rho_{(m,4)}(\alpha,p,p')$ used as an entangled resource, is given below.
\begin{equation}
\begin{aligned}
    \mathcal{F}\left[\rho_{(1,4)}(\alpha,p,p')\right] &=  \frac{2}{3}\left[\alpha^2 + 
   q^2 q'^2 \alpha \beta + ((p + p' q)^2 + 
      q (p + p' q) q' + q^2 q'^2) \beta^2\right], \\
  \mathcal{F}\left[\rho_{(2,4)}(\alpha,p,p')\right] &= \frac{2}{3} \left[\alpha^2 + q^2 q'^2 \alpha \beta + ((p + p' q)^2 + q (p + p' q) q' + q^2 q'^2) \beta^2\right], \\
    \mathcal{F}\left[\rho_{(3,4)}(\alpha,p,p')\right] &= \frac{1}{3} \left[(p' + 2 q') \alpha^2 +  2 q^2 q'^2 \alpha \beta + (p^2 (p' + 2 q') +  q^2 (p' + 2 q') +  p q (1 + p'^2 + 4 p' q' + q'^2)) \beta^2 \right], \\
    \mathcal{F}\left[\rho_{(4,4)}(\alpha,p,p')\right] &= \frac{2}{3} \left[(p'^2 + p' q' + q'^2) \alpha^2 + q^2 q'^2 \alpha \beta + ((p p' + q)^2 +  p (p p' + q) q' + p^2 q'^2) \beta]^2\right].
\end{aligned}
\end{equation}

In Fig.\,(\ref{fig13}), we extend the CQT fidelity analysis and its comparison with GMC to $n = 4$ qubits for $|\alpha|^2 = 0.3$.
In subfigures\,(\ref{fig13a}) and (\ref{fig13b}), similar to three-qubit case in Fig.~\ref{fig12}(a-b), fidelity for all-NOT ($m=4$) exhibits a clear improvement over corresponding partial-NOT ($m = 1,2,3$) cases, where $\mathcal{F}$ remains highest across the range of $p'$ values, and approaches the classical limit only at $p'=0.60$ and $p'=0.45$, respectively. 
CQT fidelity corresponding to $m=1,2,\&~3$ cases overlap and lie above the $m=0$ case until $p'=0.29$ ($p'=0.21$) for $p=0$ ($p=0.1$) and then fall below it, approaching the classical limit at $p'=0.37$ ($p'=0.30$). In contrast, for $p=0$ cases in Fig.\,(\ref{fig13c}), GMC is preserved best for the partial-NOT case $m=1,2,3$, whereas the all-NOT case gives intermediate improvement over the no-NOT case. 
In Fig.\,(\ref{fig13d}), even though any combination of NOT-gates is useful in delaying GMC death, the advantage comes only for higher values of $p'>0.1$, where $m=2~\&~3$ perform the best.
These observations further indicate that CQT fidelity is not preserved in the same way as GMC, similar to the two- and three-qubit cases. 
Moreover, fidelities in Figs.\,\ref{fig13}(a-b) for $m = 4$ and $m = 0$ stay above the classical limit until a larger value of $p'$ than where the GMC exhibits ESD. Thus, CQT with two \emph{controllers} can still be performed even after the loss of genuine four-party entanglement. 

\subsubsection{\texorpdfstring{NOT on all $n$-qubit states with no initial ADC ($p=0$)}{All-NOT for n-qubit states with no initial ADC (p=0)}}

\begin{figure}[ht!]
    \centering
    \begin{minipage}[b]{0.32\textwidth}
        \centering
        \includegraphics[width=\textwidth]{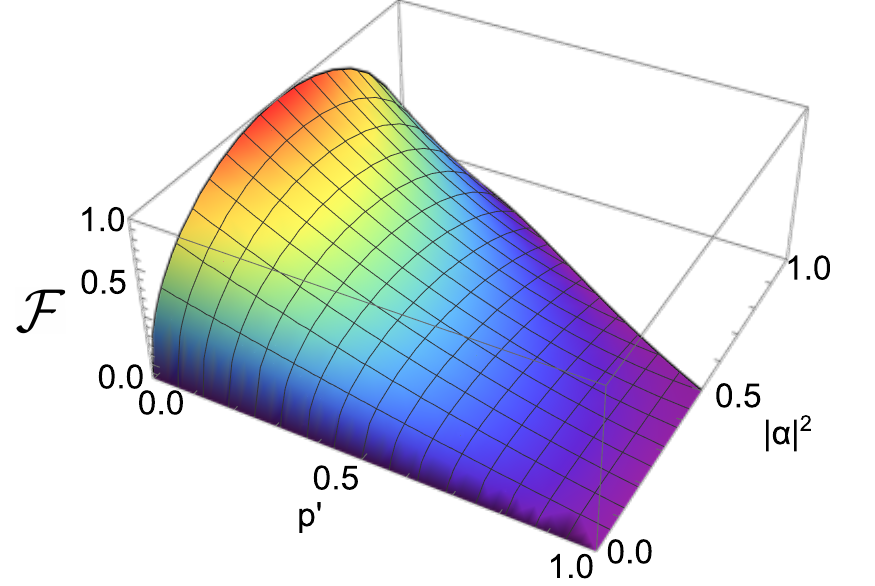} 
        \caption*{(a)}
    \end{minipage}
    \begin{minipage}[b]{0.32\textwidth}
        \centering
        \includegraphics[width=\textwidth]{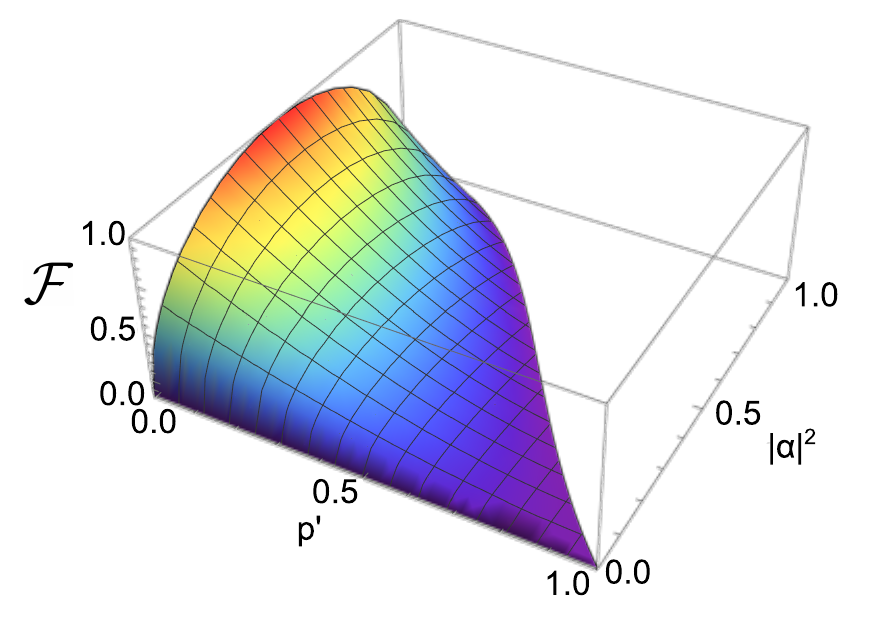} 
        \caption*{(b)}
    \end{minipage}
    \begin{minipage}[b]{0.32\textwidth}
        \centering
        \includegraphics[width=\textwidth]{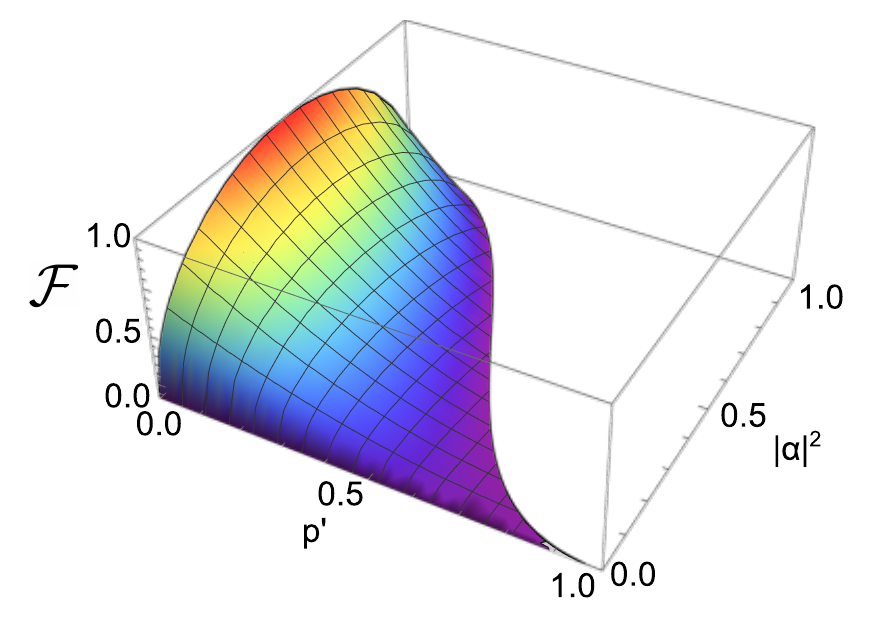} 
        \caption*{(c)}
    \end{minipage}
    \caption{Dynamics of (controlled) quantum teleportation fidelity ($\mathcal{F}$) of two-, three- and four-party entangled state with NOT gates on all qubits vs. $p'~\&~|\alpha|^2$. Only fidelity above the classical limit $\left(\mathcal{F}=2/3\right)$ was plotted. Fidelity of state (a) $\rho_{(2,2)}(\alpha,0,p')$, (b) $\rho_{(3,3)}(\alpha,0,p')$, and (c) $\rho_{(4,4)}(\alpha,0,p')$.}
    \label{fig14}
\end{figure}

The Fig.\,(\ref{fig14}) shows the effect of the NOT operation applied to all the qubits ($m=n$) of two-, three-, and four-qubit entangled states $\rho_{(m,n)}(\alpha,p,p')$ on the fidelity of (controlled) quantum teleportation. 
Specifically, we show the fidelity of $\rho_{(2,2)}(\alpha,0,p')$, $\rho_{(3,3)}(\alpha,0,p')$, and $\rho_{(4,4)}(\alpha,0,p')$ vs. $p'$ and $|\alpha|^2$ for $p=0$. 
All these subfigures exhibit behaviour similar to the no-NOT case as shown in Fig.\,(\ref{fig9}), except that the curves are mirrored about $|\alpha|^2 = 0.5$.  This clearly indicates that the crossover to the classical limit under ADC is delayed only for $|\alpha|^2 < 0.5$.
Furthermore, consistent with the trend observed in GMC, the fidelity of teleportation is preserved the longest for the two-qubit case ($\rho_{(2,2)}(\alpha,0,p')$), followed by $\rho_{(3,3)}(\alpha,0,p')$, and then $\rho_{(4,4)}(\alpha,0,p')$. 
This once again confirms the increasing vulnerability of larger GHZ-type systems in their utility for teleportation protocols. Consequently, when employing an entangled state under ADC, the optimal strategy is to flip all qubits for $|\alpha|^2 < 0.5$ at $p=0$, when using GHZ-type resources to achieve the best teleportation fidelity. These findings thus reinforce that preserving (or maximizing) entanglement is not always equivalent to preserving teleportation utility under noise: the form of the residual coherence and population matters, and unitary controls (single- or all-NOT) must be selected with the operational task (e.g., teleportation vs entanglement preservation) in mind.

\subsection{\texorpdfstring{GHZ‐symmetric $(x,y)$-parametrization}{GHZ‐symmetric (x,y)-parametrization}}

As described in the Appendix\,(\ref{Sec_Appendix}), we adopt the GHZ-symmetric $(x,y)$-parametrization to categorize the states $\rho_{(m,n)}(\alpha,p,p')$ [see Eq.\,(\ref{Eq_07})] into the well-studied SLOCC hierarchy \cite{GHZ-Sym_Eltschka2012, TripartiteEnt_Siewert2012, GHZ-SymClass_Park2014} as they evolve through the ADC. Importantly, the GHZ-symmetric two-qubit family reduces to states diagonal in the Bell basis; consequently, for two qubits, the only relevant division is between entangled and separable states \cite{TripartiteEnt_Siewert2012}.  
In contrast, the three-qubit GHZ-symmetric triangle admits four exclusive SLOCC regions — GHZ, W, biseparable, and separable. These regions are directly obtained from the $(x,y)$ coordinates of the GHZ-symmetric states \cite{GHZ-Sym_Eltschka2012, TripartiteEnt_Siewert2012}. 
A closed-form expression of three-tangle ($\tau_3$) -- a standard quantitative measure of genuine tripartite entanglement for three qubits -- exists for the GHZ-symmetric family \cite{TripartiteEnt_Siewert2012}.
Using the parametric expression for the GHZ–W boundary and the analytic prescription (twirling) in the GHZ-symmetric construction \cite{GHZ-Sym_Eltschka2012}, we evaluate 
(i) whether $(x,y)$ lies on or below the GHZ–W curve  (in this case $\tau_3(\rho^S)=0$, where $\rho^S$ is twirled state), and (ii) if $(x,y)$ lies in the GHZ region, in this case $\tau_3(\rho^S)>0$ (Using the GHZ twirling operation, any three-qubit state $\rho$ can be mapped to its GHZ-symmetric projection $\rho^S$ by averaging over the GHZ symmetry group). A nonzero value of $\tau_3(\rho^S)$ certifies GHZ-class SLOCC resources in the original state. Conversely, $\tau_3(\rho^S)=0$ together with a nonzero GMC identifies W-class resources, while the vanishing of both quantities signals biseparability or full separability.  
Thus, the GHZ projection provides a simple procedure and experimentally relevant SLOCC “symmetrization witness”. The projected coordinates $(x,y)$ locate a mixed three-qubit state within the GHZ–W–biseparable–separable hierarchy, and the analytic three-tangle formula for GHZ-symmetric states yields a certified quantitative indicator of GHZ-type entanglement relevant for teleportation protocols.  
For completeness, we note that the GHZ-symmetric description for four qubits we performed is three-dimensional $(\tilde x,\tilde y,\tilde z)$ and, while it carries useful geometric structure, its SLOCC partitioning is not yet completely characterized \cite{GHZ-SymClass_Park2014}.

\subsubsection{Two-qubit state dynamics}

\begin{figure}[ht!]
    \centering
    \includegraphics[width=0.45\linewidth]{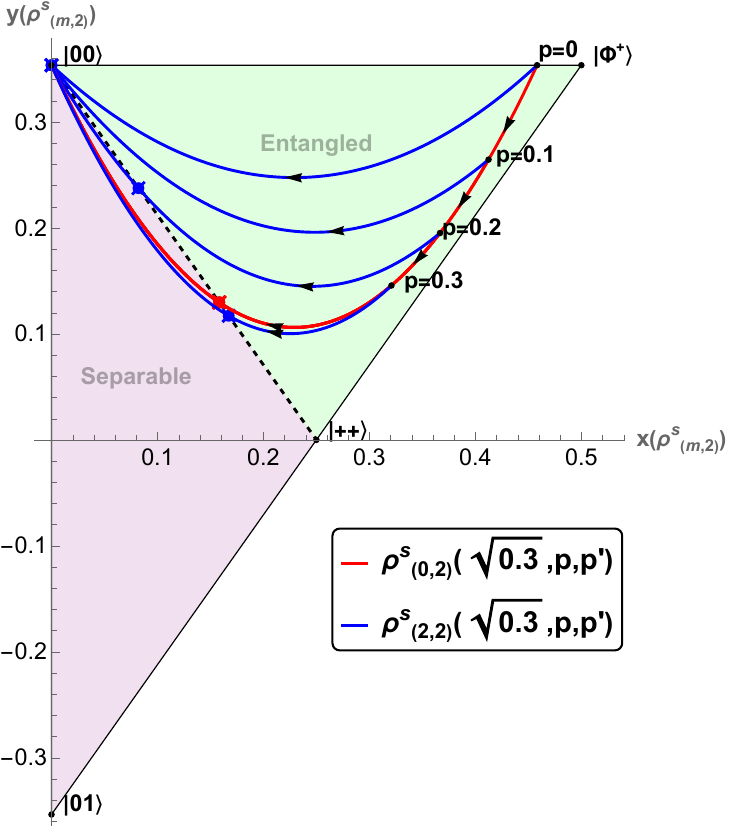}
    \caption{Two-qubit entangled state $\rho^S_{(m,2)}(\sqrt{0.3}, p, p')$ parametrization under ADC. A coloured cross (solid dot) marks the point where entanglement (teleportation fidelity) drops to zero (classical limit) along the state trajectory. As noise increases, trajectories move downward in the triangle, crossing into the separable region (dotted line), indicating loss of entanglement.}
    \label{fig_2QubitClass}
\end{figure}

Fig.\,(\ref{fig_2QubitClass}) shows the GHZ-symmetrized $(x,y)$-coordinates  of two-qubit twirled states $\rho^S_{(m,2)}(\sqrt{3},p,p')$ for different settings of the ADC parameters ($p,p'$) [as detailed in Appendix\,(\ref{Sec_2qubitsymmetrization})].
The family of these twirled states can be represented within a triangle \cite{GHZ-Sym_Eltschka2012}.
The vertices of this (full) triangle correspond to the pure GHZ‐symmetric states: the upper corners are $|\Phi^+\rangle\langle\Phi^+|$ and $|\Phi^-\rangle\langle\Phi^-|$, and the lower corner is $|01\rangle$ (which is an equal mixture $\tfrac{1}{2}(|\Psi^+\rangle\langle\Psi^+|+|\Psi^-\rangle\langle\Psi^-|)$). 
Note that Fig.\,(\ref{fig_2QubitClass}) shows only the right half of this triangle, which suffices for our study, where the left-top corner corresponds to the state $|00\rangle$.   
This triangle is partitioned into entangled (above) vs. separable (below) regions by a linear separability boundary $y_{\rm sep}=\pm(1/(2\sqrt2) - \sqrt2\,x$). The completely mixed two‐qubit state sits at the origin $x=y=0$.

Each coloured trajectory corresponds to an entangled state's evolution for different settings of NOT-gates and ADC parameters. 
The red curve corresponds to the no-NOT case and shows the entanglement evolution vs. $p$. Each blue curve corresponds to a fixed value of $p$, NOT on both qubits, and subsequent entanglement evolution vs. $p'$.
The direction of the arrow indicates increasing ADC parameter $p$ or $p'$. Notably, a `coloured cross' on the curves marks the ESD (the point where GMC drops to zero), while a `coloured dot' marks the point where teleportation fidelity drops to the classical limit. 
For $m=0,2$ cases, these two points overlap on the corresponding plots at the separability line. 
The top blue curve with $p=0$ remains in the entangled regime for $p'\in[0,1)$, indicating ADE.  As $p$ increases, the corresponding GMC vs. $p'$ curves shift downward. For $p=0.2$, the corresponding entangled state undergoes a delay of ESD as it enters into the separable region before $p'\rightarrow 1$ but after the no-NOT (red curve) case. On the other hand, for $p=0.3$, we observe hastening of ESD. 

\subsubsection{Three-qubit state dynamics}
\label{Sec_Results_3QubitClass}

\begin{figure}[ht!]
    \centering
    \includegraphics[width=0.45\linewidth]{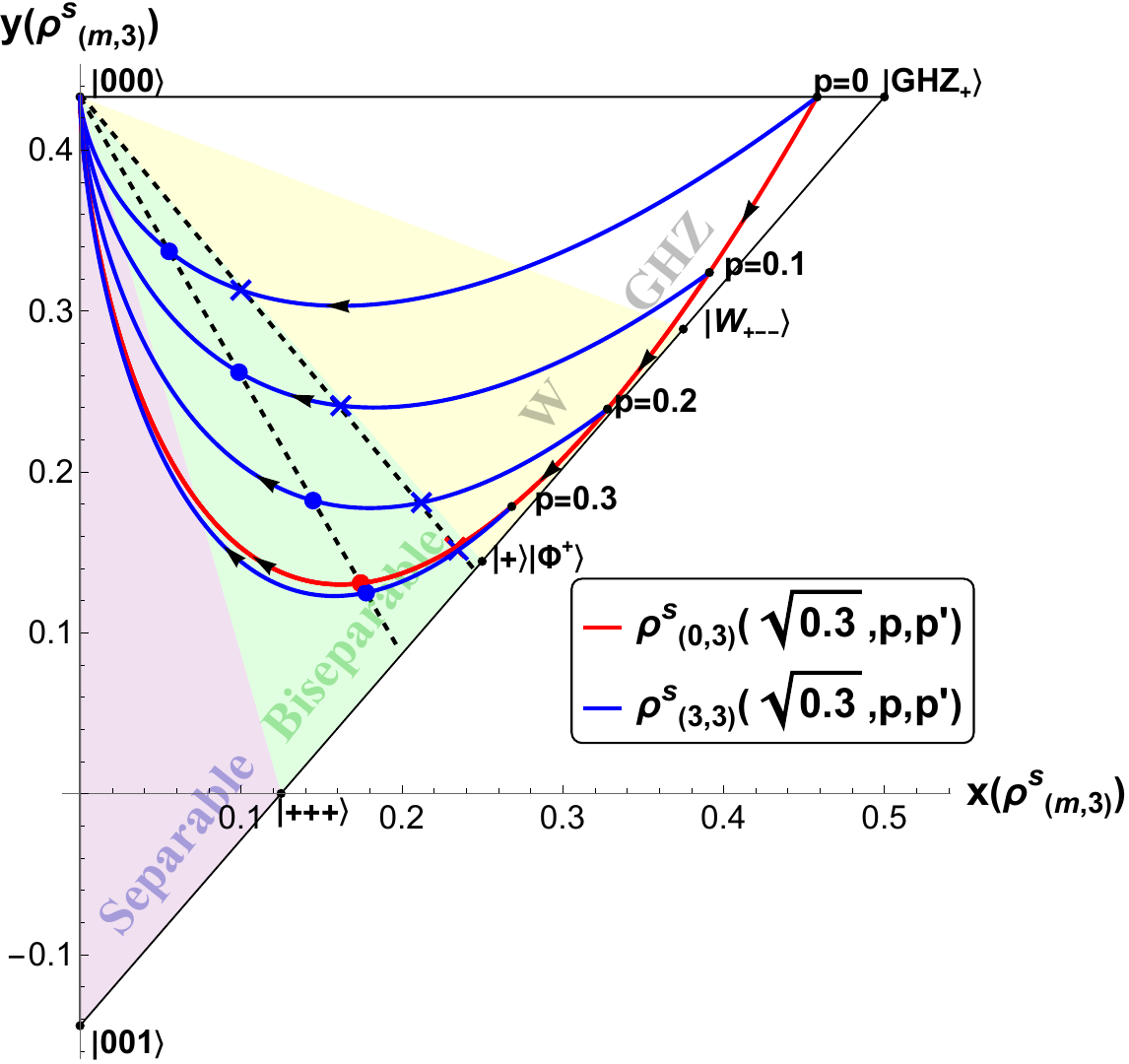}
    \caption{Three-qubit entangled state $\rho^S_{(m,3)}(\sqrt{0.3}, p, p')$ parametrization under ADC. 
    A coloured cross (solid dot) marks the point where GME (teleportation fidelity) drops to zero (classical limit) along the given state trajectory. As noise increases, trajectories move downward in the GHZ-symmetric triangle, crossing the GHZ-W, W-biseparable, and biseparable-separable boundaries. Trajectories entering into the biseparable region (cross) signal loss of GME, while the subsequent dots (in the biseparable region) indicate loss of non-classical teleportation (CQT) fidelity.} 
    \label{fig_3QubitClass}
\end{figure}

For three-qubits, there are four SLOCC-equivalent entanglement classes: fully separable, (at least one) biseparable, W‐type, and GHZ‐type, as shown in the shaded regions in Fig.\,(\ref{fig_3QubitClass}).  
Importantly, each SLOCC class forms a convex region in the GHZ-symmetric triangle. Note that Fig.\,(\ref{fig_3QubitClass}) shows only the right half of this triangle, which suffices for our study.
The boundaries between these SLOCC classes can be determined analytically.  For example, the boundary between GHZ and W classes (the “GHZ=W line”) is given by a known parametric curve \cite{GHZ-Sym_Eltschka2012, TripartiteEnt_Siewert2012}. 
To apply the GHZ-symmetrization framework to three‐qubit states $\rho_{(m,3)}(\sqrt{0.3},p,p')$ (with $m=0,3$), we compute the twirled states $\rho^S_{(m,3)}(\sqrt{0.3},p,p')$ and extract ($x,y$) coordinates.  
The resulting $(x,y)$ points fall in one of the SLOCC classes within the three-qubit GHZ-symmetric triangle, and thus reveal the entanglement class of the state [as described in the Appendix\,(\ref{Sec_3qubitsymmetrization})].

Fig.\,(\ref{fig_3QubitClass}) shows the state trajectories of $\rho^S_{(m,3)}(\sqrt{0.3},p,p')$ for different settings of the ADC parameters $p,p'$. The red and blue curves correspond to the no-NOT ($m=0$) and all-NOT ($m=3$) cases, where we plot respectively.  
The initial states lie near the GHZ corner ($|\text{GHZ}^+\rangle$), indicating GHZ‐type entanglement.  
 As noise increases, the state moves away from the GHZ corner in the GHZ region, enters the W region, followed by the biseparable and separable regions. 
The Kraus operators of single-qubit ADC [see Eq.\,(\ref{Eq_ADCKraus})], are represented by \(M_1\) (which damps off-diagonal coherence in the computational basis) and \(M_2\) (a rank-one lowering/reset operator). 
When \(M_2\) acts locally on one qubit of a GHZ superposition, it converts the \(\ket{111}\) component into single-excitation terms \(\ket{011},\ket{101},\ket{110}\), thereby populating the W-class subspace with nonzero probability and enabling a stochastic GHZ\(\to\)W transition under ADC.

The NOT on all qubits applied early on ($p=0$) preserves the GME and CQT fidelity the longest compared to the non-zero values of $p$. 
The ‘coloured-cross’ and ‘coloured-dot’ markers represent the points where the GME vanishes and the CQT fidelity approaches the classical limit, respectively. 
Notably, in every trajectory, the dot appears (in the biseparable region) after the cross (at the W-biseparable boundary), indicating that even after GME is lost, the mixed biseparable GHZ state can still support non-classical CQT fidelity, as evidenced by finite localizable concurrence \cite{CQT_Barasinski2018}.
Using the definitions provided in Section\,(\ref{Sec_LocEnt}),
we compute localizable concurrence which satisfies: $\mathcal{C}_L[\rho_{(3,3)}^{(1,2)}(\sqrt{0.3},p,p')] = \mathcal{C}_L[\rho_{(3,3)}^{(2,3)}(\sqrt{0.3},p,p')] = \mathcal{C}_L[\rho_{(3,3)}^{(1,3)}(\sqrt{0.3},p,p')]$ $\left(\mathcal{C}_L[\rho_{(0,3)}^{(1,2)}(\sqrt{0.3},p,p')] = \mathcal{C}_L[\rho_{(0,3)}^{(2,3)}(\sqrt{0.3},p,p')] = \mathcal{C}_L[\rho_{(0,3)}^{(1,3)}(\sqrt{0.3},p,p')]\right)$, 
which vanishes precisely when the $\mathcal{F}[\rho_{(3,3)}(\sqrt{0.3},p,p')]$ ($\mathcal{F}[\rho_{(0,3)}(\sqrt{0.3},p,p')]$) approaches the classical limit (the dashed line connecting the dot markers). 

\subsubsection{Four-qubit state dynamics}

The four-qubit entanglement classification is substantially more complex \cite{PhysRevA.65.052112, PhysRevA.75.022318, Cao2007, PhysRevLett.105.100507} than the three-qubit case \cite{ThreeQubitGME_Dur2000}. 
 No unique or complete SLOCC classification exists for four-qubit mixed states. Even for pure states, multiple inequivalent classifications have been proposed \cite{Eltschka_2014}, and the situation remains confusing and partially unresolved. 
Consequently, any analysis based on GHZ symmetry necessarily probes only a restricted, GHZ-like subset of four-qubit states rather than the full entanglement landscape \cite{GHZ-SymClass_Park2014}. Within this limitation, GHZ-symmetric trajectories in the \((\tilde{x},\tilde{y},\tilde{z})\) space cover only a subset of four-qubit SLOCC classes that appear \cite{GHZ-Symmetry_Park2016}.

\begin{figure}[ht!]
    \centering
    \begin{minipage}[b]{0.45\textwidth}
        \centering
        \includegraphics[width=\textwidth]{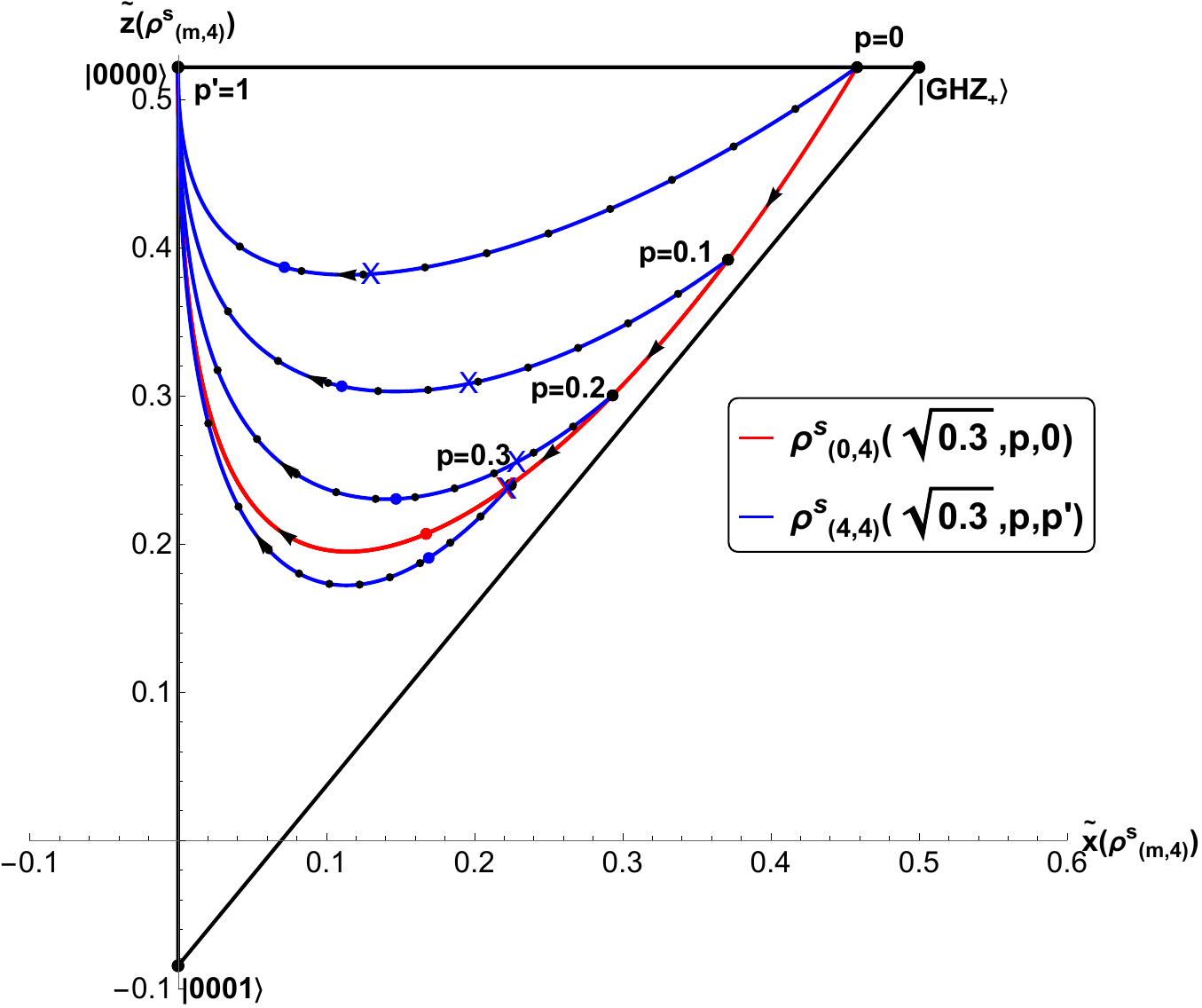} 
        \caption*{\parbox[t]{\textwidth}{\small (a) }}
    \end{minipage}
    \begin{minipage}[b]{0.45\textwidth}
        \centering
        \includegraphics[width=\textwidth]{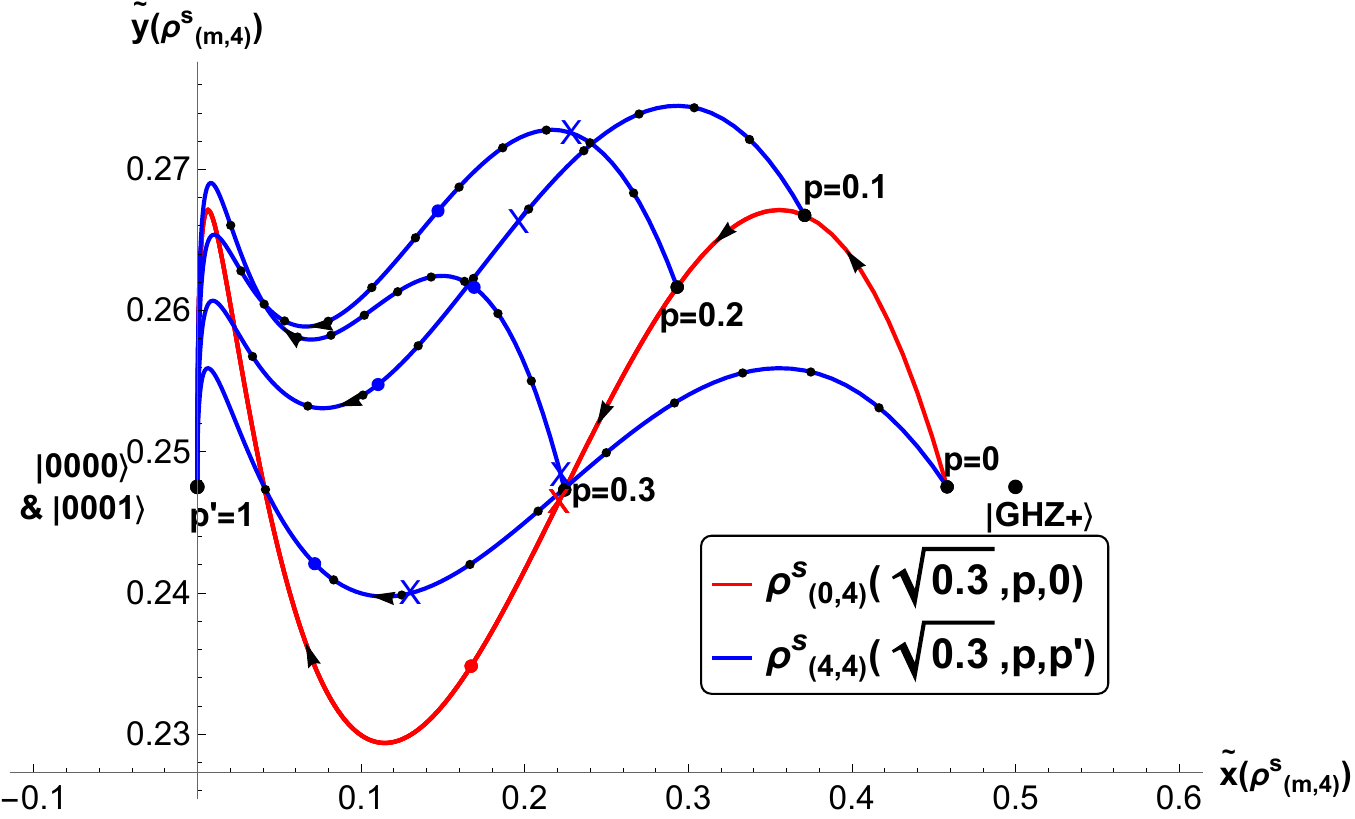} 
        \caption*{\parbox[t]{\textwidth}{\small (b)}}
    \end{minipage}
    \caption{Four-qubit entangled states $\rho^S_{(m,4)}(\sqrt{0.3}, p, p')$ parametrization under ADC. (a) $\tilde{x}$-$\tilde{z}$ orthographic view, and  (b) $\tilde{x}$-$\tilde{y}$ orthographic view. A coloured cross (solid dot) marks the point where GMC (CQT fidelity) falls to zero (classical limit $2/3$) along the given state trajectory. The red (blue) curve indicates the state trajectory for the no-NOT (NOT on all qubits) case. The small black dots sample the trajectory at intervals of $p'=0.1$. For $\rho^S_{(4,4)}(\sqrt{0.3}, p, p')$ (blue curves), the coloured cross (GMC=0) followed by a coloured dot ($\mathcal{F}<2/3$) occurs progressively closer to the starting point (NOT operations applied) as the value of $p$ increases from $0$ to $0.3$.}
    \label{fig_4QubitClass}
\end{figure}

To apply the GHZ-symmetric classification framework \cite{GHZ-SymClass_Park2014} for four‐qubit states $\rho_{(m,4)}(\sqrt{0.3},p,p')$,  $m=0,4$, we again compute the corresponding GHZ‐symmetric twirled state ($\rho^S$) and extract $(\tilde{x}, \tilde{y}, \tilde{z})$ parameters as described in Appendix\,(\ref{Sec_4qubitsymmetrization}). 
As noted, unlike the two- and three-qubit cases, no complete entanglement classification of these parameters has yet been achieved. 
However, to shed some light on the three-dimensional trajectories, we show two complementary two-dimensional projections; the $(\tilde{x},\tilde{y})$ and $(\tilde{x},\tilde{z})$ orthographic views, in Fig.\,(\ref{fig_4QubitClass}). 
Surprisingly, the $(\tilde{x}, \tilde{z})$ view reveals trajectories that behave qualitatively similar to those for two- and three-qubit GHZ-symmetric states.
As before, the direction of the arrow indicates the increasing ADC parameter $p~(p')$ on the red (blue) curve. Similarly, the coloured cross (solid dot) markers indicate the point where the GME (CQT fidelity) approaches zero (classical limit $2/3$).

As shown in Fig.\,(\ref{fig_4QubitClass} a), as the noise parameter $p$ increases, the blue curves ($m=4$) descend through the possible entanglement classification regions \cite{Multi2012, Multi2013, Multi2015}, mirroring the downward trend observed in the fewer-qubit systems.
Consistent with the three-qubit case, the CQT approaches the classical fidelity threshold only after GME has vanished.
The $(\tilde{x},\tilde{y})$ projection shows an overall progression (upward or downward) as noise increases, but with additional spread and curvature for \(m=0\) curves compared to \(m=4\) curves along the $\tilde{y}$-axis.
We conjecture that the state trajectories possibly cross the entanglement boundaries in the same sequence (i.e., cross marks followed by dot marks), but projecting the three-dimensional ($\tilde{x}$, $\tilde{y}$, $\tilde{z}$) trajectories onto the $(\tilde{x}, \tilde{y})$ plane introduces shifts in the projected crossing points. The noise parameters \(p,p'\) at which a boundary appears crossed in this two-dimensional view differ because of the hidden $\tilde{z}$-dimension in Fig.\,(\ref{fig_4QubitClass} b).

Finally, we stress that operational quantities such as GMC or CQT fidelity thresholds serve only as indirect markers of entanglement changes. While vanishing GMC implies loss of genuine four-partite entanglement and fidelity dropping below classical bound signals loss of operational usefulness, no direct mapping between these quantities and SLOCC boundaries in four-qubit GHZ-symmetric space is currently known. A definitive identification of class-transition surfaces would require a complete characterization of four-qubit GHZ-symmetric mixed-state entanglement, which remains an open problem.

\section{Conclusion and Outlook}
\label{Sec_Conclusion} 

We have presented a unified theoretical framework for studying the entanglement and (controlled) quantum teleportation fidelity dynamics of bipartite Bell-type ($n=2$), and multipartite GHZ-type ($n=3,4$) states; $|\text{GHZ}_n\rangle=\alpha |0\rangle^{\otimes n} + \beta|1\rangle^{\otimes n}$, $|\alpha|^2+|\beta|^2=1$, under an amplitude damping channels (ADC). Our results show that multipartite states are more vulnerable to entanglement (GMC) sudden death than their bipartite counterparts.
To mitigate entanglement loss, we applied local NOT operations on one or more qubits and found that they can favourably alter the entanglement decay. Remarkably, even a single-qubit NOT on a GHZ-type state (for $0<|\alpha|^2<1$) can often transform sudden death into asymptotic decay, effectively extending the GMC lifetime.
The closed-form expressions for GMC and CQT fidelity under ADC further reveal that flipping all qubits (for $|\alpha|^2 < 0.5$), generally helps preserve teleportation fidelity, while single-qubit flips are more effective in preserving GMC. 

Our findings emphasize that a higher amount of entanglement does not necessarily guarantee better teleportation fidelity. 
This observation for bipartite states also fits naturally into the established hierarchy of correlations. Teleportation fidelity is an operational indicator of nonclassicality: any purely classical (measure-and-prepare) procedure is bounded by $\mathcal{F}=2/3$ \cite{PhysRevLett.72.797}, while fidelity exceeding an LHV benchmark $\mathcal{F}_{\rm LHV}\approx 0.87$ signals Bell-nonlocality that cannot be accounted for by local hidden variables \cite{GISIN1996, HORODECKI1996}. Thus, there exists a broad intermediate regime $2/3 < \mathcal{F} \lesssim 0.87$ where a channel is useful for nonclassical teleportation but still admits an LHV description. 
In this sense, teleportation fidelity is a stronger operational criterion than entanglement alone: a state can remain substantially entangled, yet fail to deliver high teleportation fidelity because the residual populations and coherences after ADC are not arranged favourably for maximizing the fully entangled fraction. For pure bipartite states, entanglement implies Bell nonlocality by Gisin’s theorem \cite{GISIN1991}; this result admits multipartite generalizations \cite{Yu2012}. However, in the context of mixed states, the mixing can dilute entanglement without entirely destroying it, whereas the teleportation fidelity can drop down to within the classical limit. 
For completeness, this is not a contradiction of resource ordering but a manifestation of it: entanglement is necessary but not sufficient for teleportation utility under non-symmetric noise; Bell nonlocality is stronger still and requires even more stringent preservation of the relevant coherence terms. These points justify why the two-NOT (all-NOT) protocol can outperform single-NOT in teleportation fidelity, while single-NOT is often optimal for preserving entanglement under ADC.

Further, we explicitly demonstrated that even after genuine $n$-partite entanglement vanishes, the mixed biseparable states can still support non-classical CQT fidelity, which is consistent with the idea that non-zero LE is sufficient to perform CQT \cite{CQT_Barasinski2018, CQT-PRA_Barasinski2019, CQT-PRL_Barasinski2019, LocalConc_Consiglio2021}. 
Furthermore, we performed GHZ-symmetric SLOCC classification for two (three) qubit ADC-degraded states, which evolve through the entangled/separable \cite{TripartiteEnt_Siewert2012} (GHZ/W/biseparable/separable \cite{GHZ-Sym_Eltschka2012}) hierarchy. 
A similar four-qubit GHZ-symmetric framework \cite{GHZ-SymClass_Park2014} reveals how these states evolve through a possibly similar hierarchy for four-qubit GHZ-symmetric states \cite{Multi2015} as noise increases. 
Notably, the CQT fidelity drops below the classical threshold ($2/3$) only after GMC is lost for the three- and four-qubit states for no-NOT and all-NOT cases. 
While this behaviour has been experimentally observed in three-qubit systems \cite{CQT-PRL_Barasinski2019}, four-qubit CQT with two \emph{controllers} awaits experimental realization. 
The key resource for CQT is the LE \cite{CQT_Barasinski2018}, which is an averaged conditioned biseparable entanglement. It is equivalent to bipartite entanglement in the standard Bennet teleportation protocol \cite{PhysRevLett.90.097901}. 
In contrast, GMC is inherently a global property, which vanishes whenever the state becomes separable across any bipartition. Thus, CQT can be achieved even with mixed multipartite states that have zero GMC but non-zero LE.

Exploring quantum-discord-type measures is a natural next step: quantum discord is known to be more robust under ADC as it typically vanishes only in the asymptotic limit  \cite{Werlang2009}, so applying the NOT-gate protocol to multipartite discord \cite{PhysRevA.84.042109, PhysRevLett.107.190501, PhysRevLett.124.110401, Vaishali2025} could reveal new preservation effects. 
Previously, we had studied closed-form measures of discord \cite{PhysRevLett.106.120401, Ciccarello_2014} in two-qubit models \cite{Abhignan_2021, ESDM_Venkat2025} and in a qubit-qutrit system \cite{abhignan2026hierarchy}. These techniques could be directly useful for applied tasks in quantum networks (e.g., improving fidelity in CQT, secret sharing, and distributed sensing), and can be extended to higher-dimensional (qudit) channels. Taken together, the NOT-gate protocol and other decoherence-mitigation tools may offer a practically accessible toolkit for preserving useful quantum correlations in noisy multipartite states.

\section{Appendices} 
\label{Sec_Appendix}
\subsection{\texorpdfstring{Best NOT-gate strategy for $n$-qubit states with initial ADC ($p \geq 0$)}{Best NOT-gate strategy for n-qubit states with initial ADC (p ≥ 0)}}
\label{Sec_Appendix-A}

\begin{figure}[ht!]
    \centering
    \scalebox{0.7}{
    \begin{subfigure}[b]{0.46\textwidth}
        \centering
        \includegraphics[width=\textwidth]{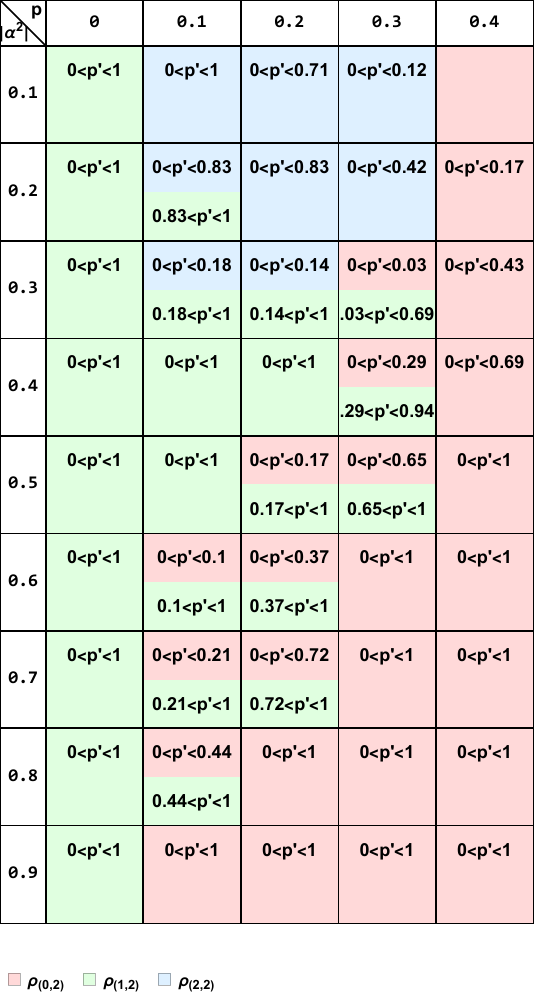} 
        \caption*{\small{(a) Evolution of state $\rho_{(m,2)}(\alpha,p,p')$.}}
    \label{Table18a}
    \end{subfigure}
    \begin{subfigure}[b]{0.46\textwidth}
        \centering
        \includegraphics[width=\textwidth]{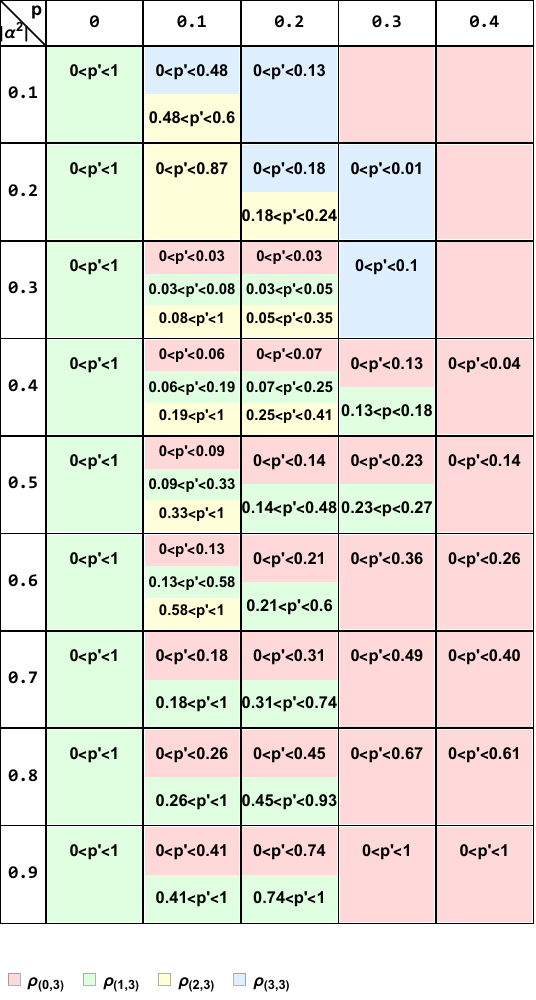}
        \caption*{\small{(b) Evolution of state $\rho_{(m,3)}(\alpha,p,p')$.}}
    \label{Table18b}
    \end{subfigure}
    \begin{subfigure}[b]{0.47\textwidth}
        \centering
        \includegraphics[width=\textwidth]{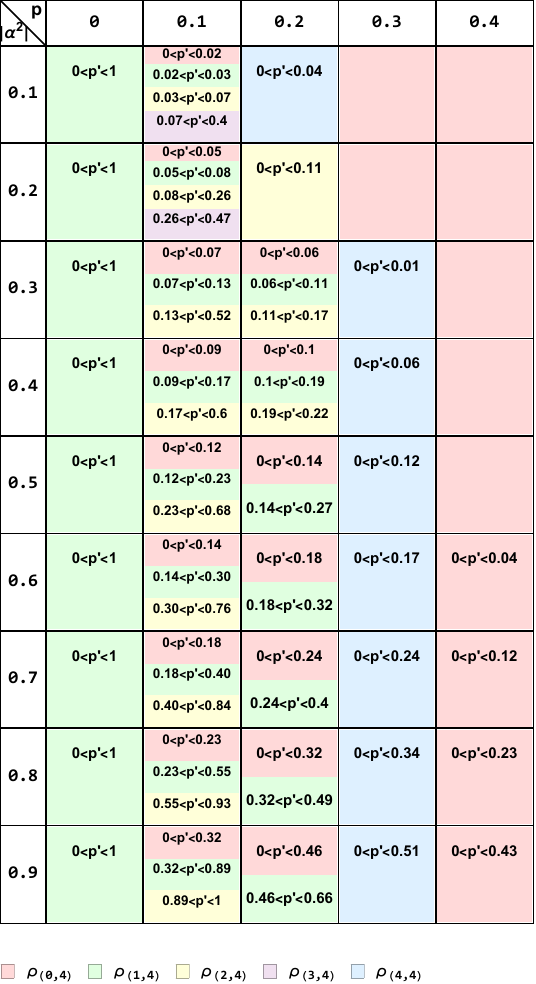} 
        \caption*{\small{(c) Evolution of state $\rho_{(m,4)}(\alpha,p,p')$.}}
    \label{Table18c}
    \end{subfigure}}
    \caption{Analysis of two-, three-, and four-qubit entangled state $\rho_{(m,n)}(\alpha,p,p')$ under the initial ADC $(p)$, NOT gates on $m$-qubits, and  subsequent ADC $(p')$.}
    \label{Table18}
\end{figure}

Tables shown in Fig.\,(\ref{Table18}) summarize the best strategy for protecting entanglement in $n$-qubit states $\rho_{(m,n)}(\alpha,p,p')$ using $m$-NOT gates applied for specific values of initial ADC ($p=0,0.1,\cdots,0.4$) and state parameters $(|\alpha|^2= 0,0.1, \cdots, 0.9$), where $p'$ is the effective range of second ADC. 
By explicitly calibrating when and how many NOT gates to apply, one can maximize the GME and its lifetime in the multi‐qubit state for the schematic described in the Fig.\,(\ref{fig_NOTScheme}).
For instance, Table~(\ref{Table18} a) shows the effect of no-, one-, and two-qubit NOT gates applied to the two-qubit states $\rho_{(m,2)}(\alpha,p,p')$. The red, green, and blue colour palettes indicate $m= 0,1$ and $2$, respectively.  
The range of $p'$ and the colour palette indicate the regime over which the shown colour palette denotes the best strategy that preserves the maximum amount of entanglement.
For example, for $|\alpha|^2=0.1$, a single-NOT gate works best at initial ADC, e.g., $p=0$ [see Fig.\,(\ref{fig06})], whereas two-NOT gates are more useful later on, i.e., at $p=0.1, 0.2, 0.3$. 
For $|\alpha|^2=0.3$, a single-NOT gate remains useful at $p=0$, whereas at $p=0.1$ a single NOT gate works best for $p'< 0.18$, and a two-qubit NOT gate for $p'> 0.18$.  As $p$ increases to $0.2, 0.3$, a two-qubit NOT gate again works best for smaller values of $\alpha$.  

Another key observation is that even though the two-qubit entangled states with $|\alpha|^2\geq 0.5$ undergo ADE, a single-NOT gate at $p=0$ remains useful in maintaining a higher amount of entanglement in the state throughout the decoherence process, which is also evident in Fig.\,(\ref{fig06}a). 
The results reported in tables\,(\ref{Table18}, b-c) for three- and four-qubit entangled states can also be interpreted similarly. The observation that NOT-gates are not only useful for states undergoing ESD but also for ADE states in preserving higher amounts of entanglement highlights the utility of such a scheme in a broader context. 

\subsection{Three‐qubit dynamics with ADC and NOT gates}
\label{Sec_Appendix-B}

Using the dynamics of multiqubit states under first ADC and NOT gates on $m$-qubits given by Eq.\,(\ref{Eq_06}), we obtain the states $\rho_{(m,3)}(\alpha,p)$ for $m=0,1,2,3$ as
\begin{equation} \label{Eq26}
\rho_{(0,3)}(\alpha,p) =
\left( \begin{array}{cccccccc}
 \alpha ^2+\beta ^2 p^3 & 0 & 0 & 0 & 0 & 0 & 0 & \alpha  \beta  q^{3/2} \\
 0 & \beta ^2 p^2 q & 0 & 0 & 0 & 0 & 0 & 0 \\
 0 & 0 & \beta ^2 p^2 q & 0 & 0 & 0 & 0 & 0 \\
 0 & 0 & 0 & \beta ^2 p q^2 & 0 & 0 & 0 & 0 \\
 0 & 0 & 0 & 0 & \beta ^2 p^2 q & 0 & 0 & 0 \\
 0 & 0 & 0 & 0 & 0 & \beta ^2 p q^2 & 0 & 0 \\
 0 & 0 & 0 & 0 & 0 & 0 & \beta ^2 p q^2 & 0 \\
 \alpha^{*}  \beta^{*}  q^{3/2} & 0 & 0 & 0 & 0 & 0 & 0 & \beta ^2 q^3 \\
\end{array} \right),
\end{equation}
\begin{equation} \label{Eq27}
\rho_{(1,3)}(\alpha,p) =
\left(\begin{array}{cccccccc}
 \beta ^2 p^2 q & 0 & 0 & 0 & 0 & 0 & 0 & 0 \\
 0 & \beta ^2 p q^2 & 0 & 0 & 0 & 0 & 0 & 0 \\
 0 & 0 & \beta ^2 p q^2 & 0 & 0 & 0 & 0 & 0 \\
 0 & 0 & 0 & \beta ^2 q^3 & \alpha  \beta  q^{3/2} & 0 & 0 & 0 \\
 0 & 0 & 0 & \alpha^{*}  \beta^{*}  q^{3/2} & |\alpha| ^2+|\beta| ^2 p^3 & 0 & 0 & 0 \\
 0 & 0 & 0 & 0 & 0 & \beta ^2 p^2 q & 0 & 0 \\
 0 & 0 & 0 & 0 & 0 & 0 & \beta ^2 p^2 q & 0 \\
 0 & 0 & 0 & 0 & 0 & 0 & 0 & \beta ^2 p q^2 \\
\end{array} \right),
\end{equation}
\begin{equation}
\rho_{(2,3)}(\alpha,p) =
\left(\begin{array}{cccccccc}
 \beta ^2 p q^2 & 0 & 0 & 0 & 0 & 0 & 0 & 0 \\
 0 & \beta ^2 q^3 & 0 & 0 & 0 & 0 & \alpha  \beta  q^{3/2} & 0 \\
 0 & 0 & \beta ^2 p^2 q & 0 & 0 & 0 & 0 & 0 \\
 0 & 0 & 0 & \beta ^2 p q^2 & 0 & 0 & 0 & 0 \\
 0 & 0 & 0 & 0 & \beta ^2 p^2 q & 0 & 0 & 0 \\
 0 & 0 & 0 & 0 & 0 & \beta ^2 p q^2 & 0 & 0 \\
 0 & \alpha^{*}  \beta^{*}  q^{3/2} & 0 & 0 & 0 & 0 & \alpha ^2+\beta ^2 p^3 & 0 \\
 0 & 0 & 0 & 0 & 0 & 0 & 0 & \beta ^2 p^2 q \\
\end{array}\right),
\end{equation}
\begin{equation}
\rho_{(3,3)}(\alpha,p) =
\left(\begin{array}{cccccccc}
 \beta ^2 q^3 & 0 & 0 & 0 & 0 & 0 & 0 & \alpha  \beta  q^{3/2} \\
 0 & \beta ^2 p q^2 & 0 & 0 & 0 & 0 & 0 & 0 \\
 0 & 0 & \beta ^2 p q^2 & 0 & 0 & 0 & 0 & 0 \\
 0 & 0 & 0 & \beta ^2 p^2 q & 0 & 0 & 0 & 0 \\
 0 & 0 & 0 & 0 & \beta ^2 p q^2 & 0 & 0 & 0 \\
 0 & 0 & 0 & 0 & 0 & \beta ^2 p^2 q & 0 & 0 \\
 0 & 0 & 0 & 0 & 0 & 0 & \beta ^2 p^2 q & 0 \\
 \alpha^{*}  \beta^{*}  q^{3/2} & 0 & 0 & 0 & 0 & 0 & 0 & \alpha ^2+\beta ^2 p^3 \\
\end{array} \right).
\end{equation}

To elucidate the action of local NOT gates on the GMC, we first identify its explicit dependence on density-matrix elements associated with GHZ-type coherence and competing populations, and then analyze how ADC modifies them. In this context, the NOT gate redistributes populations among computational-basis states, thereby modifying the subsequent entanglement dynamics. 
We adopt the standard three-qubit computational basis, labelling the qubit basis states $|b_1 b_2 b_3\rangle$ by their corresponding integer indices:
$|000\rangle \equiv |0\rangle,\; |001\rangle \equiv |1\rangle,\; |010\rangle \equiv |2\rangle,\; |011\rangle \equiv |3\rangle,\;
|100\rangle \equiv |4\rangle,\; |101\rangle \equiv |5\rangle,\; |110\rangle \equiv |6\rangle,\; |111\rangle \equiv |7\rangle$,
and define the density-matrix elements as $\rho_{ij} \equiv \langle i | \rho | j \rangle$, with $i,j = 0,\ldots,7$.

As an example, for the state in Eq.~(\ref{Eq26}), GMC is given as: $\mathrm{GMC}[\rho] = 2 \max\!\Big[ 0, \, |\rho_{07}| - \big(\sqrt{\rho_{11}\rho_{66}} + \sqrt{\rho_{22}\rho_{55}} + \sqrt{\rho_{33}\rho_{44}}\big) \Big]$
Thus, the state remains genuinely multipartite entangled as long as $|\rho_{07}| > \sqrt{\rho_{11}\rho_{66}} + \sqrt{\rho_{22}\rho_{55}} + \sqrt{\rho_{33}\rho_{44}}$.
The GHZ-type coherence is contained in the off-diagonal elements $\rho_{07}$ and $\rho_{70}$ (with $\rho_{70} = \rho_{07}^*$), while $\rho_{00}$ and $\rho_{77}$ correspond to populations of the ground and fully excited states, respectively.
Under ADC (e.g., in atomic systems), the coherence term $\rho_{07}$ decays exponentially as $\sim q^{3/2}$, where $q = 1 - p \sim e^{-t/\tau}$ and $\tau$ is the coherence time of the qubit. In parallel, population initially in $\rho_{77}$ decays as $\sim q^3$, cascading into doubly excited states ($\rho_{33}, \rho_{55}, \rho_{66}$), singly excited states ($\rho_{11}, \rho_{22}, \rho_{44}$), and ultimately the ground state ($\rho_{00}$).
The interplay between coherence suppression and population redistribution governs entanglement degradation and underlies ESD observed in many X-state models, wherein the residual coherence becomes insufficient relative to the competing population terms, causing the entanglement to vanish at a finite time \cite{ESD_Yu2004, ESD_Almeida2007}.

Applying a single NOT, for example, transforms the GHZ superposition as given in Eq.\,(\ref{Eq27}). The coherence terms now appear in $\rho_{34}$ and $\rho_{43}$, and the dominant ``GHZ-populations'' are moved to the doubly- and singly-excited sectors. Specifically, $\rho_{33} (\rho_{44})$ in Eq.\,(\ref{Eq27}) now contains the populations originating from  $|111\rangle (|000\rangle)$. GMC is now given as  $\mathrm{GMC}[\rho] = 2 \max\,[0,\; |\rho_{34}| -\left(\sqrt{\rho_{11} \rho_{66}} + \sqrt{\rho_{22}\rho_{55}} + \sqrt{\rho_{00} \rho_{77}} \right)]$. Coherence terms continue to decay at the same rate as before, but with the swapped population terms, their subsequent decay rates are modified \cite{ESDM_Rau2008}, thus altering the GMC death.
This is the crucial difference where the one-NOT gate moves population support into the double and single-excitation sector rather than leaving it concentrated in the fully excited and ground sectors. 
Previous theoretical analyses of X-states under ADC show that when population resides primarily in single-excitation entries, entanglement generally decays asymptotically rather than exhibiting a finite-time disappearance; in contrast, population concentrated in multi-excitation sectors promotes ESD \cite{ESD_Yu2009}. 

Further, using the second ADC and final NOT flips to get the state in `original configuration', we obtain the final states $\rho_{(m,3)}(\alpha,p,p')$ [see Eq.~(\ref{Eq_07})], for $m=0,1,2,3$, respectively, as 
\begin{equation}
\resizebox{\columnwidth}{!}{$
\left(
\begin{array}{cccccccc}
 \alpha ^2+\beta ^2 (p+p' q)^3 & 0 & 0 & 0 & 0 & 0 & 0 & \alpha  \beta  (q q')^{3/2} \\
 0 & \beta ^2 q q' (p+p' q)^2 & 0 & 0 & 0 & 0 & 0 & 0 \\
 0 & 0 & \beta ^2 q q' (p+p' q)^2 & 0 & 0 & 0 & 0 & 0 \\
 0 & 0 & 0 & \beta ^2 q^2 q'^2 (p+p' q) & 0 & 0 & 0 & 0 \\
 0 & 0 & 0 & 0 & \beta ^2 q q' (p+p' q)^2 & 0 & 0 & 0 \\
 0 & 0 & 0 & 0 & 0 & \beta ^2 q^2 q'^2 (p+p' q) & 0 & 0 \\
 0 & 0 & 0 & 0 & 0 & 0 & \beta ^2 q^2 q'^2 (p+p' q) & 0 \\
 \alpha^{*}  \beta^{*}  (q q')^{3/2} & 0 & 0 & 0 & 0 & 0 & 0 & \beta ^2 q^3 q'^3 \\
\end{array}
\right),
$}
\end{equation}
\begin{equation}
\resizebox{\columnwidth}{!}{$
\left(
\begin{array}{cccccccc}
 q' \left(\alpha ^2+\beta ^2 p (p+p' q)^2\right) & 0 & 0 & 0 & 0 & 0 & 0 & \alpha  \beta  (q q')^{3/2} \\
 0 & \beta ^2 p q q'^2 (p+p' q) & 0 & 0 & 0 & 0 & 0 & 0 \\
 0 & 0 & \beta ^2 p q q'^2 (p+p' q) & 0 & 0 & 0 & 0 & 0 \\
 0 & 0 & 0 & \beta ^2 p q^2 q'^3 & 0 & 0 & 0 & 0 \\
 0 & 0 & 0 & 0 & \beta ^2 p'^2 q \left(2 p^2+q^2\right)+p' \left(\alpha ^2+\beta ^2 p \left(p^2+2 q^2\right)\right)+\beta ^2 p^2 q+\beta ^2 p p'^3 q^2 & 0 & 0 & 0 \\
 0 & 0 & 0 & 0 & 0 & \beta ^2 q q' \left(p^2 p'+p \left(p'^2+1\right) q+p' q^2\right) & 0 & 0 \\
 0 & 0 & 0 & 0 & 0 & 0 & \beta ^2 q q' \left(p^2 p'+p \left(p'^2+1\right) q+p' q^2\right) & 0 \\
 \alpha^{*}  \beta^{*}  (q q')^{3/2} & 0 & 0 & 0 & 0 & 0 & 0 & \beta ^2 q^2 q'^2 (p p'+q) \\
\end{array}
\right),
$}
\end{equation}
\begin{equation}
\resizebox{\columnwidth}{!}{$
\left(
\begin{array}{cccccccc}
 q'^2 \left(\alpha ^2+\beta ^2 p^2 (p+p' q)\right) & 0 & 0 & 0 & 0 & 0 & 0 & \alpha  \beta  (q q')^{3/2} \\
 0 & \beta ^2 p^2 q q'^3 & 0 & 0 & 0 & 0 & 0 & 0 \\
 0 & 0 & q' \left(\beta ^2 p^2 p'^2 q+p' \left(\alpha ^2+\beta ^2 p \left(p^2+q^2\right)\right)+\beta ^2 p^2 q\right) & 0 & 0 & 0 & 0 & 0 \\
 0 & 0 & 0 & \beta ^2 p q q'^2 (p p'+q) & 0 & 0 & 0 & 0 \\
 0 & 0 & 0 & 0 & q' \left(\beta ^2 p^2 p'^2 q+p' \left(\alpha ^2+\beta ^2 p \left(p^2+q^2\right)\right)+\beta ^2 p^2 q\right) & 0 & 0 & 0 \\
 0 & 0 & 0 & 0 & 0 & \beta ^2 p q q'^2 (p p'+q) & 0 & 0 \\
 0 & 0 & 0 & 0 & 0 & 0 & \beta ^2 p^2 p'^3 q+p'^2 \left(\alpha ^2+\beta ^2 p \left(p^2+2 q^2\right)\right)+\beta ^2 p' q \left(2 p^2+q^2\right)+\beta ^2 p q^2 & 0 \\
 \alpha^{*}  \beta^{*}  (q q')^{3/2} & 0 & 0 & 0 & 0 & 0 & 0 & \beta ^2 q q' (p p'+q)^2 \\
\end{array}
\right),
$}
\end{equation}
\begin{equation}
\resizebox{\columnwidth}{!}{$
\left(
\begin{array}{cccccccc}
 q'^3 \left(\alpha ^2+\beta ^2 p^3\right) & 0 & 0 & 0 & 0 & 0 & 0 & \alpha  \beta  (q q')^{3/2} \\
 0 & q'^2 \left(p' \left(\alpha ^2+\beta ^2 p^3\right)+\beta ^2 p^2 q\right) & 0 & 0 & 0 & 0 & 0 & 0 \\
 0 & 0 & q'^2 \left(p' \left(\alpha ^2+\beta ^2 p^3\right)+\beta ^2 p^2 q\right) & 0 & 0 & 0 & 0 & 0 \\
 0 & 0 & 0 & q' \left(p'^2 \left(\alpha ^2+\beta ^2 p^3\right)+2 \beta ^2 p^2 p' q+\beta ^2 p q^2\right) & 0 & 0 & 0 & 0 \\
 0 & 0 & 0 & 0 & q'^2 \left(p' \left(\alpha ^2+\beta ^2 p^3\right)+\beta ^2 p^2 q\right) & 0 & 0 & 0 \\
 0 & 0 & 0 & 0 & 0 & q' \left(p'^2 \left(\alpha ^2+\beta ^2 p^3\right)+2 \beta ^2 p^2 p' q+\beta ^2 p q^2\right) & 0 & 0 \\
 0 & 0 & 0 & 0 & 0 & 0 & q' \left(p'^2 \left(\alpha ^2+\beta ^2 p^3\right)+2 \beta ^2 p^2 p' q+\beta ^2 p q^2\right) & 0 \\
 \alpha^{*}  \beta^{*}  (q q')^{3/2} & 0 & 0 & 0 & 0 & 0 & 0 & p'^3 \left(\alpha ^2+\beta ^2 p^3\right)+3 \beta ^2 p^2 p'^2 q+3 \beta ^2 p p' q^2+\beta ^2 q^3 \\
\end{array}
\right),
$}
\end{equation}
 The closed-form expressions in Eq.~(\ref{Eq_09}) and Eq.~(\ref{Eq_15}) given for these three-qubit X-states are used to obtain GMC and CQT fidelity, respectively. Aforementioned population-based picture helps intuitively understand why GMC display a better lifetime for the one-NOT protocol for a wide range of
 \(|\alpha|^2\): the NOT flip changes which matrix elements carry the coherence and population terms, and when the dominant population terms live in the single-excitation sector, the state is intrinsically more robust to ADC. This interpretation is consistent with earlier analytic and experimental studies of two-qubit X-state entanglement and sudden death under ADC.


\subsection{GHZ‐symmetric mapping of entangled states under ADC}

We are interested in classifying the behaviour of the $\rho_{(m,n)}(\alpha,p,p')$ states given in Eq.\,(\ref{fig07}) based on the GHZ‐symmetric state $x, y$-parametrization framework \cite{GHZ-Sym_Eltschka2012, TripartiteEnt_Siewert2012}. We discuss the procedure here for two-, three-, and four-qubit cases.
\subsubsection{GHZ‐symmetric mapping of two-qubit states}
\label{Sec_2qubitsymmetrization}

For two qubits, the GHZ symmetry is defined by invariance under (i) qubit swap, (ii) simultaneous Pauli‐$X$ flips $\sigma_x \otimes \sigma_x$ on both qubits, and (iii) compensating local $Z$‐rotations $U_{\rm 2qb}(\phi)=e^{i\phi\sigma_z}\otimes e^{-i\phi\sigma_z}$. Any two‐qubit arbitrary state $\rho_{(m,2)}(\alpha,p,p')$ can be projected into the GHZ‐symmetric family by applying a \emph{twirling} (symmetrization) over the GHZ symmetry group. Specifically, one forms the twirled state by symmetrization
\begin{equation}
\rho^S_{(m,2)}(\alpha,p,p') = \int_{0}^{2\pi}\frac{d\phi}{2\pi}\;U_{\rm 2qb}(\phi)\,\rho_{(m,2)}(\alpha,p,p')\,U_{\rm 2qb}(\phi)^\dagger,
\end{equation}
and similarly, averaging over the discrete flip and swap operations.  The twirled state ($\rho^S$) commutes with all GHZ symmetry operations, and thus has only four nonzero density‐matrix elements: $\rho^S_{00,00}=\rho^S_{11,11}$, $\rho^S_{01,01}=\rho^S_{10,10}$, and real off‐diagonal $\rho^S_{01,10}=\rho^S_{10,01}$.  Subject to unit trace, this two‐qubit GHZ‐symmetric form is fully characterized by two real parameters $x$ and $y$, defined by the projections onto the Bell states $\ket{\Phi^\pm}=(\ket{00}\pm\ket{11})/\sqrt{2}$) \cite{TripartiteEnt_Siewert2012}. The $x, y$-parameters for the two-qubit state are given as
\begin{equation}
\begin{aligned}
    x(\rho^S_{(m,2)}(\alpha,p,p')) & =\frac{1}{2} \Bigl[\bra{\Phi^+}\rho^S_{(m,2)}(\alpha,p,p')\ket{\Phi^+} - \bra{\Phi^-}\rho^S_{(m,2)}(\alpha,p,p')\ket{\Phi^-}\Bigr], \\
y(\rho^S_{(m,2)}(\alpha,p,p')) &=\frac{1}{\sqrt{2}} \Bigl[\bra{\Phi^+}\rho^S_{(m,2)}(\alpha,p,p')\ket{\Phi^+} + \bra{\Phi^-}\rho^S_{(m,2)}(\alpha,p,p')\ket{\Phi^-} - \tfrac12\Bigr]\,. 
\end{aligned}
\end{equation}
Physically, $x$ measures the difference in weight of the state on the two Bell states $\ket{\Phi^+}$ and $\ket{\Phi^-}$, while $y$ measures the total weight on those maximally entangled components (above the uniform background). The normalization is chosen such that $(x,y)$ coordinates follow the Euclidean metric under the Hilbert–Schmidt norm \cite{GHZ-Sym_Eltschka2012, TripartiteEnt_Siewert2012}.


\subsubsection{GHZ‐symmetric mapping of three‐qubit states}
\label{Sec_3qubitsymmetrization}

For arbitrary three qubit states $\rho_{(m,3)}(\alpha,p,p')$, the GHZ symmetry is defined by invariance under (i) any permutation of the three qubits, (ii) simultaneous bit flips $ \sigma_x\otimes \sigma_x\otimes \sigma_x$, and (iii) correlated $Z$-rotations $U_{\rm 3qb}(\phi_1,\phi_2)=e^{i\phi_1\sigma_z}\otimes e^{i\phi_2\sigma_z}\otimes e^{-i(\phi_1+\phi_2)\sigma_z}$ \cite{GHZ-Sym_Eltschka2012}. Twirling the three‐qubit state over this GHZ symmetry group (integrating over $\phi_1,\phi_2$ and summing the discrete symmetries) projects it onto a GHZ‐symmetric form as
\begin{equation}
\rho^S_{(m,3)}(\alpha,p,p')=\int dU\;U_{\rm 3qb}\rho_{(m,3)}(\alpha,p,p')U_{\rm 3qb}^\dagger,
\end{equation}
where the integral runs over the full GHZ symmetry group of unitary transformations.  The twirled or symmetrized state ($\rho^S_{(m,3)}(\alpha,p,p')$) necessarily lies in the GHZ‐symmetric two‐parameter ($x,y$) family. It is fully specified by its overlaps with the GHZ basis states $\left(\ket{GHZ^\pm}=(\ket{000}\pm\ket{111})/\sqrt{2}\right)$ as
\begin{equation}
\begin{aligned}
x(\rho^S_{(m,3)}(\alpha,p,p')) &= \frac12\Bigl[\langle GHZ^+|\rho^S_{(m,3)}(\alpha,p,p')|GHZ^+\rangle - \langle GHZ^-|\rho^S_{(m,3)}(\alpha,p,p')|GHZ^-\rangle \Bigr], \\
y(\rho^S_{(m,3)}(\alpha,p,p')) &=\frac{1}{\sqrt{3}}\Bigl[\langle GHZ^+|\rho^S_{(m,3)}(\alpha,p,p')|GHZ^+\rangle + \langle GHZ^-|\rho^S_{(m,3)}(\alpha,p,p')|GHZ^-\rangle - \tfrac14\Bigr]\,.
\end{aligned}
\end{equation}
Geometrically, the set of all GHZ‐symmetric three‐qubit density matrices forms a convex triangle in the $(x,y)$ plane. The completely mixed three‐qubit state is at the origin. The top corners of this triangle are the pure GHZ states $\ket{GHZ^+},\ket{GHZ^-}$, while points along the top horizontal edge are mixtures of $\ket{000}\bra{000}$ and $\ket{111}\bra{111}$.  In particular, the midpoint of the top edge at $(x=0,y=\tfrac{\sqrt3}{4})$ is the \emph{fully separable} mixture $\tfrac12(\ket{000}\bra{000}+\ket{111}\bra{111})$.  Points below that edge represent biseparable or W‐type states. These classifications can be explicitly observed in the Section\,(\ref{Sec_Results_3QubitClass}).


\subsubsection{GHZ‐symmetric mapping of four‐qubit states}
\label{Sec_4qubitsymmetrization}

In analogy with the two- and three-qubit cases, we consider here the GHZ-symmetric projection for arbitrary four-qubit states $\rho_{(m,4)}(\alpha,p,p')$ \cite{GHZ-SymClass_Park2014}. The full four-qubit GHZ symmetry group is defined by invariance under: (i) all permutations of the four qubits, (ii) simultaneous Pauli-\(X\) flips \(\sigma_x^{\otimes 4}\) and (iii) correlated \(Z\)-rotations of the form $U_{\rm 4qb}(\phi_1,\phi_2,\phi_3)
    = e^{i\phi_1 \sigma_z}\otimes e^{i\phi_2 \sigma_z}\otimes e^{i\phi_3 \sigma_z}\otimes e^{-i(\phi_1+\phi_2+\phi_3)\sigma_z}.$ Applying a twirling (symmetrization) over this GHZ symmetry group projects the four-qubit state into the GHZ-symmetric family as
\begin{equation}
  \rho^S_{(m,4)}(\alpha,p,p')
  = \int dU \; U_{\rm 4qb}\,\rho_{(m,4)}(\alpha,p,p')\,U_{\rm 4qb}^\dagger,
\end{equation}
where the integral runs over the continuous angles \(\phi_1, \phi_2, \phi_3\) and the sum over all qubit permutations and global \(X\)-flips.  By construction, the resulting twirled state commutes with every GHZ generator and hence is \emph{GHZ-symmetric}.

Notably, the most general GHZ-symmetric four-qubit density matrix has only the corner off-diagonals and a constrained diagonal:
\begin{equation}
\label{eq:4q-ghz-form}
\begin{aligned}
\rho_4^S &= \tilde{x}\bigl(\lvert 0000\rangle\langle 1111\rvert + \lvert 1111\rangle\langle 0000\rvert\bigr)  + \mathrm{diag}\bigl(\alpha_1,\alpha_2,\alpha_2,\alpha_3,\;\alpha_2,\alpha_3,\alpha_3,\alpha_2,\;\alpha_2,\alpha_3,\alpha_3,\alpha_2,\;\alpha_3,\alpha_2,\alpha_2,\alpha_1\bigr),
\end{aligned}
\end{equation}
with real parameters $\tilde{x},\alpha_1,\alpha_2,\alpha_3$, satisfying $\alpha_1 + 4\alpha_2 + 3\alpha_3 = \tfrac{1}{2}$.
Thus, the GHZ-symmetric family is characterized by \emph{three} independent real parameters, in contrast to the two-parameter families for two- and three-qubit GHZ symmetry \cite{GHZ-SymClass_Park2014}. To endow this parameter space with a Euclidean metric, new coordinates are defined $(\tilde x,\tilde y,\tilde z)$ via a suitable linear transformation. Geometrically, the allowed GHZ-symmetric states form a tetrahedron in $(\tilde x,\tilde y,\tilde z)$-space, with the maximally mixed state at the origin and the pure GHZ states \(\ket{GHZ_4^\pm}=(\ket{0000}\pm\ket{1111})/\sqrt2\) at opposite vertices.
The $(\tilde x,\tilde y,\tilde z)$ coordinates themselves are expressed in terms of overlaps with the GHZ-basis and the auxiliary state $\ket{\Psi_4}=(\ket{0001}+\ket{1110})/\sqrt2$ as

\begin{equation}
\begin{aligned}
\tilde{x}(\rho_{(m,4)}^S(\alpha,p,p'))  = & \frac{1}{2}\bigl(\langle GHZ_4^+|\rho_{(m,4)}^S(\alpha,p,p')|GHZ_4^+\rangle - \langle GHZ_4^-|\rho_{(m,4)}^S(\alpha,p,p')|GHZ_4^-\rangle\bigr),\\
\tilde{y}(\rho_{(m,4)}^S(\alpha,p,p')) = & \frac{N_1}{2}\bigl[\langle GHZ_4^+|\rho_{(m,4)}^S(\alpha,p,p')|GHZ_4^+\rangle + \langle GHZ_4^-|\rho_{(m,4)}^S(\alpha,p,p')|GHZ_4^-\rangle
  \\ &+ 2(\sqrt{10}+3)\,\langle \Psi_4|\rho_{(m,4)}^S (\alpha,p,p')|\Psi_4\rangle\bigr],\\
\tilde{z}(\rho_{(m,4)}^S(\alpha,p,p')) =& \frac{N_2}{2}\bigl[(\sqrt{10}+3)\bigl(\langle GHZ_4^+|\rho_{(m,4)}^S (\alpha,p,p')|GHZ_4^+\rangle - \langle GHZ_4^-|\rho_{(m,4)}^S(\alpha,p,p') |GHZ_4^-\rangle\bigr) 
\\ & - 2\,\langle \Psi_4|\rho_{(m,4)}^S(\alpha,p,p')|\Psi_4\rangle\bigr],
\end{aligned}
\end{equation}
where the normalization constants $N_1 = \sqrt{\frac{2}{3} - \frac{2\sqrt{10}}{15}}$ and $N_2 = \sqrt{\frac{14}{3} - \frac{22\sqrt{10}}{15}}$.
The full four-qubit GHZ-symmetric family is three-dimensional, and unlike the two- and three-qubit cases, no complete SLOCC entanglement classification of this three-parameter set $(\tilde x,\tilde y,\tilde z)$ has yet been achieved \cite{Multi2015}.

\acknowledgements
RS acknowledges partial financial support of the Anusandhan National Research Foundation (ANRF) grant CRG/2022/008345. VA thanks Aditi Das at Qdit Labs Pvt. Ltd. for useful discussions. 

\bibliographystyle{apsrev4-2}
\bibliography{Reference}
\end{document}